\documentclass[10pt,twocolumn,superscriptaddress,english,prb,showpacs,floatfix,aps]{revtex4-2}

\usepackage[utf8]{inputenc}
\usepackage{babel}
\usepackage[normalem]{ulem}

\usepackage{amsmath}
\usepackage{amssymb}
\usepackage{mathtools}
\usepackage{graphicx}
\usepackage{physics} 
 
\usepackage[backref=none,
bookmarksnumbered=true,
bookmarks=true,
bookmarksopen=true,
colorlinks=true,
citecolor=blue,
linkcolor=blue,
anchorcolor=green,
urlcolor=blue,unicode=false]{hyperref}


\begin{document}

\title{Dynamics of Josephson junctions beyond the tunneling limit}

\author{Jacob F.~Steiner}
\affiliation{Department of Physics and Institute for Quantum Information and Matter, California Institute of Technology, Pasadena, California 91125, USA}

\author{Larissa Melischek}
\affiliation{\mbox{Dahlem Center for Complex Quantum Systems and Fachbereich Physik, Freie Universit\"at Berlin, 14195 Berlin, Germany}}

\author{Felix von Oppen}
\affiliation{\mbox{Dahlem Center for Complex Quantum Systems, Halle-Berlin-Regensburg Cluster of Excellence CCE,}\\ and Fachbereich Physik, Freie Universit\"at Berlin, 14195 Berlin, Germany}

\begin{abstract}
The dynamics of the superconducting phase difference across a Josephson junction can be described within the resistively and capacitively shunted Josephson junction (RCSJ) model. Microscopic derivations of this model traditionally rely on the tunneling limit. Here, we present a derivation of a generalized version of the RCSJ model, which accounts for dissipative currents with nonlinear current-voltage characteristics as well as supercurrents with arbitrary current-phase relations. This requires a generalized fluctuation-dissipation theorem to describe the Langevin current, which we deduce along the lines of fluctuation theorems for mesoscopic conductors. 
Our work is motivated in particular by recent theories of the Josephson diode effect, which is not captured within the RCSJ model in the tunneling limit. 
\end{abstract}


\maketitle 

\section{Introduction}

There has been much recent interest in the nonreciprocal response of superconductors  \cite{nadeem2023superconducting,nagaosa2024nonreciprocal}. In particular, numerous experiments on various types of Josephson junctions have demonstrated bias-direction-dependent switching or retrapping currents, a phenomenon referred to as Josephson diode effect \cite{baumgartner2021supercurrent,diez2023symmetry,wu2022field,bauriedl2022supercurrent,pal2022josephson,jeon2022zero,turini2022josephson,gupta2023gate,chiles2023nonreciprocal,zhang2025evidence,mazur2024gate,ciaccia2023gate,trahms2023diode,reinhardt2024link,ghosh2024high,lotfizadeh2024superconducting,kim2024intrinsic,chen2024edelstein,kudriashov2025non,yan2025gate}.  

These experiments typically exhibit a hysteretic voltage response to an applied bias current. As the bias increases, the transition from the supercurrent to the resistive branch occurs at the switching current, which is larger than the retrapping current, at which the junction switches back into the supercurrent state when the bias current is reduced. Depending on the experiment, the dominant asymmetry with respect to the bias-current direction appears either in the switching \cite{wu2022field,baumgartner2021supercurrent,diez2023symmetry,bauriedl2022supercurrent,pal2022josephson,jeon2022zero,turini2022josephson,gupta2023gate,chiles2023nonreciprocal} or the retrapping current \cite{trahms2023diode,yan2025gate}. A dominant asymmetry in the switching current is associated with asymmetric current-phase relations and therefore requires broken time-reversal symmetry. Much effort has been devoted to elucidating the underlying physics \cite{dolcini2015topological,chen2018asymmetric,zhang2022general,kopasov2021geometry,davydova2022universal,souto2022josephson,tanaka2022theory,lu2023tunable,maiani2023nonsinusoidal,hu2023josephson,costa2023microscopic,cayao2024enhancing,meyer2024josephson,liu2024josephson,zazunov2023nonreciprocal,volkov2024josephson,lu2024varphi,wang2024efficient,cheng2024field,shaffer2025josephson,zhuang2025helical,patil2025spin,shen2025josephson,costa2025unconventional,cuozzo2025perfect,boruah2025field}, see \cite{shaffer2025theories} for a recent review. Dominant asymmetries in the retrapping current originate from asymmetric quasiparticle currents, akin to conventional diodes, and may occur in time-reversal symmetric junctions \cite{trahms2023diode,steiner2023diode}. 

Theoretical descriptions of switching and retrapping processes commonly rely on the resistively and capacitively shunted junction (RCSJ) model  \cite{stewart1968current,mccumber1968effect} for the phase dynamics of the Josephson junction. In its simplest form, this model assumes a sinusoidal current-phase relation and an Ohmic quasiparticle resistance. This predicts symmetric behavior of both the switching and the retrapping current. Nonreciprocity can be accounted for in a generalized RCSJ model \cite{stewart1974current,kautz1990noise,steiner2023diode}, which allows for arbitrary current-phase relations and nonlinear current-voltage characteristics of the quasiparticle current. While this generalized model is supported by compelling phenomenological arguments, it has not been derived microscopically along the lines of the seminal work of Ambegaokar, Eckern, and Schön \cite{ambegaokar1982quantum,eckern1984quantum}. 

In view of the importance of the generalized RCSJ model for understanding nonreciprocity in Josephson junctions, it seems desirable to complement the  phenomenological approach to the generalized RCSJ model  by a microscopic derivation. To this end, we extend the theory of Ambegaokar, Eckern, and Schön \cite{ambegaokar1982quantum,eckern1984quantum} in two fundamental ways. First, we retain the tunneling operator to all orders. This is essential because in transparent junctions higher-order processes, such as multiple Andreev reflections \cite{andreev1964thermal,klapwijk1982explanation,octavio1983subharmonic,averin1995ac,cuevas1996hamiltonian}, dominate the dissipative current at subgap voltages  \cite{trahms2023diode}. Second, to arrive at a simple, time-local equation of motion, we perform an adiabatic expansion which systematically neglects time derivatives of the voltage. We argue that this approximation is justified if the junction capacitance is sufficiently large, consistent with the semiclassical regime. The resulting generalized RCSJ equation includes a dissipative term, which involves the  quasiparticle current \cite{cuevas1996hamiltonian} evaluated at a voltage bias given by the instantaneous rate of change of the phase across the junction. 

To incorporate the effects of noise, the RCSJ model is frequently supplemented by a fluctuating current source. Within linear response, the fluctuations of the current are linked to the quasiparticle resistance through the Johnson-Nyquist fluctuation-dissipation relation. In our approach, the fluctuating current enters naturally as the variable conjugate to the quantum component of the phase difference across the junction (i.e., the component which is odd in the Keldysh contour). The current correlations are generally non-Gaussian, reflecting the full counting statistics of the tunneling current. However, 
they can be approximated as Gaussian in many physical situations as a consequence of the central-limit theorem, in particular when the junction dynamics are slow. We show that within the adiabatic regime, the correlations of the Langevin current reduce to the current correlations under a voltage bias \cite{cuevas1999shot} proportional to the instantaneous rate of change of the phase. 

Importantly, we explicitly include nonlinear dissipative currents. This implies that the usual equilibrium fluctuation-dissipation relation no longer suffices to characterize the noise. Based on a general symmetry of the action \cite{tobiska2005inelastic,saito2008symmetry}, we show that one can still relate 
the noise power of the current fluctuations to
the nonlinear dissipative current, providing a simple and intuitive extension of Johnson-Nyquist noise and supporting an earlier argument relying on detailed balance \cite{steiner2023diode}. 
Our relation only requires that temperature be sufficiently large compared to the voltage. It makes no assumptions on the ratio between the  voltage and other scales of the nonlinear current-voltage characteristic of the quasiparticle current. This fact seems not to be appreciated in the literature on mesoscopic fluctuation relations \cite{esposito2009nonequilibrium,campisi2011colloquium}. Our result can also be viewed as an extension of these fluctuation relations to superconducting transport. One useful implication of the nonlinear fluctuation-dissipation theorem is that one can use a measured current-voltage characteristic as input for the generalized RCSJ model \cite{trahms2023diode}. 

The remainder of this paper is organized as follows. We start with a physical discussion of the main results in Sec.~\ref{sec:generalized_rcsj_model}. This section can be read independently of the detailed microscopic derivation of the generalized RCSJ equation for the superconducting phase difference across the junction in Sec.~\ref{sec:microscopic_derivation}. In Sec.~\ref{sec:adiabatic}, we employ the adiabatic approximation to show how the generalized RCSJ model can be obtained explicitly and connect our results to previous discussions in the literature. Section~\ref{sec:nonlinear_fdt} discusses the nonlinear extension of the fluctuation-dissipation relation for voltage-biased Josephson junctions which relates the dissipative and noise currents entering the generalized RCSJ equation. We conclude in Sec.~\ref{sec:conclusions}. Background material and some details are relegated to appendices. 

\section{Generalized RCSJ equation and summary of main results}
\label{sec:generalized_rcsj_model}

The conventional RCSJ model assumes that current through the Josephson contact is carried by three parallel channels, a capacitive current $I_\textrm{c}=C\dot V$ associated with the junction capacitance $C$, a dissipative current $I_\textrm{d}=V/R$ associated with an Ohmic shunt resistor $R$, and a supercurrent $I_\textrm{s}=I_\textrm{c}\sin\varphi$ due to the superconducting phase difference $\varphi$ across the junction. Current conservation implies that these currents have to add to the bias current $I_\textrm{b}$ applied to the junction,
\begin{equation}
I_\textrm{c}+I_\textrm{d}+I_\textrm{s} + \xi = I_\textrm{b}.
\end{equation}
This expression includes Gaussian Johnson-Nyquist fluctuations $\xi$ of the Ohmic current characterized by  
\begin{equation}
  \langle\xi(t)\rangle = 0\quad , \quad \langle\xi(t)\xi(t')\rangle = \frac{2T}{R}\delta(t-t').  
\end{equation} 
Combining this with the Josephson relation 
\begin{equation}
    V = \frac{\hbar}{2e}\dot \varphi
    \label{eq:JosephsonRel}
\end{equation}
for the voltage drop $V$ across the junction, 
one obtains a Langevin equation
\begin{equation}
    \frac{\hbar C}{2e} \ddot \varphi + \frac{\hbar }{2eR} \dot \varphi + I_\textrm{c}\sin\varphi + \xi = I_\textrm{b}
\end{equation}
for the superconducting phase difference $\varphi$. In this conventional version, the RCSJ model
predicts reciprocal Josephson currents as neither the 
critical current $I_\textrm{c}$ nor the friction due to the Ohmic resistor depend on the directions of current or voltage. 

Nonreciprocal Josephson currents can be accounted for within a generalized RCSJ model
\begin{equation}\label{eq:generalized_rcsj}
\frac{\hbar C}{2e} \ddot \varphi +I_\textrm{d}\left(\frac{\hbar}{2e}\dot \varphi\right)+I_\textrm{s}(\varphi) + \xi = I_\textrm{b},
\end{equation}
which allows for general current-phase relations $I_\textrm{s}(\varphi)$ and dissipative currents $I_\textrm{d}(V)$. Focusing on the hysteretic response in the
weak-damping limit, it has been shown  that asymmetric switching currents originate in an asymmetric current-phase relation, $I_\textrm{s}(\varphi)\neq -I_\textrm{s}(-\varphi)$, while asymmetric retrapping currents originate in an asymmetric dissipative current, $I_\textrm{d}(V)\neq -I_\textrm{d}(-V)$ \cite{steiner2023diode}.

Prior to the recent studies of nonreciprocal Josephson junctions,
this generalized RCSJ model has been repeatedly used in the literature \cite{stewart1974current,kautz1990noise}. While it is natural to equate $I_\textrm{s} (\varphi)$ with the equilibrium current-phase relation of the junction, the microscopic forms of the dissipative current $I_\textrm{d}(V)$ and the associated noise correlations $\xi$ entering into Eq.~\eqref{eq:generalized_rcsj} are not immediately obvious. Clearly, $I_\textrm{d}(V)$ and $\xi$ have their origin in quasiparticle tunneling processes, which, in junctions with significant transparencies, involve higher-order processes such as multiple Andreev reflections. The seminal microscopic derivation of the phase action governing the junction by Ambegaokar, Eckern, and Schön \cite{ambegaokar1982quantum,eckern1984quantum} is restricted to the limit of weak tunneling. This can account for nonlinear dissipative currents, but fails to account for higher harmonics of the current-phase relation $I_\textrm{s}(\varphi)$ as well as for higher-order tunneling processes such as multiple Andreev reflections contributing to the dissipative current. Both of these features are generally important in the context of Josephson diodes.

Our main result is a microscopic derivation of the generalized RCSJ model, which takes into account processes to all orders in the tunneling amplitude. We first derive a time-\textit{non}local version of the generalized RCSJ equation, which takes the form 
\begin{equation}\label{eq:generalized_rcsj_nonlocal}
    \frac{\hbar C}{2e}\ddot\varphi(t) + \mathcal{I}([\varphi];t)+\xi(t) = I_\textrm{b},
\end{equation}
Here, $\mathcal{I}([\varphi];t)$ is the tunneling current for a given time-dependent phase bias $\varphi(t)$ across the junction, encompassing both the supercurrent as well as the dissipative current. The Langevin current $\xi(t)$ originates from fluctuations of the tunneling current and in general has non-Gaussian correlations. However, we argue that in many situations, Gaussian correlations are an excellent approximation, so that $\xi$ can be characterized by its mean and variance,
\begin{equation}
    \langle\xi(t)\rangle = 0 \quad , \quad \ev{\xi(t) \xi(t')} = \mathcal{K}([\varphi];t,t').
\end{equation}
We give explicit expressions for $\mathcal{I}([\varphi];t)$ and 
$\mathcal{K}([\varphi];t,t')$ in Sec.~\ref{sec:explicit_expressions_for_I_and_K} below. Note that we write the dependence on the phase bias $\varphi(t)$ in square brackets to emphasize that $\mathcal{I}$ and $\mathcal{K}$ depend on the full history of the phase bias. 

Time-nonlocal equations of motion for the phase difference such as Eq.~\eqref{eq:generalized_rcsj_nonlocal} are well-known in the RCSJ literature \cite{werthamer1966nonlinear,rossignol2019role,choi2022microscopic,lahiri2023nonequilibrium,lahiri2025origin}. While they describe the exact junction dynamics, memory effects can often be neglected. In particular, for a sufficiently large junction capacitance, the phase dynamics becomes adiabatic compared to microscopic time scales. In such situations, one can derive a more intuitive time-local description, with the resulting generalized RCSJ equation containing the dissipative current evaluated at the instantaneous junction voltage. 

We arrive at the time-local description by using an adiabatic approximation which systematically neglects higher time derivatives $\ddot\varphi$, $\dddot\varphi$, etc.~of the phase difference in calculating $\mathcal{I}([\varphi];t)$ and $\mathcal{K}([\varphi];t,t')$ \cite{nazarov2007full}. This gives 
\begin{subequations}    
\begin{align}
    \mathcal{I}([\varphi];t) \simeq  &\ \mathcal{I}(\varphi(t), \hbar\dot{\varphi}(t)/2e) \\ =&\ \sum_n e^{-in\varphi(t)} \mathcal{I}_n(\hbar \dot{\varphi}(t)/2e), \\
    \mathcal{K}([\varphi];t,t') \simeq&\ \mathcal{K}(\varphi(t),\hbar \dot{\varphi}(t)/2e;t-t') \\ =&\ \sum_n e^{-in\varphi(t)} \mathcal{K}_n( \hbar \dot{\varphi}(t)/2e;t-t').
\end{align}
\end{subequations}
The expansion coefficients $\mathcal{I}_n(V)$ and $\mathcal{K}_n(V;t-t')$ are equal to the Fourier components of the current and noise \cite{cuevas1996hamiltonian,cuevas1999shot} for a voltage-biased junction. Within the adiabatic approximation, we can take the noise correlator to have a white-noise form,
\begin{equation}
    \mathcal{K}(\varphi(t),\hbar \dot{\varphi}(t)/2e;t-t') \simeq \mathcal{K}(\varphi(t),\hbar \dot{\varphi}(t)/2e) \delta(t-t'),
\end{equation}
in terms of the noise power 
\begin{equation}
    \mathcal{K}(\varphi,V) = \int dt\, \mathcal{K}(\varphi,V;t) = \sum_n e^{-in\varphi} \mathcal{K}_n(\varphi,V).
\end{equation}
To make the connection to the RCSJ model, we decompose the current as
\begin{equation}
    \mathcal{I}(\varphi,V) = I_\textrm{s}(\varphi) + I_\textrm{d}(V) + I_\textrm{m}(\varphi,V).
\end{equation}
Here, the first and second terms are the supercurrent 
\begin{equation}I_\textrm{s}(\varphi) = \sum_n e^{-in\varphi} \mathcal{I}_n(0)
\label{eq:supcurrIs}
\end{equation}
and the dissipative current 
\begin{equation}I_\textrm{d}(V) = \mathcal{I}_0(V),
\label{eq:disscurrId}
\end{equation}respectively.
Typically, these two terms are retained in the RCSJ model in Eq.~\eqref{eq:generalized_rcsj}. Our derivation shows that in principle, there are also ``mixed" terms 
$I_\textrm{m} = \mathcal{I} -I_\textrm{s} - I_\textrm{d}$, which reflect that the supercurrent will in general depend on the bias voltage and the dissipative current on the phase. 

The noise correlations are related to the dissipative current by a nonlinear generalization of the Johnson-Nyquist fluctuation-dissipation theorem. Assuming $T \gg eV$ but making no assumption on the relative magnitude of $eV$ and other scales, we find that the $dc$ noise power $K_\textrm{d} (V) = \mathcal{K}_0(V)$ can be expressed through $I_\textrm{d}(V)$
\begin{equation}
    K_\textrm{d} (V) \simeq \frac{2T}{V}I_\textrm{d}(V).
\end{equation}
This relation, which holds beyond the linear-response regime, had been previously argued for based on detailed-balance arguments \cite{steiner2023diode}
and supersedes ad-hoc assumptions in earlier  discussions of the generalized RCSJ model \cite{kautz1990noise}. In time-reversal symmetric junctions, this relation extends to the mixed terms which contribute to dissipation. Concretely, we find
\begin{equation}
   \Re[\mathcal{K}_n(V)] = \frac{2T}{V} \Re[\mathcal{I}_n(V)],
\end{equation}
which relates the coefficients of the $\cos(n  \varphi)$ oscillations in the noise power and the current.  

In general, the dissipative current and its associated fluctuations will also include contributions from the electromagnetic environment of the junction \cite{ingold1992single}. While their inclusion is frequently essential to reproduce the dynamics of real junctions \cite{kautz1990noise}, we focus here on the microscopic contributions intrinsic to the junction. 

\section{Microscopic derivation of the time-nonlocal generalized RCSJ equation}
\label{sec:microscopic_derivation}

\subsection{Model}
\label{sec:model}
We model a Josephson junction between two superconducting electrodes ($a=L,R$) by the Hamiltonian $H =  H_L + H_R +  H_T + H_C$. The electrodes are described by  
\begin{eqnarray}
 H_a &=& \int d\mathbf{r}\,\sum_{\sigma}\psi^\dagger_{a,\sigma}(\mathbf{r})
\xi^{\phantom{\dagger}}_{a 
}
(-i\nabla) \psi^{\phantom{\dagger}}_{a,\sigma}(\mathbf{r})
 \nonumber\\
 &&- \lambda_a \int d\mathbf{r}\,\psi^\dagger_{a,\uparrow}(\mathbf{r})\psi^\dagger_{a,\downarrow}(\mathbf{r})\psi^{\phantom{\dagger}}_{a,\downarrow}(\mathbf{r})\psi^{\phantom{\dagger}}_{a,\uparrow}(\mathbf{r})
\end{eqnarray}
with attractive interaction of strength $\lambda_a$ and single-particle dispersion $\xi_a(\mathbf{k})$. The operators $\psi_{a,\sigma}$ are fermion fields annihilating electrons in lead $a$ with spin $\sigma$. For simplicity of presentation, we assume that spin is a good quantum number. We note that our theory readily extends to spin-orbit coupled superconductors, provided that they are well-described by mean field theory with a one-component order parameter. Tunneling of electrons between the electrodes is accounted for by 
\begin{equation} 
H_{\textrm{tun}}=\vartheta\sum_\sigma \psi^\dagger_{R,\sigma}({\bf 0}) \psi^{}_{L,\sigma}({\bf 0})+{\rm h.c.},
\end{equation}
where the tunneling amplitude $\vartheta$ is assumed to be real and point-like at the origin (as, e.g., in a scanning-tunneling-microscope setup)  \footnote{It may be possible to relax this assumption by adapting insights from Nazarov's Keldysh action for full counting statistics in terms of the scattering matrix \cite{snyman2008keldysh}.}. We denote the associated tunneling current operator by 
\begin{equation}
  \mathcal{I}_\textrm{tun} = - i e \vartheta\sum_\sigma \psi^\dagger_{R,\sigma}({\bf 0}) \psi^{}_{L,\sigma}({\bf 0})+{\rm h.c.}\,    .
\end{equation}
We also include a capacitive coupling
\begin{equation}
    H_C = \frac{1}{8C}(Q_L-Q_R)^2
    \label{eq:capac}
\end{equation}
between the electrodes, where 
\begin{equation}
    Q_a = e \int d\mathbf{r} \sum_\sigma \psi^\dagger_{a,\sigma}(\mathbf{r})
\psi^{\phantom{\dagger}}_{a,\sigma}(\mathbf{r})
\end{equation}
is the total charge in electrode $a$ and $C$ denotes the geometric capacitance of the junction. The total  Hamiltonian $H$ conserves the combined charge $Q_L+Q_R$.

\subsection{Keldysh action for the phase difference}
\label{sec:Keldysh}

We formulate the theory in terms of the Keldysh generating functional
\cite{kamenev2011field}
\begin{equation}
    \mathcal{Z}[\eta] = \int [d\overline\psi] [d\psi]\,
    \exp{iS_\eta[\overline\psi,\psi]}
\end{equation}
with action
\begin{equation}
    S_\eta[\overline\psi,\psi] = \int_\mathcal{C} dt\, \,  \bqty{\, \overline\psi(t)  i\partial_t  \psi(t) - H_\eta(\overline\psi,\psi) },
    \label{eq:actionfull}
\end{equation}
setting $\hbar = 1$. The contour-odd source field
$\eta$ couples to an observable $\cal O$ via $H_\eta = H \pm_\mathcal{C} \tfrac{1}{2}\eta(t)\cal O$, with opposite signs for the forward ($+$) and backward ($-$) parts of the Keldysh contour $\mathcal{C}$. Averages of the observable $\mathcal O$ with respect to the density matrix at time $t$ can then be obtained as functional derivatives, 
\begin{equation}
    \ev{\mathcal O(t)} =
    i
    \left. \frac{\delta }{\delta\eta(t)}\ln \mathcal{Z}[\eta]\right|_{\eta=0}.
    \label{eq:averages}
\end{equation}
Connected correlation functions of the observable $\mathcal{O}$ follow by taking higher derivatives of $\ln \mathcal{Z}$.

We assume that the electrodes are bulk s-wave superconductors, which are well described within mean-field theory. Then, the only relevant low-energy dynamical degree of freedom of the Josephson junction is the superconducting phase difference $\phi=\phi_R -\phi_L$, where $\phi_a$ denotes the phase of the order parameter of electrode $a$. (Without loss of generality, we assume equal magnitudes of the order parameters which we denote by $\Delta$.) Following standard procedures \cite{ambegaokar1982quantum,eckern1984quantum}, we reduce the full microscopic action in Eq.~\eqref{eq:actionfull} to an action for the phase $\phi$ alone (see App.~\ref{app:action} for a sketch of the derivation). Decoupling the attractive interaction between the electrons as well as the capacitive energy, integrating over the electron fields, and performing a gauge transformation to move the phase difference into the tunneling term, we arrive at \begin{equation}
    \mathcal{Z}=\int [d\phi]\, e^{iS[\phi]}
    \label{eq:ZZ0}
\end{equation}
with the Keldysh action 
\begin{equation}
    iS[\phi] = iS_0 + i\int_\mathcal{C} dt\, \frac{C}{8e^2} \dot{\phi}^2(t) + iS_\textrm{tun}[\phi].
    \label{eq:action}
\end{equation}
Here, $S_0$ contains the $\phi$-independent parts of the action, which describe superconductivity in the two electrodes within mean-field theory. The tunneling contribution to the action is given by 
\begin{equation}\label{eq:tun_action}
    iS_\textrm{tun} =   \Tr \ln\bqty{1 - g \mathcal{T}},
\end{equation}
where the trace $\Tr$ includes integration over the Keldysh contour as well as summation over the particle-hole and electrode spaces. The local equilibrium Nambu Green function $g = \textrm{diag}(g_{L},g_{R})$ describes the uncoupled electrodes at the tunneling site. The tunneling term 
\begin{equation}\label{eq:mathcal_T}
    \mathcal{T}(t) = \vartheta\tau_3 \bqty{\rho_- e^{-\frac{i}{2}\phi(t)\tau_3} + \rho_+ e^{\frac{i}{2}\phi(t)\tau_3}}
\end{equation}
describes tunneling between the electrodes from right to left and left to right, 
$\mathcal{T} =   \mathcal{T}_{RL} + \mathcal{T}_{RL}$. Here, we use the two sets of Pauli matrices $\tau_i$ and $\rho_i$, which act in the particle-hole and electrode spaces, respectively. The tunneling action can also be expressed as 
\begin{equation}\label{eq:tun_action_2}
    iS_\textrm{tun} =   \Tr \ln\bqty{1 - g_L \mathcal{T}_{LR} g_R \mathcal{T}_{RL}},
\end{equation}
where the trace no longer involves the electrode space. 
It will prove useful to decompose $\mathcal{T}$ into components $\mathcal{T}_\pm$ that transfer charge from left to right and vice versa, $\mathcal{T}(t) = \sum_{\pm} e^{\mp \frac{i}{2} \phi(t) } \mathcal{T}_\pm$, where 
\begin{align} \label{eq:mathcal_T_pm}
    \mathcal{T}_\pm  =&\ \vartheta \bqty{  \rho_\mp \frac{1 + \tau_3}{2} -  \rho_\pm  \frac{1 - \tau_3}{2}  }.
\end{align}

\subsection{Full counting statistics and generalized RCSJ equation}

References \cite{ambegaokar1982quantum,eckern1984quantum} expand the action in powers of the tunnel coupling $\vartheta$ between the electrodes. In many practical situations, the tunnel coupling is actually substantial and this expansion does not apply. Therefore, we proceed along a more general route. 

It is instructive to write the generating functional as 
\begin{equation}\label{eq:action_2}
    \mathcal{Z} = e^{iS_0} \int [d\varphi] [d\chi]\, \mathcal{Z}_\varphi [\chi],\quad \mathcal{Z}_\varphi [\chi] = e^{iS_\varphi[\chi]},
\end{equation}
where we decompose the phase field 
\begin{equation}
    \phi(t) = \varphi(t)\pm_\mathcal{C} \frac{1}{2}\chi(t)
\end{equation}
into its classical and quantum components $\varphi$ and $\chi$, respectively, and introduce $iS_\varphi[\chi] = iS[\phi]-iS_0$. 
A source field for the tunneling current across the junction may be introduced by promoting 
\begin{equation}
    \vartheta \to \left\{ \begin{array}{ccc}
    \vartheta e^{\mp_\mathcal{C} \frac{i}{2} e\eta} & ; & R \leftarrow L 
    \\
    \vartheta e^{\pm_\mathcal{C} \frac{i}{2} e\eta} & ;  & L \leftarrow R 
    \end{array}
    \right.
\end{equation}
in the tunneling Hamiltonian \footnote{This is equivalent to the introduction of a source field through $H_\eta = H \pm_\mathcal{C} \tfrac{1}{2}\eta(t)\mathcal{I}_\mathrm{tun}$ up to contact terms which arise from repeated action of derivatives on the exponential factors $\exp{\mp\pm_\mathcal{C}\tfrac{i}{2} e\eta}$. The latter arise only for cumulants of third and higher order, which are dropped below.}. By comparison with Eq.~\eqref{eq:mathcal_T} we observe that this amounts to a shift $\chi \to \chi +2e\eta$. Thus, $\chi$ acts as a counting field and $\mathcal{Z}_\varphi [\chi]$ can be interpreted as the moment generating functional for the full counting statistics of the junction subject to a time-dependent classical phase bias $\varphi(t)$. Moreover, $iS_\varphi[\chi]$ is the corresponding cumulant generating function. The coefficients of the expansion of $iS_\varphi[\chi]$ in powers of the quantum component $\chi(t)$ are just the connected correlation functions of the current across the junction subject to a phase bias $\varphi(t)$. 

The cumulant generating function is entirely due to the tunneling action in Eq.~\eqref{eq:tun_action}. Only the average current has a contribution originating from the capacitive term in the action, since the latter is linear in $\chi$,
\begin{equation}
      i\int_\mathcal{C} dt\, \frac{C}{8e^2} \dot{\phi}^2(t)
      =  i\int dt\, \frac{C}{4e^2} \dot\varphi(t) \dot\chi(t).
      \label{eq:actcapchi}
\end{equation}
This motivates us to isolate the average tunneling current $\mathcal{I}([\varphi];t)$ from the fluctuations by writing 
\begin{equation}\label{eq:expansionchi}
    i S_\varphi[\chi] =  \frac{1}{2ei} \int dt\, \bqty{\frac{C}{2e}\ddot\varphi(t) + \mathcal{I}([\varphi];t)}\chi(t) + W_\varphi[\chi].
\end{equation}
Here we use that the zeroth order in the expansion of the action in powers of $\chi$ vanishes due to cancellation between the forward and backward branches of the contour. The coefficient of the linear term in $\chi$ is the average current composed of the capacitive and the tunneling contributions. The tunneling current can be written as 
\begin{subequations}\label{eq:I_theta}    
\begin{align}
  \mathcal{I}([\varphi];t)  
  =&\  
  \left. 2ei \frac{\delta}{\delta\chi(t)} iS_\textrm{tun}[\varphi,\chi]\right|_{\chi=0} \\
  =&\ \langle \mathcal{I}_\mathrm{tun}(t)\rangle_{\varphi(t)}
\end{align}
\end{subequations}
in terms of the tunneling action. The subscript on the (quantum) expectation value indicates that it is to be evaluated for a given classical phase difference $\varphi(t)$. 

The generating functional $W_\varphi[\chi]$ describes the fluctuations of the current about its average. The second and higher cumulants of the full counting statistics can thus be obtained from 
\begin{multline}
    W_\varphi[\chi] = \sum_{n\geq 2} \frac{1}{(2ei)^n n!} \int dt_1\ldots dt_n\, \\ \times \mathcal{K}^{(n)}([\varphi];t_1,\ldots, t_n) \chi(t_1) \ldots \chi(t_n)
\end{multline}
or
\begin{equation}
    \mathcal{K}^{(n)}([\varphi];t_1,\ldots, t_n) = \left. \frac{ (2ei)^n \delta^n i S_\textrm{tun}[\varphi,\chi]}{\delta\chi(t_1)\ldots \delta\chi(t_n)} \right|_{\chi=0}. 
\end{equation}
The average tunneling current $\mathcal{I}([\varphi];t)$ as well as the tunneling current cumulants $\mathcal{K}^{(n)}([\varphi];t_1,\ldots,t_n)$ are functionals of $\varphi$, i.e., they depend on the entire history $\varphi(t)$. 

The cumulant generating function $W_\varphi$ is related to the probability distribution $P_\varphi[\xi]$ of the fluctuations $\xi(t)$ of the tunneling current by a Fourier transform,
\begin{equation}
    e^{W_\varphi[\chi]} = \int [d\xi]\, e^{-\frac{i}{2e} \int dt\, \xi(t) \chi(t)} P_\varphi[\xi].
\end{equation}
Combining this with Eqs.~\eqref{eq:action_2} and  \eqref{eq:expansionchi}, the integral over $\chi$ reduces to a $\delta$-functional and we obtain
\begin{equation}
    \mathcal{Z} =  
     \int [d\varphi]\, 
     \ev{\delta\bqty{\frac{C}{2e}\ddot\varphi(t) + \mathcal{I}([\varphi];t)+\xi(t) - I_\mathrm{b}(t)}
    }_{\xi}.
\end{equation}
Here, the average over the fluctuating current $\xi$ is defined as $\ev{\ldots}_\xi = \int [d\xi]\, \ldots P_\varphi[\xi]$ and we included a bias current $I_\mathrm{b}(t)$ by adding 
\begin{equation}
    iS_\textrm{b}[\chi] = -\frac{1}{2ei}\int dt\, I_\textrm{b}(t)\chi(t)   
\end{equation}
to the action $iS$. 
The $\delta$-functional effectively enforces current conservation. As in the RCSJ model, it includes the capacitive current, the tunneling current encompassing both quasiparticle and Cooper-pair currents, and a stochastic current $\xi$ with  connected correlation functions 
\begin{equation}
    \ev{\xi(t_1)\ldots \xi(t_n)}_{\xi}^{(c)} = \mathcal{K}^{(n)}([\varphi];t_1,\ldots, t_n).
\end{equation}
The $\delta$-functional implies that the phase difference $\varphi(t)$ across the junction is governed by the time-nonlocal Langevin equation anticipated in Eq.~\eqref{eq:generalized_rcsj_nonlocal}. It can be used, e.g., to compute the voltage across the junction by using the Josephson relation in Eq.~\eqref{eq:JosephsonRel}. 

While this clearly parallels the conventional RCSJ model, there are several differences. First, the tunneling current retains all orders in the tunneling amplitude $\vartheta$, going beyond the usual derivation of the effective action \cite{ambegaokar1982quantum,eckern1984quantum}, which focuses on the lowest-order expansion in tunneling. Our formulation can thus account for general current-phase relationships of the supercurrent as well as current-voltage characteristics of the quasiparticle current. Both of these are central to a theory of nonreciprocal Josephson currents \cite{steiner2023diode}. 

Second, the Langevin current has non-Gaussian correlations in general. However, in many physical situations, a large number of electrons pass the junction during the characteristic timescales of Eq.~\eqref{eq:generalized_rcsj_nonlocal}. Coarse-graining will then yield effectively Gaussian noise correlations as a consequence of the central limit theorem. This is equivalent to the standard semi-classical expansion to quadratic order in $\chi$ \cite{schmid1982quasiclassical,kamenev2011field}. We formulate conditions under which this coarse-graining is valid in Sec.~\ref{sec:validity}. 
In the following, we thus limit the discussion to treating $\xi$ as Gaussian noise with correlation function  
\begin{subequations}\label{eq:Kphi}
    \begin{align}
    \mathcal{K}([\varphi];t,t') = &\ \mathcal{K}^{(2)}([\varphi];t,t')\\
    =&\ 
    \frac{1}{2}\ev {\Bqty{\delta \mathcal{I}_\mathrm{tun}(t), \delta\mathcal{I}_\mathrm{tun}(t') }}_{\varphi(t)},
\end{align}
\end{subequations}
in terms of $\delta \mathcal{I}_\mathrm{tun}(t) = \mathcal{I}_\mathrm{tun}(t) - \ev{ \mathcal{I}_\mathrm{tun}(t)}_{\varphi(t)}$. Studying situations in which the non-Gaussian correlations have observable consequences is an interesting problem. 

Unlike in the conventional RCSJ model, the dynamics of the phase difference $\varphi$ has not yet been  reduced to a time-local Langevin equation. In fact, the coefficients $\mathcal{I}([\varphi];t)$ and $\mathcal{K}([\varphi];t,t')$ still depend on the phase difference $\varphi(t)$ at all previous times. We can simplify the result, when the phase dynamics are slow compared to the intrinsic time scales of the bulk superconductors. This is discussed below in Sec.~\ref{sec:adiabatic}.  

\subsection{Average current and noise correlations}
\label{sec:explicit_expressions_for_I_and_K}

We can make it explicit that the current $\mathcal{I}([\varphi];t)$ and noise correlator $\mathcal{K}([\varphi];t,t')$ correspond to the current and noise of a junction with imposed phase bias $\varphi(t)$. To this end, we express Eqs.~\eqref{eq:I_theta} and \eqref{eq:Kphi} in terms of the dressed Green function 
\begin{align}\label{eq:full_gf}
    \mathcal{G} = 
    \pqty{g^{-1} - \mathcal{T}}^{-1}, 
\end{align}
which accounts for tunneling between the electrodes as described by $\mathcal{T}$. Here, we switched to a matrix representation of the Keldysh contour by defining 
\begin{equation}\label{eq:contour_green_function_matrix}
    g = \begin{pmatrix}
        g^\mathrm{T} & g^< \\ -g^> & -g^{\tilde{\mathrm{T}}}
    \end{pmatrix}
    ,\quad  
    \mathcal{G} = \begin{pmatrix}
        \mathcal{G}^\textrm{T} & 
        \mathcal{G}^< \\
        -\mathcal{G}^> &
        -\mathcal{G}^{\tilde{\textrm{T}}}
    \end{pmatrix},
\end{equation}
as well as the contour-independent tunneling operator $\mathcal{T}(t) =   \sum_\pm e^{\mp i \frac{1}{2} \varphi(t)} \mathcal{T}_\pm$ (which now only depends on $\varphi$) 
The negative signs in the second row of $g$ and $\mathcal{G}$ are included to account for the reversed direction of integration along the backward branch of $\mathcal{C}$. As detailed in
App.~\ref{app:tun_cur_and_noise}, we find that the current and noise correlator can be expressed as
\begin{equation}\label{eq:tun_cur}
\mathcal{I}([\varphi];t)
   = - e \tr \bqty{ \mathcal{G}^<([\varphi];t,t) \mathcal{T}(t) \tau_3 \rho_3
   },
\end{equation} 
and 
\begin{multline}\label{eq:tun_noise}
    \mathcal{K}([\varphi]; t,t')
   =  -  \frac{e^2}{2}   \tr \big[ \mathcal{G}^<([\varphi];t,t') 
   \mathcal{T}(t') \tau_3 \rho_3 \\ \times \mathcal{G}^>([\varphi];t',t) \mathcal{T}(t) \tau_3 \rho_3
   \big] 
   + (t \leftrightarrow t')\ ,
\end{multline}
respectively. The trace $\tr$ is taken over the particle-hole and electrode spaces. These expressions reproduce familiar expressions \cite{cuevas1996hamiltonian,cuevas1999shot} 
for the expectation value of the current  
and the symmetrized, connected current-current correlation function in the presence of an imposed phase bias $\varphi(t)$. 

\section{Adiabatic approximation}
\label{sec:adiabatic}

A strictly adiabatic approximation in the limit of slow phase dynamics would replace $\mathcal{I}([\varphi];t)$ by the current for a phase difference set to $\varphi(t)$ at \textit{all} times. This approximation retains only the Josephson contribution to the action, but neglects quasiparticle dissipation. We account for the latter by including 
the dependence on $\dot\varphi(t)$ (or, equivalently, the instantaneous voltage $V(t)=\frac{\hbar}{2e}\dot\varphi(t)$ across the junction), corresponding to the next order in a gradient expansion. We neglect $\ddot\varphi(t)$ and higher derivatives. This is justified if the rate of change of the voltage is slow compared to the microscopic time scales of the superconducting electrodes. This requires a sufficiently large capacitance $C$ of the junction as discussed further in Sec.~\ref{sec:validity}.

\subsection{Green function}

The adiabatic approximation can be implemented systematically by passing to the Wigner representation
\begin{equation}\label{eq:wigner_trf}
    \mathcal{G}([\varphi];\tau,\omega) = \int d(\delta t)\, e^{i\omega\delta t} \mathcal{G}([\varphi];\tau+\tfrac{\delta t}{2},\tau- \tfrac{\delta t}{2}) 
\end{equation}
of the Green function. Here we introduce the mean and relative times $\tau = \tfrac{1}{2}(t+t')$ and $\delta t = t-t'$, respectively. We consider the Dyson equation for the full Green function
\begin{equation}\label{eq:dyson}
\mathcal{G} = g + g\mathcal{T} \mathcal{G}.  
\end{equation} 
In the Wigner representation, convolutions become Moyal products $\bqty{AB} (\tau,\omega) =  A (\tau,\omega) \exp\{\frac{i}{2} \mathcal{D}\} B(\tau,\omega)$ with 
\begin{equation}
    \mathcal{D} = 
    \overset{\leftarrow}{\partial}_\omega
    \overset{\rightarrow}{\partial}_\tau
    -
    \overset{\leftarrow}{\partial}_\tau
    \overset{\rightarrow}{\partial}_\omega.
\end{equation}
We observe that the Wigner representation  \begin{equation}
    g(\tau,\omega) = g(\omega) 
\end{equation}
of the uncoupled Green function is independent of the time $\tau$ and the Wigner representation 
\begin{equation}
\mathcal{T} (\tau,\omega) = \mathcal{T} (\tau)
\end{equation}
of the tunneling term is independent of frequency.  
This simplifies the Moyal product   
\begin{equation}
     \bqty{g \mathcal{T} }  (\tau,\omega)  = \sum_\pm g(\omega) e^{ \frac{i}{2} 
    \overset{\leftarrow}{\partial}_\omega
    \overset{\rightarrow}{\partial}_\tau} \mathcal{T}_\pm e^{\mp \tfrac{i}{2}\varphi(\tau)} . 
\end{equation}
Within an adiabatic approximation, which neglects $\ddot{\varphi}$ and higher derivatives, this becomes
\begin{align}\label{eq:adiabatic_0}
     \bqty{g\mathcal{T} }  (\tau,\omega)    \simeq&\ \sum_\pm g(\omega) 
     e^{\pm\frac{1}{4}\dot\varphi(\tau)\overset{\leftarrow}{\partial}_\omega}  \mathcal{T}_\pm  e^{\mp \tfrac{i}{2}\varphi(\tau)} 
      \nonumber  
     \\ 
    = &\ \sum_\pm e^{\mp \tfrac{i}{2}\varphi(\tau)} g(\omega \pm \tfrac{1}{4}\dot{\varphi}(\tau) ) \mathcal{T}_\pm   . 
\end{align}
Here, the exponential in the Moyal product effectively acts as a translation operator in frequency space. The resulting expression depends only on the instantaneous phase $\varphi(\tau)$ and its time derivative $\dot{\varphi}(\tau)$ (i.e., the instantaneous voltage). As time enters only through $\varphi(\tau)$, one can systematically implement the adiabatic approximation by approximating the Moyal derivative as
\begin{equation}
    \mathcal{D} \simeq  
     \dot{\varphi}(\tau)\bqty{ \overset{\leftarrow}{\partial}_\omega
    \overset{\rightarrow}{\partial}_{\varphi(\tau)}
    -
    \overset{\leftarrow}{\partial}_{\varphi(\tau)}
    \overset{\rightarrow}{\partial}_\omega }. 
    \label{eq:fcalTWigner}
\end{equation}
Applying this order by order in the expansion for $\mathcal{G} = \sum_{m=0} (g\mathcal{T} )^mg$, the full Green function no longer depends on the entire history of $\varphi$, but only on the instantaneous phase and its velocity, 
\begin{equation}
    \mathcal{G}([\varphi];\tau,\omega) \simeq \mathcal{G}(\varphi(\tau),\dot{\varphi}(\tau)/2e;\omega). 
\end{equation}
The function $\mathcal{G}(\varphi,\dot{\varphi}/2e;\omega)$ on the right hand side depends \textit{parametrically} on $\varphi$ and $\dot{\varphi}$. Combining  Eq.~\eqref{eq:adiabatic_0} with the expansion of $\mathcal{G}$, one can write
\begin{equation}\label{eq:gf_floquet_exp}
    \mathcal{G} (\varphi ,\dot{\varphi} /2e;\omega) = \sum_m e^{-\frac{i}{2} m \varphi } \mathcal{G}_m(\dot{\varphi}/2e;\omega- m \tfrac{\dot{\varphi}}{4}).  
\end{equation} 
Inserting this Fourier series into the Dyson equation yields a set of equations for the coefficients $\mathcal{G}_m$, 
\begin{multline}\label{eq:dyson_for_coefficients}
\mathcal{G}_m(\dot{\varphi}/2e;\omega) = g(\omega)\delta_{m,0} \\ +  g(\omega +m \tfrac{\dot{\varphi}}{2}) \sum_\pm  \mathcal{T}_\pm \mathcal{G}_{m\mp 1}(\dot{\varphi}/2e;\omega). 
\end{multline}
Setting $\dot{\varphi}/2e = V$, this reduces to the Floquet expansion of the Dyson equation for a voltage-biased junction. The coefficients $\mathcal{G}_m$ can thus be obtained from calculations for voltage-biased Josephson junctions and are well known in the literature \cite{cuevas1996hamiltonian,cuevas1999shot}. 

\subsection{Tunneling current  \texorpdfstring{$\mathcal{I}$}{I}}

We can now give explicit expressions within the adiabatic approximation for the supercurrent and the dissipative current as defined in Eqs.~\eqref{eq:supcurrIs} and \eqref{eq:disscurrId}, respectively. By virtue of the same arguments as in the previous section, the average current becomes a function only of the instantaneous phase and its instantaneous velocity, i.e., 
\begin{equation}
   \mathcal{I}([\varphi];t) 
  \simeq \mathcal{I}(\varphi(t),\dot{\varphi}(t)/2e). 
\end{equation}
We can expand $\mathcal{I}$ into a Fourier series
\begin{equation}\label{eq:I_fourier_series}
    \mathcal{I}(\varphi,\dot{\varphi}/2e) = \sum_n  e^{-in\varphi} \mathcal{I}_{n}(\dot\varphi/2e),
\end{equation}
where we use that due to the trace in Eq.~\eqref{eq:tun_cur}, the expansion involves only even powers of $e^{\mp\frac{i}{2}\varphi}$. From Eqs.~\eqref{eq:tun_cur} and \eqref{eq:gf_floquet_exp}, the expansion coefficients are  
\begin{multline}
    \mathcal{I}_{n}(\dot\varphi/2e) \\  = - e \sum_{m = \pm} m \int \frac{d\omega}{2\pi}\, \tr \left\{ \mathcal{G}_{2n - m}^<(\dot{\varphi}/2e;\omega) \mathcal{T}^{}_m 
   \right\},
\end{multline} 
where we use that $\mathcal{T}_\pm\rho_3 \tau_3 = \pm \mathcal{T}_\pm$. These coefficients are identical to those used to expand the current in voltage-biased junctions \cite{cuevas1996hamiltonian}. 

We can now insert these expressions into Eqs.~\eqref{eq:supcurrIs} and \eqref{eq:disscurrId} to obtain the supercurrent as well as the dissipative current in the generalized RCSJ model Eq.~\eqref{eq:generalized_rcsj}. For nonzero $\dot{\varphi}$, the supercurrent oscillates at multiples of the Josephson frequency $\dot\varphi$, which is set here by the instantaneous voltage. For vanishing $\dot{\varphi}$, the supercurrent remains finite, and is determined by $\varphi$. The dissipative current is due to the generation of quasiparticles by tunneling. Importantly, it is nonvanishing for $eV \sim \dot{\varphi}/2 < 2\Delta$ due to multiple Andreev reflections. These arise only beyond the tunneling regime. Thus, for high junction transparencies there may be appreciable dissipation due to quasiparticle tunneling for voltages much smaller than the superconducting gap.

\subsection{Current-current correlation function \texorpdfstring{$K$}{K}}

Within the adiabatic approximation, the explicit time dependence of the
noise correlator only involves the time difference $\delta t = t- t'$. The mean time $\tau = \tfrac{1}{2}(t+t')$ enters only implicitly through $\varphi$ and $\dot\varphi$, 
\begin{equation}
    \mathcal{K}([\varphi];t,t') \simeq 
    \mathcal{K}(\varphi(\tau),\dot{\varphi}(\tau)/2e; \delta t).
\end{equation}
To find an explicit expression, one  considers the Wigner transform of the noise correlator. Expanding $\varphi(\tau\pm \delta t /2) \simeq \varphi(\tau) \pm  \delta t \dot{\varphi}(\tau) /2$ in $\mathcal{T}$ and inserting the expression for the Green function in the adiabatic approximation, we obtain 
\begin{equation}\label{eq:K_fourier_series}
\mathcal{K}(\varphi,\dot{\varphi}/2e;\omega)= \sum_n e^{-in\varphi} \mathcal{K}_n(\dot{\varphi}/2e;\omega),
\end{equation} 
where
\begin{multline}
    \mathcal{K}_n(\dot{\varphi}/2e;\omega) 
    = - \frac{e^2}{2}\sum_{l}\sum_{m_1m_2} m_1 m_2 \sum_\pm \int \frac{d\varepsilon}{2\pi} \\ \times \tr\big\{\mathcal{G}_{2n - l }^<(\dot{\varphi}/2e;\varepsilon+ (l - m_2 -n)\tfrac{\dot{\varphi}}{2} )\\ 
    \times \mathcal{T}^{}_{m_1}  \mathcal{G}_{l-m_1- m_2}^>(\dot{\varphi}/2e;\varepsilon \pm\omega )\mathcal{T}^{}_{m_2} \big\}.
\end{multline}    
As for the average current, this includes only even powers of $e^{\mp \frac{i}{2}\varphi}$ due to the trace. 

We arrive at a completely time-local version of the generalized RCSJ model when neglecting the correlation time of the fluctuating current. This amounts to neglecting the $\omega$ dependence of $\mathcal{K}_n(\dot{\varphi}/2e;\omega)$ or, equivalently, approximating the dependence on relative time by a $\delta$-function,
\begin{equation}
    \mathcal{K}(\varphi,\dot{\varphi}/2e;t-t') \simeq
     \mathcal{K}(\varphi,\dot{\varphi}/2e) \delta(t-t'), 
\end{equation}
where we use the  abbreviated notation 
$ \mathcal{K}(\varphi,\dot{\varphi}/2e) = \mathcal{K}(\varphi,\dot{\varphi}/2e;\omega = 0) $. This white-noise approximation is appropriate when the junction dynamics is slow compared to the correlation time of the noise set by $\Delta^{-1}$, and thus consistent with the previous Gaussian and adiabatic approximations.  

\subsection{Validity of generalized RCSJ model}
\label{sec:validity}

We have now completed the derivation of the time-local version of the generalized RCSJ model as given in Eq.~\eqref{eq:generalized_rcsj}. We are left with discussing the conditions for the validity of the approximations entering into the derivation. We first consider the adiabatic approximation. Using that the microscopic time scale is controlled by the gap $\Delta$, it is valid provided that 
$\ddot \varphi/\dot \varphi \ll \Delta$. The left-hand side is controlled by the characteristic time scales of the RCSJ equation, i.e., the plasma frequency 
\begin{equation}
    \omega_\mathrm{pl} = \sqrt{E_C E_J}
\end{equation}
($E_C=e^2/C$ and $E_J$ are the charging and characteristic Josephson energies of the junction, respectively) and the RC time 
\begin{equation}
    \tau_\mathrm{RC} = RC.
\end{equation}
Here, $R$ is a typical resistance of the junction controlling the damping at subgap voltages. We conclude that the adiabatic approximation is valid for
\begin{equation}
    \label{eq:condadiabatic}
    \mathrm{max}\{\omega_\mathrm{pl},\tau_\mathrm{RC}^{-1}\} \ll \Delta.
\end{equation}
This implies that the adiabatic approximation is valid for sufficiently large junction capacitance \footnote{In the overdamped regime, the velocity is proportional to the instantaneous force. This leads to $\ddot{ \varphi}/\dot{\varphi} \sim \omega_\textrm{pl}^2 \tau_{RC}$ which should be small compared to  $\omega_\textrm{pl}$ in the overdamped regime. However, $\omega_\textrm{pl}^2 \tau_{RC} \sim E_J R / (e^2 /h) \sim \Delta R/R_N$ independent of $C$, with $R_N$ the resistance in the normal state. This leads to a contradiction unless $R$ is much smaller than $R_N$. This is rather unrealistic if $R$ represents the resistance from quasiparticle tunneling only but plausible if the external electromagnetic environment is included \cite{kautz1990noise}.}.  

Next, we consider the Gaussian approximation for the statistics of the Langevin current $\xi$. In line with the central limit theorem, it holds when the fluctuations $\delta N$ in the number of electrons passing the junction during a time interval $1/\mathrm{max}\{\omega_\mathrm{pl},\tau_\mathrm{RC}^{-1}\}$ is large compared to unity. Anticipating the fluctuation-dissipation theorem in Eq.~\eqref{eq:generalized_fdt},
we estimate
\begin{equation}
  e^2 \langle (\delta N)^2 \rangle  \sim 2T G/\mathrm{max}\{\omega_\mathrm{pl},\tau_\mathrm{RC}^{-1}\}, 
\end{equation}
where $G\sim 1/R$ denotes a typical conductance of the junction. Thus, we find the condition
\begin{equation} 
   \frac{T}{E_C} \mathrm{max}\{1 , \omega_\mathrm{pl}\tau_\mathrm{RC} \} \gg 1 
\end{equation}
for the validity of the Gaussian approximation. 
Moreover, the correlation function of the Langevin current can be taken as $\delta$-correlated provided that 
condition Eq.~\eqref{eq:condadiabatic} is satisfied. 

Within the adiabatic approximation, all degrees of freedom other than the phase are assumed to be in equilibrium. In some cases, other slow degrees of freedom need to be accounted for, e.g., the occupations of Andreev levels \cite{liu2024giant} or other subgap states \cite{martin2014nonequilibrium,ruby2015tunneling}. This can be achieved by introducing equations tracking these occupations and  their coupling to the phase variable.

\section{ Nonlinear fluctuation-dissipation theorem}
\label{sec:nonlinear_fdt}

\begin{figure*}[ht!]
    \centering
    \includegraphics[width=\linewidth]{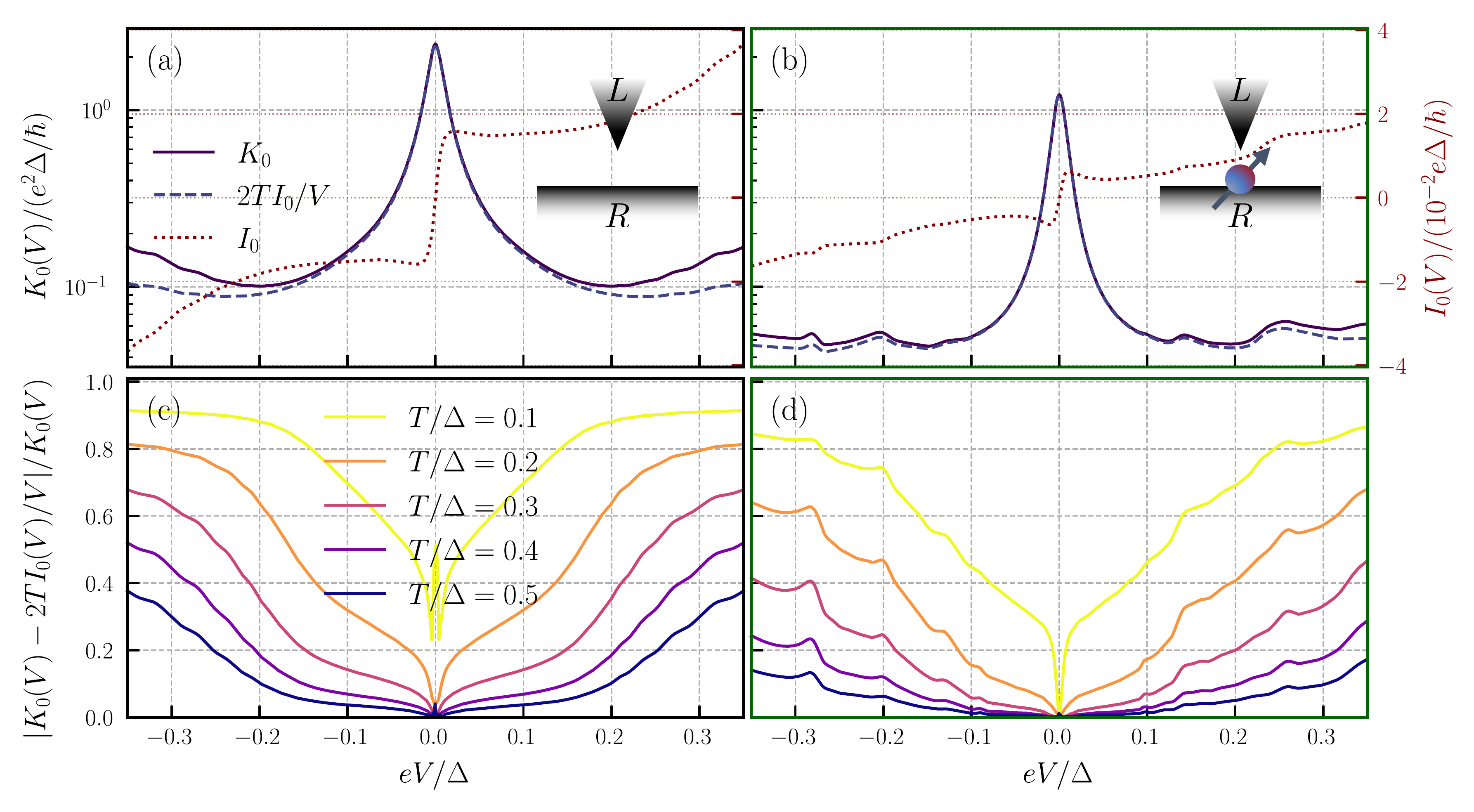}
    \caption{Nonlinear fluctuation-dissipation relation in junctions without [(a,c); black frame] and with magnetic adatom [(b,d), green frame]. We assume that the Kondo temperature is small compared to $\Delta$, so that Kondo correlations can be neglected. Panels (a) and (b) illustrate that the nonlinear features in $K_0(V)$ (solid line), which are especially rich in the presence of a magnetic adatom, are well reproduced by  $2TI_0(V)/V$ (dashed line) over a voltage range that far transcends the linear regime of the current $I_0(V)$ (red, dotted line). Only at voltages of the order of half the temperature (here: $T = 0.5 \Delta$) the quantitative agreement breaks down, while qualitative features are still reproduced. This behavior can be systematically studied by considering the relative deviation $\abs{K_0(V)-2TI_0(V)/V}/K_0(V)$ for a variety of temperatures as shown in panels (c) and (d). One observes that the voltage at which the nonlinear fluctuation-dissipation relation is violated grows with increasing temperature. Model: The plain tunneling junction (a,c) is modeled by $h_{L,R} = \xi + \Delta \tau_1$, while the magnetic adatom junction  (b,d) has $h_L = \xi + \Delta \tau_1$ and $h_R = h_L + (w\tau_3 - JS)\delta(\mathbf{r})$ with $\pi \nu_0 w = 1 = \pi \nu_0 J$ ($\nu_0$ is the normal state local density of states at the Fermi level). Parameters common to all panels: $\pi \nu_0 \vartheta = 0.5$, $\eta = 0.01\Delta$. This corresponds to normal state transparencies $\theta = 0.64$ [(a,b)] and $\theta=0.41$ [(c,d)]. We note that the peaks at $eV = 0$ in (c,d) are numerical artifacts originating from the truncation of the Floquet Green function approach at vanishing frequency (i.e., voltage). In fact, the equilibrium fluctuation-dissipation relation guarantees that the noise power exactly equals $2T \partial_V I_0(V)$ here.}\label{fig:nonlinear_fluctuation_dissipation}
\end{figure*}

The noise power $\mathcal{K}(\varphi,\dot{\varphi}/2e = V)$ is intricately linked to the dissipative current $\mathcal{I}(\varphi,\dot{\varphi}/2e = V)$ 
through a nonlinear fluctuation-dissipation relation. This follows from the symmetry 
\begin{equation}\label{eq:fdt_symmetry}
    iS_{\varphi}[\chi(t)] =   iS_\varphi[- \chi(-t) - 2ieV/T] \Big\vert_{\varphi_0 \to -\varphi_0},
\end{equation}
of the action for fixed $\varphi(t) = \varphi_0 + 2eV t$, which has been established on general grounds \cite{tobiska2005inelastic,saito2008symmetry}. We provide an explicit derivation of this relation based on Eq.~\eqref{eq:tun_action_2} in App.~\ref{app:fluctuation_dissipation_1}. The relation holds provided that the microscopic Hamiltonian is time-reversal symmetric. The relation between the dissipative current and the Gaussian noise fluctuations relevant to the Langevin current 
can be derived from the implications of Eq.~\eqref{eq:fdt_symmetry} for the
symmetrized action 
\begin{equation}
iS_{+,\varphi}[\chi(t)]= \frac{1}{2}\{iS_{\varphi}[\chi(t)] + iS_{\varphi}[\chi(-t)]\vert_{\varphi_0 \to -\varphi_0}\}.
\end{equation}
This generates the parts of $\mathcal{I}$ and $\mathcal{K}$, which are even in the phase,
\begin{subequations}
\begin{align}
    \mathcal{I}_+(\varphi,V) =&\ \frac{1}{2} \bqty{\mathcal{I}(\varphi ,V) + \mathcal{I}(-\varphi ,V)},\\ 
    \mathcal{K}_+(\varphi,V;\delta t) =&\ \frac{1}{2} \bqty{\mathcal{K}(\varphi,V;\delta t) + \mathcal{K}(-\varphi,V;\delta t)}. 
\end{align}
\end{subequations}
Equation \eqref{eq:fdt_symmetry} implies that the symmetrized action satisfies
\begin{equation}
iS_{+,\varphi}[\chi(t)] =  iS_{+,\varphi}[-\chi(t) - 2ieV/T].
\end{equation} For $eV \ll T$, we can expand both sides to second order in their arguments. This (as well as a generalization to higher-order correlation functions) is discussed in App.~\ref{app:fluctuation_dissipation_2}. We arrive at
\begin{multline}
    \frac{2T}{V}  \int dt\, \chi(t) \mathcal{I}_+(\varphi(t),V) \\ = \int dt\,  \chi(t)  \int dt'\,  \mathcal{K}_+(\varphi(\tfrac{t+t'}{2}),V;t-t').
\end{multline}
As we are interested in timescales large compared to microscopic times, we can replace the argument of $\chi$ on the right hand side with the central time. We then obtain a relation involving only the zero frequency noise,
\begin{equation}
    \frac{2T}{V}\mathcal{I}_+(\varphi(t),V) =  \mathcal{K}_+(\varphi(t),V).
\end{equation}
Using the Fourier expansions in Eqs.~\eqref{eq:I_fourier_series} and \eqref{eq:K_fourier_series}, we obtain 
\begin{equation}\label{eq:generalized_fdt}
    \frac{2T}{V} \Re\bqty{\mathcal{I}_n(V)} = \Re\bqty{\mathcal{K}_n(V)},
\end{equation} 
which constitutes the nonlinear fluctuation-dissipation theorem. 

A few comments are in order: 
(i) The derivation of the fluctuation-dissipation relation Eq.~\eqref{eq:generalized_fdt} required only that $eV \ll T$ but made no assumptions on the magnitude of $eV$ compared to other energy scales (in particular $\Delta$ or fractions thereof). It is therefore valid beyond voltage ranges where the current and the noise are Ohmic. Thus, it remains valid in the presence of nonlinear features arising, e.g., from multiple Andreev reflections \cite{zazunov2023nonreciprocal} or tunneling into subgap states \cite{ruby2015tunneling}. 
(ii) For the $dc$ current and noise power ($n=0$),  Eq.~\eqref{eq:generalized_fdt} can be interpreted as a generalization of the Johnson-Nyquist relation $K_\textrm{d} = 2TG$, which replaces the zero-bias conductance by the nonlinear conductance $G(V) = I_\textrm{d}(V)/V$. While Eq.~\eqref{eq:fdt_symmetry} requires time-reversal symmetry, $K_\textrm{d} = 2TI_\textrm{d}(V)/V$ holds without any such assumption. We show this in  App.~\ref{app:fluctuation_dissipation_3} by an order-by-order calculation. (iii) The $n>0$ terms in Eq.~\eqref{eq:generalized_fdt} correspond to the $\cos(n 2eV t)$ Fourier components of the time-dependent tunneling current in a voltage-biased situation. These are just the dissipative current components in a time-reversal symmetric junction \cite{cuevas1996hamiltonian}.  (iv) A phenomenological approach based on demanding detailed balance in the effective Fokker-Planck equation also yields $K_\textrm{d} = 2TI_\textrm{d}(V)/V$ (see Supplemental Material Sec.~S4 of Ref.~\cite{steiner2023diode} for more details). (v) 
Expansion to higher orders gives many more relations between $\mathcal{K}^{(n)}$ and $\mathcal{K}^{(n+1)}$ involving also the sinusoidal components of the higher-order correlators. (vi) In the limit of zero voltage, Eq.~\eqref{eq:generalized_fdt} reduces to the equilibrium fluctuation-dissipation theorem $2T \partial_V\Re\bqty{\mathcal{I}_n(V)}_{V\to 0} = \Re\bqty{\mathcal{K}_n(V)}_{V \to 0}$ (since $\Re\bqty{\mathcal{I}_n(0)} = 0$). 

We illustrate the validity of the nonlinear fluctuation-dissipation theorem in Fig.~\ref{fig:nonlinear_fluctuation_dissipation}. We compare $\mathcal{K}_0(V)$ and $2T\mathcal{I}_0(V)/V$ far beyond the linear-response regime for both, a plain scanning-tunneling-microscope junction [Fig.~\ref{fig:nonlinear_fluctuation_dissipation}(a,c)] and a junction hosting a magnetic adatom [Fig.~\ref{fig:nonlinear_fluctuation_dissipation}(b,d)]. Qualitative features of the noise power $\mathcal{K}_0(V)$ are reproduced by $2T\mathcal{I}_0(V)/V$ up to voltages far outside the linear regime of the current $\mathcal{I}_0(V)$ [Fig.~\ref{fig:nonlinear_fluctuation_dissipation}(a,b)]. In particular, peaks in the noise power due to tunneling processes which involve Yu-Shiba-Rusinov states \cite{yu1965bound,shiba1968classical,rusinov1969superconcductivity} are captured [Fig.~\ref{fig:nonlinear_fluctuation_dissipation}(b)]. Note that these features are not symmetric in the voltage. In both types of junctions, we find excellent quantitative agreement between $\mathcal{K}_0(V)$ and $2T\mathcal{I}_0(V)/V$ for voltages up to $40\%$ of the temperature [Fig.~\ref{fig:nonlinear_fluctuation_dissipation}(c,d)]. We also observe that the true noise power lies above the approximation provided by $2T\mathcal{I}_0(V)/V$. 

\section{Conclusion}
\label{sec:conclusions}

The RCSJ model has long played a central role in the theory of Josephson junctions \cite{stewart1968current,mccumber1968effect}. A microscopic derivation of this model was given in seminal work by Ambegaokar, Eckern, and Sch\"on \cite{ambegaokar1982quantum,eckern1984quantum}. Here, we extended this derivation to include generalized versions of the model, which account for nonlinear dissipative currents and general current-phase relations \cite{stewart1974current,kautz1990noise,steiner2023diode}. Our derivation does not rely on an expansion in the tunneling amplitude across the junction and thus accounts for higher-order contributions to the current such as multiple Andreev reflections. The central observation of our approach is that the contribution of the tunneling term to the action can be interpreted as the generating functional of the full counting statistics of a voltage-biased junction, with the quantum component of the phase effectively acting as a counting field. When combining this with an adiabatic approximation and a central-limit argument, we recover the generalized RCSJ model. 

Given that the dissipative current is non-Ohmic in the generalized RCSJ
model, the correlation function of the Langevin current can no longer be obtained from the conventional fluctuation-dissipation theorem. While the latter is limited to the linear-response regime, the dissipative current of Josephson junctions is frequently highly nonlinear. We show that in the limit $T\gg eV$, one can derive a generalized nonlinear fluctuation-dissipation theorem, which controls the correlations of the Langevin current. This connects to recent discussions of fluctuation theorems in mesoscopic transport \cite{tobiska2005inelastic,saito2008symmetry}.

Our derivation also clarifies the nature of the dissipative current. In the tunneling limit, the dissipative current is strongly suppressed as long as the phase dynamics is slow on the scale of the superconducting gap, $\dot\varphi \ll \Delta$. For larger junction transparencies, dissipation for slow junction dynamics may be dominated by the quasiparticle generation due to multiple Andreev processes. 

Recent interest in the RCSJ model has centered around the nonreciprocal response of Josephson junctions. In the tunneling limit, Josephson junctions necessarily exhibit reciprocal behavior. To include nonreciprocal behavior such as the Josephson diode effect, one has to include higher harmonics of the current-phase relation as well as the possibly nonreciprocal (diode-like) behavior of the dissipative current. Our derivation of the generalized RCSJ model can account for both of these effects and thus provides a microscopic basis for recent applications of the generalized RCSJ model to the Josephson diode effect \cite{misaki2021theory,trahms2023diode,steiner2023diode,volkov2024josephson,seleznev2024influence,souto2024tuning,monroe2024phase,wang2025josephson}.

Our derivation assumes a point-like contact between $s$-wave superconducting electrodes. It should be straightforward to extend the derivation to more general junctions with extended  contacts or non-$s$-wave electrodes. Essentially, one would just need to use the appropriate local Green function $g$ of the electrodes as well as the appropriate tunneling term $\mathcal{T}$. An interesting aspect of our derivation is that in general, the phase dynamics involves a non-Gaussian Langevin current. While one expects that this is frequently well approximated by a Gaussian Langevin source, it is interesting to look for situations in which deviations from this approximation have observable consequences. Similarly, one may want to explore the interplay of retardation effects (due to memory in the current kernel or colored noise) and nonreciprocity if the assumption of adiabaticity is relaxed.

\begin{acknowledgments}
We thank Aritra Lahiri and Rosa L\'opez for discussions. Research in Berlin was supported by CRC 183 (project C03) of the Deutsche Forschungsgemeinschaft. J.F.S. acknowledges the support of the AFOSR MURI program, under Agreement No. FA9550-22-1-0339. 
\end{acknowledgments}

\appendix

\onecolumngrid

\section{Derivation of the Keldysh action for the phase difference}
\label{app:action}

We sketch the derivation of Eq.~\eqref{eq:action}, following Refs.~\cite{ambegaokar1982quantum,eckern1984quantum}. We decouple the attractive interaction by introducing the complex order-parameter fields $\Delta_a$,
\begin{multline}
    \exp{
        -i \lambda_a\int_\mathcal{C} dt\, \int d\mathbf{r}\, \overline{\psi}_{a,\uparrow}\overline{\psi}_{a,\downarrow}\psi^{}_{a,\downarrow}\psi^{}_{a,\uparrow} 
    } \\ 
    =   \int [d\Delta^{}_a][d\Delta^*_a]\, 
        \exp{ i\int_\mathcal{C} dt\, \int  d\mathbf{r}\, \bqty{\frac{1}{ \lambda_a} 
        \abs{\Delta_a}^2
        -\Delta^*_a \psi^{}_{a,\downarrow}\psi^{}_{a,\uparrow}
        -
        \Delta^{}_a\overline{\psi}_{a,\uparrow}
        \overline{\psi}_{a,\downarrow}} } 
     ,
\end{multline}
and the capacitive energy by introducing the voltage field $V$,
\begin{equation}
    \exp{-i\int_\mathcal{C} dt\, \frac{1}{2C}\pqty{\frac{Q_L-Q_R}{2}}^2}
    = \int [dV]\, 
    \exp{i\int_\mathcal{C} dt\, \bqty{
    \frac{1}{2}C V^2
     - \frac{1}{2}V(Q_L-Q_R) 
    } }.
\end{equation}
The generating functional (here in the absence of sources, $\mathcal{Z} = \mathcal{Z}[\eta = 0] = 1$) then takes the form
\begin{equation}\label{eq:partition_function}
    \mathcal{Z}
    = \int [dV]\prod_a [d\psi_a][d\overline{\psi}_a][d\Delta_a][d\Delta^*_a]\, 
    \exp{ i\int_\mathcal{C} dt\, 
    \bqty{ 
    \frac{1}{2}CV^2 +   
    \int  d\mathbf{r}\, \pqty{ \sum_a \frac{1}{\lambda_a} \abs{ \Delta_a}^2
    + \int_\mathcal{C} dt'\, \overline{\Psi}\,  \bqty{g_\mathcal{C}^{-1} - \mathcal{T}_\mathcal{C}} \Psi' } }  }
    .
\end{equation}
We introduced the Nambu spinor $\Psi =[\psi_{L,\uparrow},\overline{\psi}_{L,\downarrow},\psi_{R,\uparrow},\overline{\psi}_{R,\downarrow}]^T$ and the Nambu Green function of the uncoupled electrodes through 
\begin{equation}
     g_\mathcal{C}^{} = \textrm{diag}(g^{}_{\mathcal{C},L},g^{}_{\mathcal{C},R}),\quad g_{\mathcal{C},a}^{-1}(t,t') = \Bqty{ i\partial_t -\bqty{ h_a(-i\nabla\tau_3)\tau_3  + \Delta_a \tau_+ + \Delta_a^* \tau_-  + \frac{1}{2}eV \tau_3 (\rho_3)_{aa} }}\delta_\mathcal{C}(t,t'),
\end{equation}
where $\tau_i$ denote Pauli matrices in particle-hole space, and $\rho_i$ denote Pauli matrices in electrode space. In the appendix we distinguish operators which are defined on the Keldysh contour with the subscript $\mathcal{C}$ from their Keldysh matrix counterparts. For ease of notation, in the main text we do not make this distinction. We leave the boundary terms implicit which encode the initial occupations of the superconducting leads and couple the forward and backward branches of the contour. These terms are accounted for when inverting $g_\mathcal{C}^{}$, leading to Eq.~\eqref{eq:contour_green_function_matrix_eq} \cite{kamenev2011field}. 
In Eq.~\eqref{eq:partition_function} we also introduced the tunneling operator 
\begin{equation}
    \mathcal{T}_\mathcal{C}(t,t';\mathbf{r}) = \delta_\mathcal{C}(t,t') \delta(\mathbf{r}) \sum_\pm \mathcal{T}_\pm,
\end{equation}
with coefficients $\mathcal{T}_\pm$ defined in Eq.~\eqref{eq:mathcal_T_pm}. $\delta_\mathcal{C}(t,t')$ is the generalization of the Dirac delta function to the Keldysh contour, enforcing coinciding times on the same branch of the contour (explicitly: $\delta_\mathcal{C}(t,t') = \delta (t-t')$ if $t,t'$ are on the forward branch, $\delta_\mathcal{C}(t,t') = - \delta (t-t')$ if $t,t'$ are on the backward branch, and zero otherwise). 
We can now perform the integral over the fermion field $\Psi$ and find
\begin{equation}
    \mathcal{Z}
    = \int [dV]\prod_a [d\Delta_a][d\Delta^*_a]\, 
    \exp{ i\int_\mathcal{C} dt\, 
    \left[ \frac{1}{2}CV^2 +  
    \int  d\mathbf{r} \sum_a \frac{1}{\lambda_a} \abs{ \Delta_a}^2
    \right] + \Tr_\mathcal{C} \ln \bqty{ -i\pqty{g_\mathcal{C}^{-1} - \mathcal{T}_\mathcal{C} } } },
\end{equation} 
where the trace $\mathrm{Tr}_\mathcal{C}$ is taken over the space, contour, electrode, and particle-hole degrees of freedom.

We simplify the multiple functional integral by making a few physically motivated assumptions. First, we assume a sufficiently weak Josephson link, so that self-inductance effects can be neglected. This justifies that we did not include a vector potential. We further assume that the superconducting electrodes are well described by mean-field theory. Writing $\Delta_a = |\Delta_a|\exp( i \phi_a )$, we can then replace the magnitude $|\Delta_a|$ of the order parameter by its mean-field value, which we assume to be independent of space and time. Moreover, the rigidity of the order-parameter phases $\phi_a$ implies that they are spatially uniform within each electrode. 

We are left with functional integrals over the voltage field $V$ as well as the phase fields $\phi_L$ and $\phi_R$, all of which are functions of time. These integrals can be simplified by  performing a gauge transformation
\begin{equation}
    \mathcal{U} = \mathcal{U}_L\mathcal{U}_R, \quad \mathcal{U}_a(t,t')  = \exp{-\frac{i}{2}\tau_3 p_a \phi_a(t)}\delta_\mathcal{C}(t,t').
\end{equation}
where $p_a = [1 + (\rho_3)_{aa}]/2$. 
This eliminates the phases of the order parameter at the expense of introducing additional terms into the single-particle Green function and the tunneling amplitudes. In effect, one implements the replacements
\begin{equation}
    g_{\mathcal{C},a}^{-1} \quad\longrightarrow\quad \mathcal{U}_a g_{\mathcal{C},a}^{-1} \mathcal{U}_a^\dagger  = \Bqty{  i\partial_t -\bqty{ h_a(-i\nabla\tau_3)\tau_3  + \abs{\Delta_a} \tau_1  +  e\Phi_a(t) \tau_3 } } \delta_\mathcal{C}(t,t'),
\end{equation}
as well as
\begin{equation}
    \mathcal{T}_\mathcal{C}\quad \longrightarrow \quad
   \mathcal{U} \mathcal{T}_\mathcal{C}  \mathcal{U}^\dagger = \delta_\mathcal{C}(t,t') \delta(\mathbf{r}) \sum_\pm e^{\mp \frac{i}{2} (\phi_R(t) - \phi_L(t)) } \mathcal{T}_\pm.
\end{equation}
Here, we defined the electric potentials in the left and right electrodes,
\begin{equation}
     e\Phi_L = \frac{1}{2} \pqty{ eV + \partial_t \phi_L },\quad  
     e\Phi_R =   \frac{1}{2}  \pqty{ -  eV + \partial_t \phi_R}.
\end{equation}
The finite compressibility of superconductors constrains both $\Phi_L$ and $\Phi_R$ to zero by bulk energies. Formally, this can be confirmed by expanding the action of the uncoupled electrodes to quadratic order in $\Phi_L$ and $\Phi_R$. 
This implies the conditions
\begin{equation}
     \partial_t \phi_L = -eV, \quad 
     \partial_t \phi_R = eV,
\end{equation}
so that the phase difference
\begin{equation}
    \phi = \phi_R - \phi_L
\end{equation}
across the junction satisfies the Josephson relation
\begin{equation}
     \partial_t \phi = 2eV.
\end{equation}
This condition eliminates the integral over the voltage field $V$, and we are left with a functional integral over the phase difference $\phi$ only.

Finally, it is convenient to introduce $\mathcal{Z}_0 = 1$ with 
\begin{equation}
  \mathcal{Z}_{0} = \prod_{a\in \{L,R\}} \int [d\psi_a][d\overline{\psi}_a]\, \exp{
  i\int_\mathcal{C} dt\, \int d\mathbf{r} \sum_\sigma \overline{\psi}^{}_{a,\sigma} 
  \bqty{
  i\partial_t - h_a (-i\nabla)
  }
  \psi_{a,\sigma}}
\end{equation}
describing the normal state of the uncoupled junction as a reference point. Then, we can write
\begin{equation}
\mathcal{Z} = \frac{\mathcal{Z}}{\mathcal{Z}_0}
= \int [d\phi]
\exp{
i \int_\mathcal{C} dt\, \frac{C}{8e^2}(\partial_t\phi)^2
+ \Tr_\mathcal{C} \ln \bqty{ - i  \pqty{g_\mathcal{C}^{-1} - \mathcal{T}_\mathcal{C} } } 
- \Tr_\mathcal{C} \ln \bqty{- i g_{\mathcal{C},0}^{-1} }
},
\end{equation}
where the Green function $g^{}_{\mathcal{C},0}$  describes the uncoupled, normal state electrodes. As we set $\Phi_L=\Phi_R=0$,  $g^{}_{\mathcal{C},0}$ and $g_{\mathcal{C}}^{}$ differ only by the absence or presence of the  pairing terms, respectively. We can write
\begin{equation}
\mathcal{Z}
= \exp{
i S_0 
 } \int [d\phi]\,
\exp {
i \int_\mathcal{C} dt\, \frac{C}{8e^2}(\partial_t\phi)^2
+ \Tr_\mathcal{C} \ln \bqty{ 1 - g_\mathcal{C}^{}\mathcal{T}_\mathcal{C} }
 },
\end{equation}
where $iS_0 = \Tr_\mathcal{C}\ln \bqty{g^{}_{\mathcal{C},0} g_\mathcal{C}^{-1}}$ describes superconductivity of the uncoupled electrodes within mean-field theory, while the functional integral governs the phase dynamics of the Josephson junction. We note that due to the local structure of $\mathcal{T}$, only the local Green function $\bra{\mathbf{r} = \mathbf{0}} g_\mathcal{C}^{} \ket{\mathbf{r} = \mathbf{0}}$ enters the phase action. Thus, we can take $g_\mathcal{C}^{}$ and $\mathcal{T}_\mathcal{C}$ to refer to the local quantities, with traces no longer running over space. This concludes the derivation of the action in Eq.~\eqref{eq:action} of the main text.

\section{Current and noise kernel}
\label{app:tun_cur_and_noise}

In this appendix we derive the expressions for the tunneling current and noise in terms of the dressed Green function, Eq.~\eqref{eq:tun_cur} and \eqref{eq:tun_noise}. 

We first comment on the inversion of $g^{-1}_\mathcal{C}$. Proper inclusion of boundary terms gives \cite{kamenev2011field}
\begin{equation}\label{eq:contour_green_function_matrix_eq}
    g^{}_\mathcal{C} (t,t') = \begin{pmatrix}
        g^\mathrm{T}(t-t') & g^<(t-t') \\ g^>(t-t') & g^{\tilde{\mathrm{T}}}(t-t')
    \end{pmatrix}_{ij} = \bqty{\lambda_3 g (t,t') }_{ij},
\end{equation}
where $i = 1$ ($j=1$) if $t$ ($t'$) is on the forward contour, and $i = 2$ ($j=2$) if $t$ ($t'$) is on the backward contour. The $\lambda_i$ are Pauli matrices that we take to act in this forward and backward contour space. We also introduced the time-ordered and anti-time-ordered Green functions, 
\begin{align}
    g^\mathrm{T} =  \Theta(t-t')g^>(t-t') + \Theta(t'-t)g^<(t-t'),\quad 
    g^{\tilde{\mathrm{T}}} =  \Theta(t'-t)g^>(t-t') + \Theta(t-t')g^<(t-t'),
\end{align}
respectively, as well as the lesser ($g^<$), and greater Green function ($g^>$), where $\Theta$ is the Heaviside step function. Assuming the superconducting electrodes are in equilibrium at the initial time, the lesser and greater Green functions satisfy the usual frequency space decompositions 
\begin{equation}
    g^<(\omega) = -n_F(\omega)[g^r(\omega) - g^a(\omega)],\quad g^<(\omega) = [1-n_F(\omega)][g^r(\omega) - g^a(\omega)],
\end{equation}
in terms of the retarded and advanced Green functions 
\begin{equation}
    g^{r/a} (\omega) =  \bra{\mathbf{0}}( \omega \pm i 0 - h_\textrm{BdG})^{-1} \ket{\mathbf{0}},
\end{equation}
with $h_\textrm{BdG} = \textrm{diag}(h_{\textrm{BdG},L},h_{\textrm{BdG},R})$, $h_{\textrm{BdG},a} = h_a(-i\nabla\tau_3)\tau_3 + \abs{\Delta_a}\tau_1$. 

When evaluating convolutions on the Keldysh contour, there is an additional negative sign if the time is on the backward contour. In the second equality in Eq.~\eqref{eq:contour_green_function_matrix_eq}, we absorbed this sign into the definition of the matrix Green function $g$ (see also Eq.~\eqref{eq:contour_green_function_matrix} in the main text). From now on (with the exception of App.~\ref{app:fluctuation_dissipation}), all convolutions in contour space are expressed as convolutions in time and matrix multiplication. The tunneling operator is now 
\begin{equation}
    \mathcal{T}(t) = \sum_\pm \mathcal{T}_\pm \begin{pmatrix}
         e^{\mp\frac{i}{2}\phi_+(t)} & 0 \\ 0 & 
          e^{\mp\frac{i}{2}\phi_-(t)}
    \end{pmatrix},
\end{equation} 
where $\phi_\pm (t) = \phi(t \in \mathcal{C}_\pm)$ are the forward and backward copies of the phase difference. 

To proceed, we expand the tunneling operator $\mathcal{T}$ to quadratic order in the contour-odd phase difference $\chi (t) = \phi_+ (t) - \phi_-(t) $. This gives $\mathcal{T} = \mathcal{T}_0 + \mathcal{T}_1 + \frac{1}{2}\mathcal{T}_2 + \ldots$, where $\mathcal{T}_0(t) = \mathcal{T}(t) \vert_{\chi=0}$, 
\begin{align}   
    \mathcal{T}_1(t) =&\ - \frac{i}{4} \mathcal{T}_0 (t) \lambda_3 \tau_3 \rho_3   \chi(t),    
\end{align}
and $\mathcal{T}_2 (t) = - \tfrac{1}{16} \mathcal{T}_0 (t) \chi^2(t)$. 
We also define the dressed Green function $    \mathcal{G} =\bqty{g^{-1} - \mathcal{T}_0}^{-1}$. Note that $\mathcal{G}$ is a functional of the classical phase difference $\varphi(t)$, $\mathcal{G}=\mathcal{G}([\varphi];t,t')$. To quadratic order in $\chi$, we can then expand
\begin{equation}
      \Tr\ln\bqty{1- g\mathcal{T}} = 
    - \Tr\bqty{\mathcal{G} \mathcal{T}_1 + \tfrac{1}{2}\mathcal{G}\mathcal{T}_2  + \tfrac{1}{2}\mathcal{G} \mathcal{T}_1\mathcal{G} \mathcal{T}_1 } + \ldots ,
    \label{eq:trexpansion}    
\end{equation}
in Eq.~\eqref{eq:action}. Here, we used that the zeroth order in $\chi$ vanishes due to cancellation between the forward and backward branches of the contour.
The $\mathcal{T}_2$ term vanishes for the same reason. 
By comparison with Eq.~\eqref{eq:expansionchi}, we can read off the tunneling current
\begin{equation}
    \mathcal{I} ([\varphi];t) =  -  \frac{e}{2} \tr\Bqty{ \bqty{\lambda_3\mathcal{G}(t,t)}  \mathcal{T}_0(t)   \tau_3 \rho_3 },
\end{equation}
and the fluctuation kernel
\begin{equation}
    \mathcal{K}([\varphi];t,t')  = - \frac{e^2}{4} \tr\Bqty{\bqty{\lambda_3\mathcal{G}(t,t')}  \mathcal{T}_0(t') \tau_3 \rho_3 \bqty{\lambda_3\mathcal{G}(t',t)}\mathcal{T}_0(t) \tau_3 \rho_3 }. 
\end{equation}
After expanding in contour space and dropping the subscript of $\mathcal{T}_0$, we find the expressions Eq.~\eqref{eq:tun_cur} and \eqref{eq:tun_noise} in the main text. 
We used that the equal-time argument should be understood as $(t\mp_\mathcal{C} 0_+, t)$ on the forward and backward contour such that both the time-ordered and anti-time-ordered components reduce to the lesser component at equal time. 

\section{Fluctuation-dissipation theorem}
\label{app:fluctuation_dissipation}

\subsection{Proof of \texorpdfstring{Eq.~\eqref{eq:fdt_symmetry}}{Eq. (66)}}
\label{app:fluctuation_dissipation_1}

We want to prove Eq.~\eqref{eq:fdt_symmetry}. We do not follow Refs.~\cite{tobiska2005inelastic,saito2008symmetry} but present a calculation based on the Green function expression of the tunneling action in Eq.~\eqref{eq:tun_action_2},
\begin{equation}
    iS_\textrm{tun}[\varphi,\chi] = \Tr \ln \pqty{1- g_L \mathcal{T}_{LR} g_R \mathcal{T}_{RL} },
\end{equation}
where the tunneling matrices are explicitly given by
\begin{equation}
    \mathcal{T}_{RL} = \pm_\mathcal{C} \vartheta \tau_3 e^{-\frac{i}{2}\pqty{\varphi \pm_\mathcal{C} \frac{\chi}{2}}\tau_3 } \quad , \quad \mathcal{T}_{LR} = \pm_\mathcal{C} \vartheta \tau_3 e^{\frac{i}{2}\pqty{\varphi \pm_\mathcal{C} \frac{\chi}{2}}\tau_3 }.
\end{equation}
(For the present purpose, we find it more convenient to absorb the contour sign $\pm_\mathcal{C}$ into $\mathcal{T}$ rather than $g$, different from the previous section and the main text.) We observe that these tunneling operators satisfy
\begin{equation}
    \mathcal{T}^{}_{RL} = \mathcal{T}_{RL}^\top \quad , \quad \mathcal{T}^{}_{LR} = \mathcal{T}_{LR}^\top,
\end{equation}
i.e., they are symmetric under transposition (in time, contour, and particle-hole space). Moreover, they obey 
\begin{equation}
\mathcal{T}_{RL}(\varphi,\chi) = \mathcal{T}_{LR}(-\varphi,-\chi).
\end{equation}
These facts encode microreversibility of the tunneling Hamiltonian.  

Now consider the shifted quantity
\begin{equation}
     iS_\textrm{tun}[-\varphi(-t)+iF,-\chi(-t) + iX] = \Tr \ln \pqty{1- g^{}_L \mathcal{T}'_{RL} g^{}_R \mathcal{T}_{LR}'  },
\end{equation}
where we used shorthands 
\begin{subequations}
\begin{align}
    \mathcal{T}_{RL}'  =&\ \mathcal{T}^{}_{RL}(\varphi(-t),\chi(-t))  e^{- \frac{1}{2}(F \pm_\mathcal{C} X/2) \tau_3 },\\ 
    \mathcal{T}_{LR}' =&\ \mathcal{T}^{}_{LR}(\varphi(-t),\chi(-t))  e^{\frac{1}{2}(F \pm_\mathcal{C} X/2) \tau_3 }.
\end{align}    
\end{subequations}
We aim to show that this expression reduces to $iS_\textrm{tun}[\varphi(t),\chi(t)]$ for an appropriate choice of $F,X$. First, to restore the order of the electrode labels, transpose under the trace (again, the transposition includes time arguments) and let all time arguments to minus time (such that the time-transposition is undone for the equilibrium components which depend only on the difference of their time arguments). This gives 
\begin{equation}
     iS_\textrm{tun}[-\varphi(-t)+iF,-\chi(-t) + iX]   = \Tr \ln \pqty{1-    \mathcal{T}'_{LR}  g_R^\texttt{t} \mathcal{T}'_{RL}  g_L^\texttt{t}  }, 
\end{equation}
where the transpose $g_a^\texttt{t}$ is now only in the  contour and particle-hole spaces, and $\mathcal{T}'_{RL},\ \mathcal{T}'_{LR}$ are now evaluated for $\varphi(t)$ and $\chi(t)$. Assuming that the system is time-reversal symmetric, we can choose the Bogoliubov-de Gennes Hamiltonian to be symmetric under transposition. This implies that the time-ordered, lesser, greater, and anti-time-ordered Green functions are symmetric under transposition in particle-hole space. Then, the transpose effectively only acts in contour space, where it exchanges the lesser and greater components. To bring the contour matrix structure back to its original form, we use the Kubo-Martin-Schwinger condition $g^>(\omega) = - e^{\omega/T} g^<(\omega)$. In the time domain, this translates to
\begin{equation}
   g^\texttt{t}(t,t') = 
   \begin{pmatrix}
        g^{\mathrm{T}} (t,t') & g^> (t,t') \\    
        g^<(t,t') & g^{\tilde{\mathrm{T}}} (t,t') 
    \end{pmatrix} 
    = \begin{pmatrix}
        e^{i\overset{\rightarrow}{\partial}_t /2T } & 0 \\    
        0 & - e^{-i\overset{\rightarrow}{\partial}_t /2T} 
    \end{pmatrix}
    \begin{pmatrix}
        g^{\mathrm{T}} (t,t') & g^< (t,t') \\    
        g^>(t,t') & g^{\tilde{\mathrm{T}}} (t,t') 
    \end{pmatrix}
    \begin{pmatrix}
        e^{i\overset{\leftarrow}{\partial}_{t'} /2T} & 0 \\    
        0 & - e^{-i\overset{\leftarrow}{\partial}_{t'} /2T} 
    \end{pmatrix}.
\end{equation}
We integrate by parts to have the $\partial_{t'}$  time derivatives act to the right instead (i.e., on the tunneling operator and the next Green function). This gives the factors
\begin{equation}
    \pqty{\pm_\mathcal{C} e^{\mp_\mathcal{C} i\partial_{t} /2T}} \mathcal{T}_{RL}'(t) \pqty{\pm_\mathcal{C} e^{\pm_\mathcal{C} i\partial_{t} /2T}} g_L(t,t') \simeq \mathcal{T}_{RL}^{}(t)  e^{ -\frac{1}{4T}\pqty{\pm_\mathcal{C} \dot{\varphi}  + \frac{1}{2} \dot{\chi}(t) }\tau_3 } e^{- \frac{1}{2}(F \pm_\mathcal{C} X/2) \tau_3 } g_L(t,t') ,
\end{equation}
and, after using cyclicity of the trace inside the log, a similar expression for $\mathcal{T}_{LR} g_R$ (with flipped overall sign in the exponential). In view of the adiabatic approximation, we can choose $X = - \dot{\varphi}/T$ and $F = - \dot{\chi}/4T$ to eliminate the exponential factors. In the following, we neglect $\dot{\chi}/4T$, i.e., we focus on correlations on times large compared to $T^{-1}$. Fixing $\varphi(t) = \varphi_0 + 2eVt$, and using cyclicity once more, we obtain the relation
\begin{equation}\label{eq:fdt_symmetry_2}
    iS_\textrm{tun}[\varphi_0 + 2eVt ,\chi(t)] =
    iS_\textrm{tun}[-\varphi_0 + 2eVt,-\chi(-t) - i2eV/T ]. 
\end{equation}
Equation \eqref{eq:fdt_symmetry} then follows by noting that the capacitive term vanishes for $\varphi(t)$, which is linear in time. 

This proof straightforwardly extends to spin-orbit coupled superconductors with time-reversal symmetry. Choosing time reversal to act as $U \mathrm{K}$, with $U$ unitary, time-reversal implies for the retarded and advanced Green functions $g^{r,a}(t,t') = U^\dagger (g^{r,a})^\texttt{t}(t,t')U$. By extension this holds also for the lesser, greater, time-ordered, and anti-time-ordered Green functions. For the tunneling operators time-reversal symmetry implies that $\mathcal{T}^{}_{RL}(\varphi,\chi) = U^\dagger \mathcal{T}^\texttt{t}_{LR}(-\varphi,-\chi) U$ and that $\mathcal{T}^{}_{RL}(\varphi+iF,\chi+iX) =  \mathcal{T}^{}_{RL}(\varphi,\chi) e^{\frac{1}{2}(F \pm_\mathcal{C} X/2) \tau_3 }$. Using these identities, one can follow the same steps as above to find that Eq.~\eqref{eq:fdt_symmetry_2} still holds.

\subsection{Tower of fluctuation-dissipation relations}
\label{app:fluctuation_dissipation_2}

Define the even and odd under time-reversal quantities 
\begin{equation}
iS^\pm_\textrm{tun} [\chi(t)] = \frac{1}{2}\pqty{iS_\textrm{tun}[\varphi_0 + 2eVt,\chi(t)] \pm iS_\textrm{tun}[-\varphi_0 + 2eVt,\chi(-t)]}.  
\end{equation}
It is straightforward to see that $iS^+_\textrm{tun}$ ($iS^-_\textrm{tun}$) generates the (anti)-symmetrized cumulants
\begin{subequations}
\begin{align}
    \mathcal{I}_\pm(\varphi(t),V) =&\ \frac{1}{2} \bqty{\mathcal{I}(\varphi(t),V) \pm \mathcal{I}(-\varphi(t),V)},\\ 
    \mathcal{K}_\pm(\varphi(t),V;t-t') =&\ \frac{1}{2} \bqty{\mathcal{K}(\varphi(t),V;t-t') \pm \mathcal{K}(-\varphi(t),V;t'-t)}, 
\end{align}
\end{subequations}
and similar for higher orders. (Note that we are still exclusively setting $\varphi(t) = \varphi_0 + 2eV t$ here and in the following.) Using Eq.~\eqref{eq:fdt_symmetry_2}, we obtain  
\begin{equation}
    iS^\pm_\varphi[\chi(t)] = \pm iS^\pm_\varphi[-\chi(+t)-i2eV/T]. 
\end{equation}
Expand both sides in their respective second arguments. Furthermore, expand the right hand side for small $eV/T$ up to second order. The normalization condition gives $iS^\pm_\varphi[0] = 0$ as well as $iS^\pm_\varphi[-i2eV/T] = 0$. These conditions imply that terms in the expansion without any factors of $\chi$ should be dropped. This gives
\begin{multline}
    \sum_{n \geq 1} \frac{1}{(2ei)^n n! } \int_{t_1,\ldots,t_n} \chi(t_1) \ldots \chi(t_n) \mathcal{K}^{(n)}_\pm (t_1,\ldots,t_n)= \pm \sum_{n \geq 1} \frac{(-1)^n}{(2ei)^n n! } \int_{t_1,\ldots,t_n}  \chi(t_1) \ldots \chi(t_{n-2}) \\ \times \bqty{ \chi(t_{n-1})\chi(t_n) + n \chi(t_{n-1}) \frac{2eiV}{T} + \frac{1}{2} n(n-1) \pqty{\frac{2eiV}{T}}^2 + \ldots } \mathcal{K}^{(n)}_\pm (t_1,\ldots,t_n),
\end{multline}
where on the right hand side we used that $\mathcal{K}^{(n)}$ is completely symmetric in its time arguments and we used the shorthand $\int_{t_1,\ldots,t_n} = \int dt_1 \ldots dt_n$. Due to the normalization condition $iS^\pm_\varphi[-i2eV/T] = 0$ it is understood that the second and third term are zero for $n<2$ and $n<3$, respectively. Thus, shifting $n \to n+1$ in the second and $n \to n+2$ in the third term on the right hand side, and equating terms of equal order in $\chi$ under the integral, we obtain 
\begin{multline}\label{eq:fdt_expansion}
    \bqty{1 \mp (-1)^n} \mathcal{K}^{(n)}_\pm (t_1,\ldots,t_n)=      \frac{ V}{T} \int dt_{n+1}\, \mathcal{K}^{(n+1)}_\pm (t_1,\ldots,t_n,t_{n+1}) \\ - \frac{1}{2}\pqty{\frac{V}{T}}^2 \int dt_{n+1}\int dt_{n+2}\, \mathcal{K}^{(n+2)}_\pm (t_1,\ldots,t_n,t_{n+1},t_{n+2}) + \ldots.
\end{multline}
Consider first the time-reversal symmetric part (upper sign) and odd $n$. Then, we obtain to lowest order in $V/T$
\begin{align} \label{eq:fdt_trs_odd_n}
    \frac{2T}{V} \mathcal{K}^{(n)}_+ (t_1,\ldots,t_n) = \int dt_{n+1}\, \mathcal{K}^{(n+1)}_+ (t_1,\ldots,t_n,t_{n+1}).
\end{align}
For even $n$, the left hand side of Eq.~\eqref{eq:fdt_expansion} vanishes, but so does the right: it is proportional to Eq.~\eqref{eq:fdt_trs_odd_n} integrated over time. The time-reversal anti-symmetric cumulants satisfy the equivalent relation but for even $n$ instead. 

Focus on $n=1$ for which the left hand side involves the current and the right hand side the noise fluctuations. Treating the microscopic timescale as fast, we may replace the integral over $t_2$ with an integral over relative time $t_1 - t_2$. Then the right hand side is just the zero frequency noise power. Explicitly, we obtain 
\begin{align}
    \frac{2T}{V}  \mathcal{I}_+ (\varphi(t),V) =  \mathcal{K}_+ (\varphi(t),V).
\end{align}
Plugging in the Floquet expansions for $\mathcal{I}$ and $\mathcal{K}$, we obtain our final result
\begin{equation}
    \frac{2T}{V} \Re\bqty{\mathcal{I}_n(V)} = \Re\bqty{\mathcal{K}_n(V)}.
\end{equation}

\subsection{Nonlinear fluctuation-dissipation relation without time-reversal symmetry}
\label{app:fluctuation_dissipation_3}

Here, we show that the nonlinear fluctuation-dissipation relation \begin{equation}
    \frac{2T}{V}\mathcal{I}_0(V) = \mathcal{K}_0(V)
\end{equation}
relating the $dc$ current and noise holds even in the absence of time-reversal symmetry. In this situation, our previous approach does not work since we can no longer relate the transposed Green functions and tunneling operators. Instead, we focus directly on the Green function expressions for $2T\mathcal{I}_0/V$ and $\mathcal{K}_0$ and show that they agree to all orders in tunneling up to corrections of order $eV/T$. We exploit the cyclic structure in the $dc$ components: the same number of $\mathcal{T}_+$ and $\mathcal{T}_-$ operators have to feature in all terms in the expansions of $\mathcal{I}_0$ and $\mathcal{K}_0$ for the overall phase dependence to cancel. One can use cyclicity to sum many equivalent sequences of the $\mathcal{T}_+$ and $\mathcal{T}_-$ tunneling processes.

We expand the Dyson Eq.~\eqref{eq:dyson} as $\mathcal{G} = g + g \mathcal{T} g + g \mathcal{T} g \mathcal{T} g  + \ldots$. Performing the Moyal products order by order, we find for $\mathcal{I}_0(V)$ at order $\vartheta^{2r}$,
\begin{equation}\label{eq:I_2r_app}
    \mathcal{I}^{(2r)}_0(V) = - e \int \frac{d\varepsilon}{2\pi} \sum_{m_1...m_{2r}} m_{2r}  \tr\bqty{
    g(\varepsilon_1) \mathcal{T}_{m_1}  ... \mathcal{T}_{m_{2r-1}}g(\varepsilon_{2r}) \mathcal{T}_{m_{2r}} }^<  \, \delta_{\sum_{i=1}^{2r} m_i,0},
\end{equation}
where $\varepsilon_j = \varepsilon + eV \sum_{i=j}^{2r-1} m_i$ for $j<2r$, $\varepsilon_{2r} = \varepsilon$, and $m_i = \pm 1$. 

It is useful to view the energy arguments $\varepsilon_j$ as a discrete walk of length $2r$ on a ladder, with periodic boundary condition (i.e., the walk ends on the same rung if the step $2r$ to $1$ is included). The constrained sum over all $m_i$ corresponds to a sum over all distinct such walks. Note that due to the periodicity, each walk is $2r$-fold degenerate in the sense that there are $2r$ equivalent walks if the cyclicity of the trace is taken into account. Technically, some walks have periodic substructure, and hence have lower degeneracy. For example, $+-+-$ and $-+-+$ are only two-fold rather than four-fold degenerate. In order to evaluate sums over equivalent walks below, we write them as a sum over the starting point within a given walk and introduce a normalization factor for walks with lower degeneracy to account for overcounting. E.g., for $(m_1m_2m_3m_4) \in \{(+-+-),(-+-+) \}$, we rewrite $\sum f(m_1,m_2,m_3,m_4) = \frac{1}{2} \sum_{j = 1}^4 f (m_{1+j},m_{2+j},m_{3+j},m_{4+j})$ where addition is understood modulo $4$. 

Next, we expand the product of Green functions using Langreth's rule $[g_1 g_2]^< = g_1^r g^< + g^< g^a$ iteratively. We use the shorthands $R_i = g^r(\varepsilon_i) \mathcal{T}_{m_i}$, $A_i = g^a(\varepsilon_i) \mathcal{T}_{m_i}$, and $n_i = n_F(\varepsilon_i)$. The product of Green functions becomes 
\begin{equation}\label{eq:langreth_lesser}
      m_{2r}\tr{R_1...R_{2r} n_{2r} + ... + R_1 ...R_{j}A_{j+1} ... A_{2r} (n_{j} - n_{j+1}) + ... - A_1 ... A_{2r} n_1},
\end{equation}
where we expressed the equilibrium lesser Green function as $g^< = -n_F(g^r-g^a)$.
It will prove useful to perform the sum over equivalent walks for the first ($R...R$) and last ($A...A$) term. By the cyclic property of the trace, we can write them as 
\begin{equation}\label{eq:RRR_AAA_lesser}
     \sum_{m_1...m_{2r}}' \bqty{ \tr{R_1...R_{2r}} N_R  - \tr{A_1...A_{2r}} N_A } \, \delta_{\sum_{j=1}^{2r} m_j,0},
\end{equation}
Here, $j = 2r+1 = 1$, and the prime on the sum denotes summation only over inequivalent walks. We also defined the quantities
\begin{equation}
    N_R = \sum_{j=1}^{2r} m_j n_j,\quad N_A = \sum_{j=1}^{2r} m_j n_{j+1}.
\end{equation}

We will now show that the leading-order terms in temperature of $\mathcal{K}_0(V)$ can be brought to the same form. At order $\vartheta^{2r}$, the $dc$ noise is 
\begin{multline}
    \mathcal{K}^{(2r)}_0(V) =  
    - e^2  \int \frac{d\varepsilon}{2\pi} \sum_{l=1}^{2r-1}
    \\ \times
    \sum_{m_1...m_{2r}} m_l m_{2r} \tr\Bqty{\bqty{
    g(\varepsilon_1) \mathcal{T}_{m_1}  ... \mathcal{T}_{m_{l-1}}g(\varepsilon_{l}) \mathcal{T}_{m_{l}} }^<   \bqty{g(\varepsilon_{l+1}) \mathcal{T}_{m_{l+1}}  ... \mathcal{T}_{m_{2r-1}}g(\varepsilon_{2r}) \mathcal{T}_{m_{2r}}}^>    } \, \delta_{\sum_{j=1}^{2r} m_j,0} .
\end{multline}
The first square bracket can be expanded as in Eq.~\eqref{eq:langreth_lesser}. The second square bracket can also be expanded in a similar way, but letting $-n_j \to (1-n_j)$. Both brackets thus involve two different types of terms: those that are products of only retarded or only advanced Green functions, and those that involve both Green-function species. The former come with a single factor of $n_j$ or $1-n_j$, while the latter always carry the difference $n_{j} - n_{j+1}  = n_F(\varepsilon_j)-n_F(\varepsilon_j - m_j eV)$. This is of order $eV/T$ and can thus be dropped against the terms of the first type. With this, we can write the integrand as (keeping the condition $\sum_i m_i = 0$ implicit)
\begin{multline}
    \sum_{m_1...m_{2r}}\sum_{l=1}^{2r-1}
    \tr{\bqty{ R_1...R_{l} n_{l}  - A_1 ... A_{l} n_1} \tau_3\rho_3 \bqty{R_{l+1}...R_{2r} (1-n_{2r})  - A_{l+1} ... A_{2r} (1- n_{l+1}) } \tau_3\rho_3 } \\ 
    = 
    \sum_{m_1...m_{2r}}\sum_{l=1}^{2r-1} m_{l} m_{2r} 
    \tr\big\{ R_1 ... R_{2r}  n_{l} (1-n_{2r})  + A_1 ... A_{2r} n_1 (1- n_{l+1}) \\
    - R_{1}...R_{l}A_{l+1}...A_{2r} n_l(1-n_{l+1})  
    - A_{1} ... A_{l}R_{l+1}...R_{2r} n_1 (1- n_{2r})   \big\}. 
\end{multline}
To bring the last term into the form $R...RA...A$ we first use the cyclic property of the trace. Then, we exploit the summation over equivalent walks to relabel $l+1 \to 1$, $l+2 \to 2$, ..., $2r \to 2r - l$, $1 \to 2r - l$, ..., $l \to 2r$. Finally, we set $l' = 2r - l$ and swap the relevant terms in the sum over $l$. This gives 
\begin{equation}\label{eq:langreth_lesser_greater_2}
    \sum_{m_1...m_{2r}}\sum_{l=1}^{2r-1} m_{l} m_{2r} 
    \tr{ R_1 ... R_{2r}  n_{l} (1-n_{2r})  + A_1 ... A_{2r} n_1 (1- n_{l+1}) 
    - 2 R_{1}...R_{l}A_{l+1}...A_{2r} n_l(1-n_{l+1})   }.     
\end{equation}

We now isolate the leading-order terms in temperature in this expression. Consider first the last term in Eq.~\eqref{eq:langreth_lesser_greater_2}. Using
\begin{align}
    -2m_l n_l(1-n_{l+1})
    =&\ \frac{2T}{eV}( n_l - n_{l+1} ) + \order{\frac{eV}{T}},
\end{align}
and including the sum over $l$, this, up to the prefactor, precisely reproduces the mixed terms in Eq.~\eqref{eq:langreth_lesser}.  
Consider now the $R...R$ and $A...A$ terms in Eq.~\eqref{eq:langreth_lesser_greater_2}. Performing the sum over equivalent walks, we can bring them to the form 
\begin{equation}\label{eq:RRR_AAA_noise_term}
    -\sum_{m_1...m_{2r}}'   \Bqty{ \tr{R_1...R_{2r}} \bqty{ \sum_{j=1}^{2r} n_j (1-n_j) + N_R^2 } +  \tr{A_1...A_{2r}} \bqty{ \sum_{j=1}^{2r} n_j (1-n_j) + N_A^2 }    },
\end{equation}
Note that due to the constraint $\sum_i m_i = 0$, both $N_R$ and $N_A$ are of order $eV/T$, and thus contribute to Eq.~\eqref{eq:RRR_AAA_noise_term} only at order $(eV/T)^2$. Finally, noting that 
\begin{align}
    -\sum_{j=1}^{2r} n_j (1-n_j) =&\   \frac{2T}{eV} N_R + \order{\frac{e^2V^2}{T^2}} = - \frac{2T}{eV} N_A + \order{\frac{e^2V^2}{T^2}},
\end{align}
and comparing to Eq.~\eqref{eq:RRR_AAA_lesser}, we can conclude that 
\begin{equation}
    \mathcal{K}_0^{(2r)} (V) =  \frac{2T}{V} \mathcal{I}_0^{(2r)}(V) + \order{\frac{eV}{T}}.
\end{equation}
This establishes the desired nonlinear fluctuation-dissipation theorem order-by-order.

\twocolumngrid 
\bibliography{library}

\begin{thebibliography}{90}%
\makeatletter
\providecommand \@ifxundefined [1]{%
 \@ifx{#1\undefined}
}%
\providecommand \@ifnum [1]{%
 \ifnum #1\expandafter \@firstoftwo
 \else \expandafter \@secondoftwo
 \fi
}%
\providecommand \@ifx [1]{%
 \ifx #1\expandafter \@firstoftwo
 \else \expandafter \@secondoftwo
 \fi
}%
\providecommand \natexlab [1]{#1}%
\providecommand \enquote  [1]{``#1''}%
\providecommand \bibnamefont  [1]{#1}%
\providecommand \bibfnamefont [1]{#1}%
\providecommand \citenamefont [1]{#1}%
\providecommand \href@noop [0]{\@secondoftwo}%
\providecommand \href [0]{\begingroup \@sanitize@url \@href}%
\providecommand \@href[1]{\@@startlink{#1}\@@href}%
\providecommand \@@href[1]{\endgroup#1\@@endlink}%
\providecommand \@sanitize@url [0]{\catcode `\\12\catcode `\$12\catcode `\&12\catcode `\#12\catcode `\^12\catcode `\_12\catcode `\%12\relax}%
\providecommand \@@startlink[1]{}%
\providecommand \@@endlink[0]{}%
\providecommand \url  [0]{\begingroup\@sanitize@url \@url }%
\providecommand \@url [1]{\endgroup\@href {#1}{\urlprefix }}%
\providecommand \urlprefix  [0]{URL }%
\providecommand \Eprint [0]{\href }%
\providecommand \doibase [0]{https://doi.org/}%
\providecommand \selectlanguage [0]{\@gobble}%
\providecommand \bibinfo  [0]{\@secondoftwo}%
\providecommand \bibfield  [0]{\@secondoftwo}%
\providecommand \translation [1]{[#1]}%
\providecommand \BibitemOpen [0]{}%
\providecommand \bibitemStop [0]{}%
\providecommand \bibitemNoStop [0]{.\EOS\space}%
\providecommand \EOS [0]{\spacefactor3000\relax}%
\providecommand \BibitemShut  [1]{\csname bibitem#1\endcsname}%
\let\auto@bib@innerbib\@empty
\bibitem [{\citenamefont {Nadeem}\ \emph {et~al.}(2023)\citenamefont {Nadeem}, \citenamefont {Fuhrer},\ and\ \citenamefont {Wang}}]{nadeem2023superconducting}%
  \BibitemOpen
  \bibfield  {author} {\bibinfo {author} {\bibfnamefont {M.}~\bibnamefont {Nadeem}}, \bibinfo {author} {\bibfnamefont {M.~S.}\ \bibnamefont {Fuhrer}},\ and\ \bibinfo {author} {\bibfnamefont {X.}~\bibnamefont {Wang}},\ }\bibfield  {title} {\bibinfo {title} {The superconducting diode effect},\ }\href {https://doi.org/10.1038/s42254-023-00632-w} {\bibfield  {journal} {\bibinfo  {journal} {Nat. Rev. Phys.}\ }\textbf {\bibinfo {volume} {5}},\ \bibinfo {pages} {558} (\bibinfo {year} {2023})}\BibitemShut {NoStop}%
\bibitem [{\citenamefont {Nagaosa}\ and\ \citenamefont {Yanase}(2024)}]{nagaosa2024nonreciprocal}%
  \BibitemOpen
  \bibfield  {author} {\bibinfo {author} {\bibfnamefont {N.}~\bibnamefont {Nagaosa}}\ and\ \bibinfo {author} {\bibfnamefont {Y.}~\bibnamefont {Yanase}},\ }\bibfield  {title} {\bibinfo {title} {Nonreciprocal transport and optical phenomena in quantum materials},\ }\href {https://doi.org/10.1146/annurev-conmatphys-032822-033734} {\bibfield  {journal} {\bibinfo  {journal} {Annu. Rev. Condens. Matter Phys.}\ }\textbf {\bibinfo {volume} {15}},\ \bibinfo {pages} {63} (\bibinfo {year} {2024})}\BibitemShut {NoStop}%
\bibitem [{\citenamefont {Baumgartner}\ \emph {et~al.}(2021)\citenamefont {Baumgartner}, \citenamefont {Fuchs}, \citenamefont {Costa}, \citenamefont {Reinhardt}, \citenamefont {Gronin}, \citenamefont {Gardner}, \citenamefont {Lindemann}, \citenamefont {Manfra}, \citenamefont {Junior}, \citenamefont {Kochan}, \citenamefont {Fabian}, \citenamefont {Paradiso},\ and\ \citenamefont {Strunk}}]{baumgartner2021supercurrent}%
  \BibitemOpen
  \bibfield  {author} {\bibinfo {author} {\bibfnamefont {C.}~\bibnamefont {Baumgartner}}, \bibinfo {author} {\bibfnamefont {L.}~\bibnamefont {Fuchs}}, \bibinfo {author} {\bibfnamefont {A.}~\bibnamefont {Costa}}, \bibinfo {author} {\bibfnamefont {S.}~\bibnamefont {Reinhardt}}, \bibinfo {author} {\bibfnamefont {S.}~\bibnamefont {Gronin}}, \bibinfo {author} {\bibfnamefont {G.~C.}\ \bibnamefont {Gardner}}, \bibinfo {author} {\bibfnamefont {T.}~\bibnamefont {Lindemann}}, \bibinfo {author} {\bibfnamefont {M.~J.}\ \bibnamefont {Manfra}}, \bibinfo {author} {\bibfnamefont {P.~E.~F.}\ \bibnamefont {Junior}}, \bibinfo {author} {\bibfnamefont {D.}~\bibnamefont {Kochan}}, \bibinfo {author} {\bibfnamefont {J.}~\bibnamefont {Fabian}}, \bibinfo {author} {\bibfnamefont {N.}~\bibnamefont {Paradiso}},\ and\ \bibinfo {author} {\bibfnamefont {C.}~\bibnamefont {Strunk}},\ }\bibfield  {title} {\bibinfo {title} {{Supercurrent rectification and magnetochiral effects in symmetric Josephson junctions}},\ }\href
  {https://doi.org/10.1038/s41565-021-01009-9} {\bibfield  {journal} {\bibinfo  {journal} {Nat. Nanotechnol.}\ }\textbf {\bibinfo {volume} {17}},\ \bibinfo {pages} {39} (\bibinfo {year} {2021})}\BibitemShut {NoStop}%
\bibitem [{\citenamefont {Diez-Merida}\ \emph {et~al.}(2023)\citenamefont {Diez-Merida}, \citenamefont {D{\'\i}ez-Carl{\'o}n}, \citenamefont {Yang}, \citenamefont {Xie}, \citenamefont {Gao}, \citenamefont {Senior}, \citenamefont {Watanabe}, \citenamefont {Taniguchi}, \citenamefont {Lu}, \citenamefont {Higginbotham} \emph {et~al.}}]{diez2023symmetry}%
  \BibitemOpen
  \bibfield  {author} {\bibinfo {author} {\bibfnamefont {J.}~\bibnamefont {Diez-Merida}}, \bibinfo {author} {\bibfnamefont {A.}~\bibnamefont {D{\'\i}ez-Carl{\'o}n}}, \bibinfo {author} {\bibfnamefont {S.}~\bibnamefont {Yang}}, \bibinfo {author} {\bibfnamefont {Y.-M.}\ \bibnamefont {Xie}}, \bibinfo {author} {\bibfnamefont {X.-J.}\ \bibnamefont {Gao}}, \bibinfo {author} {\bibfnamefont {J.}~\bibnamefont {Senior}}, \bibinfo {author} {\bibfnamefont {K.}~\bibnamefont {Watanabe}}, \bibinfo {author} {\bibfnamefont {T.}~\bibnamefont {Taniguchi}}, \bibinfo {author} {\bibfnamefont {X.}~\bibnamefont {Lu}}, \bibinfo {author} {\bibfnamefont {A.~P.}\ \bibnamefont {Higginbotham}}, \emph {et~al.},\ }\bibfield  {title} {\bibinfo {title} {{Symmetry-broken Josephson junctions and superconducting diodes in magic-angle twisted bilayer graphene}},\ }\href {https://doi.org/10.1038/s41467-023-38005-7} {\bibfield  {journal} {\bibinfo  {journal} {Nat. Commun.}\ }\textbf {\bibinfo {volume} {14}},\ \bibinfo {pages} {2396} (\bibinfo {year}
  {2023})}\BibitemShut {NoStop}%
\bibitem [{\citenamefont {Wu}\ \emph {et~al.}(2022)\citenamefont {Wu}, \citenamefont {Wang}, \citenamefont {Xu}, \citenamefont {Sivakumar}, \citenamefont {Pasco}, \citenamefont {Filippozzi}, \citenamefont {Parkin}, \citenamefont {Zeng}, \citenamefont {McQueen},\ and\ \citenamefont {Ali}}]{wu2022field}%
  \BibitemOpen
  \bibfield  {author} {\bibinfo {author} {\bibfnamefont {H.}~\bibnamefont {Wu}}, \bibinfo {author} {\bibfnamefont {Y.}~\bibnamefont {Wang}}, \bibinfo {author} {\bibfnamefont {Y.}~\bibnamefont {Xu}}, \bibinfo {author} {\bibfnamefont {P.~K.}\ \bibnamefont {Sivakumar}}, \bibinfo {author} {\bibfnamefont {C.}~\bibnamefont {Pasco}}, \bibinfo {author} {\bibfnamefont {U.}~\bibnamefont {Filippozzi}}, \bibinfo {author} {\bibfnamefont {S.~S.~P.}\ \bibnamefont {Parkin}}, \bibinfo {author} {\bibfnamefont {Y.-J.}\ \bibnamefont {Zeng}}, \bibinfo {author} {\bibfnamefont {T.}~\bibnamefont {McQueen}},\ and\ \bibinfo {author} {\bibfnamefont {M.~N.}\ \bibnamefont {Ali}},\ }\bibfield  {title} {\bibinfo {title} {{The field-free Josephson diode in a van der Waals heterostructure}},\ }\href {https://doi.org/10.1038/s41586-022-04504-8} {\bibfield  {journal} {\bibinfo  {journal} {Nature}\ }\textbf {\bibinfo {volume} {604}},\ \bibinfo {pages} {653} (\bibinfo {year} {2022})}\BibitemShut {NoStop}%
\bibitem [{\citenamefont {Bauriedl}\ \emph {et~al.}(2022)\citenamefont {Bauriedl}, \citenamefont {Bäuml}, \citenamefont {Fuchs}, \citenamefont {Baumgartner}, \citenamefont {Paulik}, \citenamefont {Bauer}, \citenamefont {Lin}, \citenamefont {Lupton}, \citenamefont {Taniguchi}, \citenamefont {Watanabe}, \citenamefont {Strunk},\ and\ \citenamefont {Paradiso}}]{bauriedl2022supercurrent}%
  \BibitemOpen
  \bibfield  {author} {\bibinfo {author} {\bibfnamefont {L.}~\bibnamefont {Bauriedl}}, \bibinfo {author} {\bibfnamefont {C.}~\bibnamefont {Bäuml}}, \bibinfo {author} {\bibfnamefont {L.}~\bibnamefont {Fuchs}}, \bibinfo {author} {\bibfnamefont {C.}~\bibnamefont {Baumgartner}}, \bibinfo {author} {\bibfnamefont {N.}~\bibnamefont {Paulik}}, \bibinfo {author} {\bibfnamefont {J.~M.}\ \bibnamefont {Bauer}}, \bibinfo {author} {\bibfnamefont {K.-Q.}\ \bibnamefont {Lin}}, \bibinfo {author} {\bibfnamefont {J.~M.}\ \bibnamefont {Lupton}}, \bibinfo {author} {\bibfnamefont {T.}~\bibnamefont {Taniguchi}}, \bibinfo {author} {\bibfnamefont {K.}~\bibnamefont {Watanabe}}, \bibinfo {author} {\bibfnamefont {C.}~\bibnamefont {Strunk}},\ and\ \bibinfo {author} {\bibfnamefont {N.}~\bibnamefont {Paradiso}},\ }\bibfield  {title} {\bibinfo {title} {{Supercurrent diode effect and magnetochiral anisotropy in few-layer {NbSe}$_2$}},\ }\href {{https://doi.org/10.1038%2Fs41467-022-31954-5}} {\bibfield  {journal} {\bibinfo  {journal} {Nat.
  Commun.}\ }\textbf {\bibinfo {volume} {13}},\ \bibinfo {pages} {4266} (\bibinfo {year} {2022})}\BibitemShut {NoStop}%
\bibitem [{\citenamefont {Pal}\ \emph {et~al.}(2022)\citenamefont {Pal}, \citenamefont {Chakraborty}, \citenamefont {Sivakumar}, \citenamefont {Davydova}, \citenamefont {Gopi}, \citenamefont {Pandeya}, \citenamefont {Krieger}, \citenamefont {Zhang}, \citenamefont {Date}, \citenamefont {Ju}, \citenamefont {Yuan}, \citenamefont {Schr{\"o}ter}, \citenamefont {Fu},\ and\ \citenamefont {Parkin}}]{pal2022josephson}%
  \BibitemOpen
  \bibfield  {author} {\bibinfo {author} {\bibfnamefont {B.}~\bibnamefont {Pal}}, \bibinfo {author} {\bibfnamefont {A.}~\bibnamefont {Chakraborty}}, \bibinfo {author} {\bibfnamefont {P.~K.}\ \bibnamefont {Sivakumar}}, \bibinfo {author} {\bibfnamefont {M.}~\bibnamefont {Davydova}}, \bibinfo {author} {\bibfnamefont {A.~K.}\ \bibnamefont {Gopi}}, \bibinfo {author} {\bibfnamefont {A.~K.}\ \bibnamefont {Pandeya}}, \bibinfo {author} {\bibfnamefont {J.~A.}\ \bibnamefont {Krieger}}, \bibinfo {author} {\bibfnamefont {Y.}~\bibnamefont {Zhang}}, \bibinfo {author} {\bibfnamefont {M.}~\bibnamefont {Date}}, \bibinfo {author} {\bibfnamefont {S.}~\bibnamefont {Ju}}, \bibinfo {author} {\bibfnamefont {N.}~\bibnamefont {Yuan}}, \bibinfo {author} {\bibfnamefont {N.~B.~M.}\ \bibnamefont {Schr{\"o}ter}}, \bibinfo {author} {\bibfnamefont {L.}~\bibnamefont {Fu}},\ and\ \bibinfo {author} {\bibfnamefont {S.~S.~P.}\ \bibnamefont {Parkin}},\ }\bibfield  {title} {\bibinfo {title} {Josephson diode effect from {{Cooper}} pair momentum in a
  topological semimetal},\ }\href {https://doi.org/10.1038/s41567-022-01699-5} {\bibfield  {journal} {\bibinfo  {journal} {Nat. Phys.}\ }\textbf {\bibinfo {volume} {18}},\ \bibinfo {pages} {1228} (\bibinfo {year} {2022})}\BibitemShut {NoStop}%
\bibitem [{\citenamefont {Jeon}\ \emph {et~al.}(2022)\citenamefont {Jeon}, \citenamefont {Kim}, \citenamefont {Yoon}, \citenamefont {Jeon}, \citenamefont {Han}, \citenamefont {Cottet}, \citenamefont {Kontos},\ and\ \citenamefont {Parkin}}]{jeon2022zero}%
  \BibitemOpen
  \bibfield  {author} {\bibinfo {author} {\bibfnamefont {K.-R.}\ \bibnamefont {Jeon}}, \bibinfo {author} {\bibfnamefont {J.-K.}\ \bibnamefont {Kim}}, \bibinfo {author} {\bibfnamefont {J.}~\bibnamefont {Yoon}}, \bibinfo {author} {\bibfnamefont {J.-C.}\ \bibnamefont {Jeon}}, \bibinfo {author} {\bibfnamefont {H.}~\bibnamefont {Han}}, \bibinfo {author} {\bibfnamefont {A.}~\bibnamefont {Cottet}}, \bibinfo {author} {\bibfnamefont {T.}~\bibnamefont {Kontos}},\ and\ \bibinfo {author} {\bibfnamefont {S.~S.~P.}\ \bibnamefont {Parkin}},\ }\bibfield  {title} {\bibinfo {title} {Zero-field polarity-reversible {{Josephson}} supercurrent diodes enabled by a proximity-magnetized {{Pt}} barrier},\ }\href {https://doi.org/10.1038/s41563-022-01300-7} {\bibfield  {journal} {\bibinfo  {journal} {Nat. Mat.}\ }\textbf {\bibinfo {volume} {21}},\ \bibinfo {pages} {1008} (\bibinfo {year} {2022})}\BibitemShut {NoStop}%
\bibitem [{\citenamefont {Turini}\ \emph {et~al.}(2022)\citenamefont {Turini}, \citenamefont {Salimian}, \citenamefont {Carrega}, \citenamefont {Iorio}, \citenamefont {Strambini}, \citenamefont {Giazotto}, \citenamefont {Zannier}, \citenamefont {Sorba},\ and\ \citenamefont {Heun}}]{turini2022josephson}%
  \BibitemOpen
  \bibfield  {author} {\bibinfo {author} {\bibfnamefont {B.}~\bibnamefont {Turini}}, \bibinfo {author} {\bibfnamefont {S.}~\bibnamefont {Salimian}}, \bibinfo {author} {\bibfnamefont {M.}~\bibnamefont {Carrega}}, \bibinfo {author} {\bibfnamefont {A.}~\bibnamefont {Iorio}}, \bibinfo {author} {\bibfnamefont {E.}~\bibnamefont {Strambini}}, \bibinfo {author} {\bibfnamefont {F.}~\bibnamefont {Giazotto}}, \bibinfo {author} {\bibfnamefont {V.}~\bibnamefont {Zannier}}, \bibinfo {author} {\bibfnamefont {L.}~\bibnamefont {Sorba}},\ and\ \bibinfo {author} {\bibfnamefont {S.}~\bibnamefont {Heun}},\ }\bibfield  {title} {\bibinfo {title} {Josephson diode effect in high-mobility insb nanoflags},\ }\href {https://www.ncbi.nlm.nih.gov/pmc/articles/PMC9650771} {\bibfield  {journal} {\bibinfo  {journal} {Nano Lett.}\ }\textbf {\bibinfo {volume} {22}},\ \bibinfo {pages} {8502} (\bibinfo {year} {2022})}\BibitemShut {NoStop}%
\bibitem [{\citenamefont {Gupta}\ \emph {et~al.}(2023)\citenamefont {Gupta}, \citenamefont {Graziano}, \citenamefont {Pendharkar}, \citenamefont {Dong}, \citenamefont {Dempsey}, \citenamefont {Palmstr{\o}m},\ and\ \citenamefont {Pribiag}}]{gupta2023gate}%
  \BibitemOpen
  \bibfield  {author} {\bibinfo {author} {\bibfnamefont {M.}~\bibnamefont {Gupta}}, \bibinfo {author} {\bibfnamefont {G.~V.}\ \bibnamefont {Graziano}}, \bibinfo {author} {\bibfnamefont {M.}~\bibnamefont {Pendharkar}}, \bibinfo {author} {\bibfnamefont {J.~T.}\ \bibnamefont {Dong}}, \bibinfo {author} {\bibfnamefont {C.~P.}\ \bibnamefont {Dempsey}}, \bibinfo {author} {\bibfnamefont {C.}~\bibnamefont {Palmstr{\o}m}},\ and\ \bibinfo {author} {\bibfnamefont {V.~S.}\ \bibnamefont {Pribiag}},\ }\bibfield  {title} {\bibinfo {title} {Gate-tunable superconducting diode effect in a three-terminal josephson device},\ }\href {https://doi.org/10.1038/s41467-023-38856-0} {\bibfield  {journal} {\bibinfo  {journal} {Nat. Commun.}\ }\textbf {\bibinfo {volume} {14}},\ \bibinfo {pages} {3078} (\bibinfo {year} {2023})}\BibitemShut {NoStop}%
\bibitem [{\citenamefont {Chiles}\ \emph {et~al.}(2023)\citenamefont {Chiles}, \citenamefont {Arnault}, \citenamefont {Chen}, \citenamefont {Larson}, \citenamefont {Zhao}, \citenamefont {Watanabe}, \citenamefont {Taniguchi}, \citenamefont {Amet},\ and\ \citenamefont {Finkelstein}}]{chiles2023nonreciprocal}%
  \BibitemOpen
  \bibfield  {author} {\bibinfo {author} {\bibfnamefont {J.}~\bibnamefont {Chiles}}, \bibinfo {author} {\bibfnamefont {E.~G.}\ \bibnamefont {Arnault}}, \bibinfo {author} {\bibfnamefont {C.-C.}\ \bibnamefont {Chen}}, \bibinfo {author} {\bibfnamefont {T.~F.}\ \bibnamefont {Larson}}, \bibinfo {author} {\bibfnamefont {L.}~\bibnamefont {Zhao}}, \bibinfo {author} {\bibfnamefont {K.}~\bibnamefont {Watanabe}}, \bibinfo {author} {\bibfnamefont {T.}~\bibnamefont {Taniguchi}}, \bibinfo {author} {\bibfnamefont {F.}~\bibnamefont {Amet}},\ and\ \bibinfo {author} {\bibfnamefont {G.}~\bibnamefont {Finkelstein}},\ }\bibfield  {title} {\bibinfo {title} {{Nonreciprocal supercurrents in a field-free graphene Josephson triode}},\ }\href {https://www.osti.gov/biblio/1994760} {\bibfield  {journal} {\bibinfo  {journal} {Nano Lett.}\ }\textbf {\bibinfo {volume} {23}},\ \bibinfo {pages} {5257} (\bibinfo {year} {2023})}\BibitemShut {NoStop}%
\bibitem [{\citenamefont {Zhang}\ \emph {et~al.}(2025)\citenamefont {Zhang}, \citenamefont {Li}, \citenamefont {Aguilar}, \citenamefont {Zhang}, \citenamefont {Pendharkar}, \citenamefont {Dempsey}, \citenamefont {Lee}, \citenamefont {Harrington}, \citenamefont {Tan}, \citenamefont {Meyer}, \citenamefont {Houzet}, \citenamefont {Palmstrom},\ and\ \citenamefont {Frolov}}]{zhang2025evidence}%
  \BibitemOpen
  \bibfield  {author} {\bibinfo {author} {\bibfnamefont {B.}~\bibnamefont {Zhang}}, \bibinfo {author} {\bibfnamefont {Z.}~\bibnamefont {Li}}, \bibinfo {author} {\bibfnamefont {V.}~\bibnamefont {Aguilar}}, \bibinfo {author} {\bibfnamefont {P.}~\bibnamefont {Zhang}}, \bibinfo {author} {\bibfnamefont {M.}~\bibnamefont {Pendharkar}}, \bibinfo {author} {\bibfnamefont {C.}~\bibnamefont {Dempsey}}, \bibinfo {author} {\bibfnamefont {J.~S.}\ \bibnamefont {Lee}}, \bibinfo {author} {\bibfnamefont {S.~D.}\ \bibnamefont {Harrington}}, \bibinfo {author} {\bibfnamefont {S.}~\bibnamefont {Tan}}, \bibinfo {author} {\bibfnamefont {J.~S.}\ \bibnamefont {Meyer}}, \bibinfo {author} {\bibfnamefont {M.}~\bibnamefont {Houzet}}, \bibinfo {author} {\bibfnamefont {C.~J.}\ \bibnamefont {Palmstrom}},\ and\ \bibinfo {author} {\bibfnamefont {S.~M.}\ \bibnamefont {Frolov}},\ }\bibfield  {title} {\bibinfo {title} {{Evidence of $\phi_0$-Josephson junction from skewed diffraction patterns in Sn-InSb nanowires}},\ }\href
  {https://doi.org/10.21468/SciPostPhys.18.1.013} {\bibfield  {journal} {\bibinfo  {journal} {SciPost Phys.}\ }\textbf {\bibinfo {volume} {18}},\ \bibinfo {pages} {013} (\bibinfo {year} {2025})}\BibitemShut {NoStop}%
\bibitem [{\citenamefont {Mazur}\ \emph {et~al.}(2024)\citenamefont {Mazur}, \citenamefont {van Loo}, \citenamefont {van Driel}, \citenamefont {Wang}, \citenamefont {Badawy}, \citenamefont {Gazibegovic}, \citenamefont {Bakkers},\ and\ \citenamefont {Kouwenhoven}}]{mazur2024gate}%
  \BibitemOpen
  \bibfield  {author} {\bibinfo {author} {\bibfnamefont {G.}~\bibnamefont {Mazur}}, \bibinfo {author} {\bibfnamefont {N.}~\bibnamefont {van Loo}}, \bibinfo {author} {\bibfnamefont {D.}~\bibnamefont {van Driel}}, \bibinfo {author} {\bibfnamefont {J.-Y.}\ \bibnamefont {Wang}}, \bibinfo {author} {\bibfnamefont {G.}~\bibnamefont {Badawy}}, \bibinfo {author} {\bibfnamefont {S.}~\bibnamefont {Gazibegovic}}, \bibinfo {author} {\bibfnamefont {E.}~\bibnamefont {Bakkers}},\ and\ \bibinfo {author} {\bibfnamefont {L.}~\bibnamefont {Kouwenhoven}},\ }\bibfield  {title} {\bibinfo {title} {{Gate-tunable Josephson diode}},\ }\href {https://doi.org/10.1103/PhysRevApplied.22.054034} {\bibfield  {journal} {\bibinfo  {journal} {Phys. Rev. Appl.}\ }\textbf {\bibinfo {volume} {22}},\ \bibinfo {pages} {054034} (\bibinfo {year} {2024})}\BibitemShut {NoStop}%
\bibitem [{\citenamefont {Ciaccia}\ \emph {et~al.}(2023)\citenamefont {Ciaccia}, \citenamefont {Haller}, \citenamefont {Drachmann}, \citenamefont {Lindemann}, \citenamefont {Manfra}, \citenamefont {Schrade},\ and\ \citenamefont {Sch{\"o}nenberger}}]{ciaccia2023gate}%
  \BibitemOpen
  \bibfield  {author} {\bibinfo {author} {\bibfnamefont {C.}~\bibnamefont {Ciaccia}}, \bibinfo {author} {\bibfnamefont {R.}~\bibnamefont {Haller}}, \bibinfo {author} {\bibfnamefont {A.~C.}\ \bibnamefont {Drachmann}}, \bibinfo {author} {\bibfnamefont {T.}~\bibnamefont {Lindemann}}, \bibinfo {author} {\bibfnamefont {M.~J.}\ \bibnamefont {Manfra}}, \bibinfo {author} {\bibfnamefont {C.}~\bibnamefont {Schrade}},\ and\ \bibinfo {author} {\bibfnamefont {C.}~\bibnamefont {Sch{\"o}nenberger}},\ }\bibfield  {title} {\bibinfo {title} {{Gate-tunable Josephson diode in proximitized InAs supercurrent interferometers}},\ }\href {https://doi.org/10.1103/physrevresearch.5.033131} {\bibfield  {journal} {\bibinfo  {journal} {Phys. Rev. Res.}\ }\textbf {\bibinfo {volume} {5}},\ \bibinfo {pages} {033131} (\bibinfo {year} {2023})}\BibitemShut {NoStop}%
\bibitem [{\citenamefont {Trahms}\ \emph {et~al.}(2023)\citenamefont {Trahms}, \citenamefont {Melischek}, \citenamefont {Steiner}, \citenamefont {Mahendru}, \citenamefont {Tamir}, \citenamefont {Bogdanoff}, \citenamefont {Peters}, \citenamefont {Reecht}, \citenamefont {Winkelmann}, \citenamefont {von Oppen},\ and\ \citenamefont {Franke}}]{trahms2023diode}%
  \BibitemOpen
  \bibfield  {author} {\bibinfo {author} {\bibfnamefont {M.}~\bibnamefont {Trahms}}, \bibinfo {author} {\bibfnamefont {L.}~\bibnamefont {Melischek}}, \bibinfo {author} {\bibfnamefont {J.~F.}\ \bibnamefont {Steiner}}, \bibinfo {author} {\bibfnamefont {B.}~\bibnamefont {Mahendru}}, \bibinfo {author} {\bibfnamefont {I.}~\bibnamefont {Tamir}}, \bibinfo {author} {\bibfnamefont {N.}~\bibnamefont {Bogdanoff}}, \bibinfo {author} {\bibfnamefont {O.}~\bibnamefont {Peters}}, \bibinfo {author} {\bibfnamefont {G.}~\bibnamefont {Reecht}}, \bibinfo {author} {\bibfnamefont {C.~B.}\ \bibnamefont {Winkelmann}}, \bibinfo {author} {\bibfnamefont {F.}~\bibnamefont {von Oppen}},\ and\ \bibinfo {author} {\bibfnamefont {K.~J.}\ \bibnamefont {Franke}},\ }\bibfield  {title} {\bibinfo {title} {{Diode effect in Josephson junctions with a single magnetic atom}},\ }\href {https://doi.org/10.1038/s41586-023-05743-z} {\bibfield  {journal} {\bibinfo  {journal} {Nature}\ }\textbf {\bibinfo {volume} {615}},\ \bibinfo {pages} {628} (\bibinfo
  {year} {2023})}\BibitemShut {NoStop}%
\bibitem [{\citenamefont {Reinhardt}\ \emph {et~al.}(2024)\citenamefont {Reinhardt}, \citenamefont {Ascherl}, \citenamefont {Costa}, \citenamefont {Berger}, \citenamefont {Gronin}, \citenamefont {Gardner}, \citenamefont {Lindemann}, \citenamefont {Manfra}, \citenamefont {Fabian}, \citenamefont {Kochan} \emph {et~al.}}]{reinhardt2024link}%
  \BibitemOpen
  \bibfield  {author} {\bibinfo {author} {\bibfnamefont {S.}~\bibnamefont {Reinhardt}}, \bibinfo {author} {\bibfnamefont {T.}~\bibnamefont {Ascherl}}, \bibinfo {author} {\bibfnamefont {A.}~\bibnamefont {Costa}}, \bibinfo {author} {\bibfnamefont {J.}~\bibnamefont {Berger}}, \bibinfo {author} {\bibfnamefont {S.}~\bibnamefont {Gronin}}, \bibinfo {author} {\bibfnamefont {G.~C.}\ \bibnamefont {Gardner}}, \bibinfo {author} {\bibfnamefont {T.}~\bibnamefont {Lindemann}}, \bibinfo {author} {\bibfnamefont {M.~J.}\ \bibnamefont {Manfra}}, \bibinfo {author} {\bibfnamefont {J.}~\bibnamefont {Fabian}}, \bibinfo {author} {\bibfnamefont {D.}~\bibnamefont {Kochan}}, \emph {et~al.},\ }\bibfield  {title} {\bibinfo {title} {{Link between supercurrent diode and anomalous Josephson effect revealed by gate-controlled interferometry}},\ }\href {https://doi.org/10.1038/s41467-024-48741-z} {\bibfield  {journal} {\bibinfo  {journal} {Nat. Commun.}\ }\textbf {\bibinfo {volume} {15}},\ \bibinfo {pages} {4413} (\bibinfo {year}
  {2024})}\BibitemShut {NoStop}%
\bibitem [{\citenamefont {Ghosh}\ \emph {et~al.}(2024)\citenamefont {Ghosh}, \citenamefont {Patil}, \citenamefont {Basu}, \citenamefont {Kuldeep}, \citenamefont {Dutta}, \citenamefont {Jangade}, \citenamefont {Kulkarni}, \citenamefont {Thamizhavel}, \citenamefont {Steiner}, \citenamefont {von Oppen} \emph {et~al.}}]{ghosh2024high}%
  \BibitemOpen
  \bibfield  {author} {\bibinfo {author} {\bibfnamefont {S.}~\bibnamefont {Ghosh}}, \bibinfo {author} {\bibfnamefont {V.}~\bibnamefont {Patil}}, \bibinfo {author} {\bibfnamefont {A.}~\bibnamefont {Basu}}, \bibinfo {author} {\bibnamefont {Kuldeep}}, \bibinfo {author} {\bibfnamefont {A.}~\bibnamefont {Dutta}}, \bibinfo {author} {\bibfnamefont {D.~A.}\ \bibnamefont {Jangade}}, \bibinfo {author} {\bibfnamefont {R.}~\bibnamefont {Kulkarni}}, \bibinfo {author} {\bibfnamefont {A.}~\bibnamefont {Thamizhavel}}, \bibinfo {author} {\bibfnamefont {J.~F.}\ \bibnamefont {Steiner}}, \bibinfo {author} {\bibfnamefont {F.}~\bibnamefont {von Oppen}}, \emph {et~al.},\ }\bibfield  {title} {\bibinfo {title} {{High-temperature Josephson diode}},\ }\href {https://doi.org/10.1038/s41563-024-01804-4} {\bibfield  {journal} {\bibinfo  {journal} {Nat. Mater.}\ }\textbf {\bibinfo {volume} {23}},\ \bibinfo {pages} {612} (\bibinfo {year} {2024})}\BibitemShut {NoStop}%
\bibitem [{\citenamefont {Lotfizadeh}\ \emph {et~al.}(2024)\citenamefont {Lotfizadeh}, \citenamefont {Schiela}, \citenamefont {Pekerten}, \citenamefont {Yu}, \citenamefont {Elfeky}, \citenamefont {Strickland}, \citenamefont {Matos-Abiague},\ and\ \citenamefont {Shabani}}]{lotfizadeh2024superconducting}%
  \BibitemOpen
  \bibfield  {author} {\bibinfo {author} {\bibfnamefont {N.}~\bibnamefont {Lotfizadeh}}, \bibinfo {author} {\bibfnamefont {W.~F.}\ \bibnamefont {Schiela}}, \bibinfo {author} {\bibfnamefont {B.}~\bibnamefont {Pekerten}}, \bibinfo {author} {\bibfnamefont {P.}~\bibnamefont {Yu}}, \bibinfo {author} {\bibfnamefont {B.~H.}\ \bibnamefont {Elfeky}}, \bibinfo {author} {\bibfnamefont {W.~M.}\ \bibnamefont {Strickland}}, \bibinfo {author} {\bibfnamefont {A.}~\bibnamefont {Matos-Abiague}},\ and\ \bibinfo {author} {\bibfnamefont {J.}~\bibnamefont {Shabani}},\ }\bibfield  {title} {\bibinfo {title} {{Superconducting diode effect sign change in epitaxial Al-InAs Josephson junctions}},\ }\href {https://doi.org/10.1038/s42005-024-01618-5} {\bibfield  {journal} {\bibinfo  {journal} {Commun. Phys.}\ }\textbf {\bibinfo {volume} {7}},\ \bibinfo {pages} {120} (\bibinfo {year} {2024})}\BibitemShut {NoStop}%
\bibitem [{\citenamefont {Kim}\ \emph {et~al.}(2024)\citenamefont {Kim}, \citenamefont {Jeon}, \citenamefont {Sivakumar}, \citenamefont {Jeon}, \citenamefont {Koerner}, \citenamefont {Woltersdorf},\ and\ \citenamefont {Parkin}}]{kim2024intrinsic}%
  \BibitemOpen
  \bibfield  {author} {\bibinfo {author} {\bibfnamefont {J.-K.}\ \bibnamefont {Kim}}, \bibinfo {author} {\bibfnamefont {K.-R.}\ \bibnamefont {Jeon}}, \bibinfo {author} {\bibfnamefont {P.~K.}\ \bibnamefont {Sivakumar}}, \bibinfo {author} {\bibfnamefont {J.}~\bibnamefont {Jeon}}, \bibinfo {author} {\bibfnamefont {C.}~\bibnamefont {Koerner}}, \bibinfo {author} {\bibfnamefont {G.}~\bibnamefont {Woltersdorf}},\ and\ \bibinfo {author} {\bibfnamefont {S.~S.}\ \bibnamefont {Parkin}},\ }\bibfield  {title} {\bibinfo {title} {{Intrinsic supercurrent non-reciprocity coupled to the crystal structure of a van der Waals Josephson barrier}},\ }\href {https://doi.org/10.1038/s41467-024-45298-9} {\bibfield  {journal} {\bibinfo  {journal} {Nat. Commun.}\ }\textbf {\bibinfo {volume} {15}},\ \bibinfo {pages} {1120} (\bibinfo {year} {2024})}\BibitemShut {NoStop}%
\bibitem [{\citenamefont {Chen}\ \emph {et~al.}(2024)\citenamefont {Chen}, \citenamefont {Wang}, \citenamefont {Ye}, \citenamefont {Wang}, \citenamefont {Zhou}, \citenamefont {Tang}, \citenamefont {Wang}, \citenamefont {Wang}, \citenamefont {Zhang}, \citenamefont {Mei} \emph {et~al.}}]{chen2024edelstein}%
  \BibitemOpen
  \bibfield  {author} {\bibinfo {author} {\bibfnamefont {P.}~\bibnamefont {Chen}}, \bibinfo {author} {\bibfnamefont {G.}~\bibnamefont {Wang}}, \bibinfo {author} {\bibfnamefont {B.}~\bibnamefont {Ye}}, \bibinfo {author} {\bibfnamefont {J.}~\bibnamefont {Wang}}, \bibinfo {author} {\bibfnamefont {L.}~\bibnamefont {Zhou}}, \bibinfo {author} {\bibfnamefont {Z.}~\bibnamefont {Tang}}, \bibinfo {author} {\bibfnamefont {L.}~\bibnamefont {Wang}}, \bibinfo {author} {\bibfnamefont {J.}~\bibnamefont {Wang}}, \bibinfo {author} {\bibfnamefont {W.}~\bibnamefont {Zhang}}, \bibinfo {author} {\bibfnamefont {J.}~\bibnamefont {Mei}}, \emph {et~al.},\ }\bibfield  {title} {\bibinfo {title} {{Edelstein effect induced superconducting diode effect in inversion symmetry breaking MoTe2 Josephson junctions}},\ }\href {https://doi.org/10.1002/adfm.202311229} {\bibfield  {journal} {\bibinfo  {journal} {Adv. Funct. Mater.}\ }\textbf {\bibinfo {volume} {34}},\ \bibinfo {pages} {2311229} (\bibinfo {year} {2024})}\BibitemShut {NoStop}%
\bibitem [{\citenamefont {Kudriashov}\ \emph {et~al.}(2025)\citenamefont {Kudriashov}, \citenamefont {Zhou}, \citenamefont {Hovhannisyan}, \citenamefont {Frolov}, \citenamefont {Elesin}, \citenamefont {Wang}, \citenamefont {Zharkova}, \citenamefont {Taniguchi}, \citenamefont {Watanabe}, \citenamefont {Liu} \emph {et~al.}}]{kudriashov2025non}%
  \BibitemOpen
  \bibfield  {author} {\bibinfo {author} {\bibfnamefont {A.}~\bibnamefont {Kudriashov}}, \bibinfo {author} {\bibfnamefont {X.}~\bibnamefont {Zhou}}, \bibinfo {author} {\bibfnamefont {R.~A.}\ \bibnamefont {Hovhannisyan}}, \bibinfo {author} {\bibfnamefont {A.~S.}\ \bibnamefont {Frolov}}, \bibinfo {author} {\bibfnamefont {L.}~\bibnamefont {Elesin}}, \bibinfo {author} {\bibfnamefont {Y.~B.}\ \bibnamefont {Wang}}, \bibinfo {author} {\bibfnamefont {E.~V.}\ \bibnamefont {Zharkova}}, \bibinfo {author} {\bibfnamefont {T.}~\bibnamefont {Taniguchi}}, \bibinfo {author} {\bibfnamefont {K.}~\bibnamefont {Watanabe}}, \bibinfo {author} {\bibfnamefont {Z.}~\bibnamefont {Liu}}, \emph {et~al.},\ }\bibfield  {title} {\bibinfo {title} {{Non-Majorana origin of anomalous current-phase relation and Josephson diode effect in Bi$_2$Se$_3$/NbSe$_2$ Josephson junctions}},\ }\href {https://doi.org/10.1126/sciadv.1602390} {\bibfield  {journal} {\bibinfo  {journal} {Sci. Adv.}\ }\textbf {\bibinfo {volume} {11}},\ \bibinfo {pages}
  {eadw6925} (\bibinfo {year} {2025})}\BibitemShut {NoStop}%
\bibitem [{\citenamefont {Yan}\ \emph {et~al.}(2025)\citenamefont {Yan}, \citenamefont {Luo}, \citenamefont {Su}, \citenamefont {Gao}, \citenamefont {Wu}, \citenamefont {Pan}, \citenamefont {Zhao}, \citenamefont {Wang},\ and\ \citenamefont {Xu}}]{yan2025gate}%
  \BibitemOpen
  \bibfield  {author} {\bibinfo {author} {\bibfnamefont {S.}~\bibnamefont {Yan}}, \bibinfo {author} {\bibfnamefont {Y.}~\bibnamefont {Luo}}, \bibinfo {author} {\bibfnamefont {H.}~\bibnamefont {Su}}, \bibinfo {author} {\bibfnamefont {H.}~\bibnamefont {Gao}}, \bibinfo {author} {\bibfnamefont {X.}~\bibnamefont {Wu}}, \bibinfo {author} {\bibfnamefont {D.}~\bibnamefont {Pan}}, \bibinfo {author} {\bibfnamefont {J.}~\bibnamefont {Zhao}}, \bibinfo {author} {\bibfnamefont {J.-Y.}\ \bibnamefont {Wang}},\ and\ \bibinfo {author} {\bibfnamefont {H.}~\bibnamefont {Xu}},\ }\bibfield  {title} {\bibinfo {title} {{Gate Tunable Josephson Diode Effect in Josephson Junctions Made from InAs Nanosheets}},\ }\href {https://doi.org/https://doi.org/10.1002/adfm.202503401} {\bibfield  {journal} {\bibinfo  {journal} {Adv. Funct. Mater.}\ ,\ \bibinfo {pages} {2503401}} (\bibinfo {year} {2025})}\BibitemShut {NoStop}%
\bibitem [{\citenamefont {Dolcini}\ \emph {et~al.}(2015)\citenamefont {Dolcini}, \citenamefont {Houzet},\ and\ \citenamefont {Meyer}}]{dolcini2015topological}%
  \BibitemOpen
  \bibfield  {author} {\bibinfo {author} {\bibfnamefont {F.}~\bibnamefont {Dolcini}}, \bibinfo {author} {\bibfnamefont {M.}~\bibnamefont {Houzet}},\ and\ \bibinfo {author} {\bibfnamefont {J.~S.}\ \bibnamefont {Meyer}},\ }\bibfield  {title} {\bibinfo {title} {{Topological Josephson ${\ensuremath{\phi}}_{0}$ junctions}},\ }\href {https://doi.org/10.1103/PhysRevB.92.035428} {\bibfield  {journal} {\bibinfo  {journal} {Phys. Rev. B}\ }\textbf {\bibinfo {volume} {92}},\ \bibinfo {pages} {035428} (\bibinfo {year} {2015})}\BibitemShut {NoStop}%
\bibitem [{\citenamefont {Chen}\ \emph {et~al.}(2018)\citenamefont {Chen}, \citenamefont {He}, \citenamefont {Ali}, \citenamefont {Lee}, \citenamefont {Fong},\ and\ \citenamefont {Law}}]{chen2018asymmetric}%
  \BibitemOpen
  \bibfield  {author} {\bibinfo {author} {\bibfnamefont {C.-Z.}\ \bibnamefont {Chen}}, \bibinfo {author} {\bibfnamefont {J.~J.}\ \bibnamefont {He}}, \bibinfo {author} {\bibfnamefont {M.~N.}\ \bibnamefont {Ali}}, \bibinfo {author} {\bibfnamefont {G.-H.}\ \bibnamefont {Lee}}, \bibinfo {author} {\bibfnamefont {K.~C.}\ \bibnamefont {Fong}},\ and\ \bibinfo {author} {\bibfnamefont {K.~T.}\ \bibnamefont {Law}},\ }\bibfield  {title} {\bibinfo {title} {{Asymmetric Josephson effect in inversion symmetry breaking topological materials}},\ }\href {https://doi.org/10.1103/PhysRevB.98.075430} {\bibfield  {journal} {\bibinfo  {journal} {Phys. Rev. B}\ }\textbf {\bibinfo {volume} {98}},\ \bibinfo {pages} {075430} (\bibinfo {year} {2018})}\BibitemShut {NoStop}%
\bibitem [{\citenamefont {Zhang}\ \emph {et~al.}(2022)\citenamefont {Zhang}, \citenamefont {Gu}, \citenamefont {Li}, \citenamefont {Hu},\ and\ \citenamefont {Jiang}}]{zhang2022general}%
  \BibitemOpen
  \bibfield  {author} {\bibinfo {author} {\bibfnamefont {Y.}~\bibnamefont {Zhang}}, \bibinfo {author} {\bibfnamefont {Y.}~\bibnamefont {Gu}}, \bibinfo {author} {\bibfnamefont {P.}~\bibnamefont {Li}}, \bibinfo {author} {\bibfnamefont {J.}~\bibnamefont {Hu}},\ and\ \bibinfo {author} {\bibfnamefont {K.}~\bibnamefont {Jiang}},\ }\bibfield  {title} {\bibinfo {title} {{General theory of Josephson diodes}},\ }\href {https://link.aps.org/pdf/10.1103/PhysRevX.12.041013} {\bibfield  {journal} {\bibinfo  {journal} {Phys. Rev. X}\ }\textbf {\bibinfo {volume} {12}},\ \bibinfo {pages} {041013} (\bibinfo {year} {2022})}\BibitemShut {NoStop}%
\bibitem [{\citenamefont {Kopasov}\ \emph {et~al.}(2021)\citenamefont {Kopasov}, \citenamefont {Kutlin},\ and\ \citenamefont {Mel'nikov}}]{kopasov2021geometry}%
  \BibitemOpen
  \bibfield  {author} {\bibinfo {author} {\bibfnamefont {A.~A.}\ \bibnamefont {Kopasov}}, \bibinfo {author} {\bibfnamefont {A.~G.}\ \bibnamefont {Kutlin}},\ and\ \bibinfo {author} {\bibfnamefont {A.~S.}\ \bibnamefont {Mel'nikov}},\ }\bibfield  {title} {\bibinfo {title} {Geometry controlled superconducting diode and anomalous {{Josephson}} effect triggered by the topological phase transition in curved proximitized nanowires},\ }\href {https://doi.org/10.1103/PhysRevB.103.144520} {\bibfield  {journal} {\bibinfo  {journal} {Phys. Rev. B}\ }\textbf {\bibinfo {volume} {103}},\ \bibinfo {pages} {144520} (\bibinfo {year} {2021})}\BibitemShut {NoStop}%
\bibitem [{\citenamefont {Davydova}\ \emph {et~al.}(2022)\citenamefont {Davydova}, \citenamefont {Prembabu},\ and\ \citenamefont {Fu}}]{davydova2022universal}%
  \BibitemOpen
  \bibfield  {author} {\bibinfo {author} {\bibfnamefont {M.}~\bibnamefont {Davydova}}, \bibinfo {author} {\bibfnamefont {S.}~\bibnamefont {Prembabu}},\ and\ \bibinfo {author} {\bibfnamefont {L.}~\bibnamefont {Fu}},\ }\bibfield  {title} {\bibinfo {title} {Universal {{Josephson}} diode effect},\ }\href {https://doi.org/10.1126/sciadv.abo0309} {\bibfield  {journal} {\bibinfo  {journal} {Sci. Adv.}\ }\textbf {\bibinfo {volume} {8}},\ \bibinfo {pages} {eabo0309} (\bibinfo {year} {2022})}\BibitemShut {NoStop}%
\bibitem [{\citenamefont {Souto}\ \emph {et~al.}(2022)\citenamefont {Souto}, \citenamefont {Leijnse},\ and\ \citenamefont {Schrade}}]{souto2022josephson}%
  \BibitemOpen
  \bibfield  {author} {\bibinfo {author} {\bibfnamefont {R.~S.}\ \bibnamefont {Souto}}, \bibinfo {author} {\bibfnamefont {M.}~\bibnamefont {Leijnse}},\ and\ \bibinfo {author} {\bibfnamefont {C.}~\bibnamefont {Schrade}},\ }\bibfield  {title} {\bibinfo {title} {Josephson diode effect in supercurrent interferometers},\ }\href {https://doi.org/https://doi.org/10.1103/PhysRevLett.129.267702} {\bibfield  {journal} {\bibinfo  {journal} {Phys. Rev. Lett.}\ }\textbf {\bibinfo {volume} {129}},\ \bibinfo {pages} {267702} (\bibinfo {year} {2022})}\BibitemShut {NoStop}%
\bibitem [{\citenamefont {Tanaka}\ \emph {et~al.}(2022)\citenamefont {Tanaka}, \citenamefont {Lu},\ and\ \citenamefont {Nagaosa}}]{tanaka2022theory}%
  \BibitemOpen
  \bibfield  {author} {\bibinfo {author} {\bibfnamefont {Y.}~\bibnamefont {Tanaka}}, \bibinfo {author} {\bibfnamefont {B.}~\bibnamefont {Lu}},\ and\ \bibinfo {author} {\bibfnamefont {N.}~\bibnamefont {Nagaosa}},\ }\bibfield  {title} {\bibinfo {title} {Theory of giant diode effect in d-wave superconductor junctions on the surface of a topological insulator},\ }\href {https://doi.org/https://doi.org/10.1103/PhysRevB.106.214524} {\bibfield  {journal} {\bibinfo  {journal} {Phys. Rev. B}\ }\textbf {\bibinfo {volume} {106}},\ \bibinfo {pages} {214524} (\bibinfo {year} {2022})}\BibitemShut {NoStop}%
\bibitem [{\citenamefont {Lu}\ \emph {et~al.}(2023)\citenamefont {Lu}, \citenamefont {Ikegaya}, \citenamefont {Burset}, \citenamefont {Tanaka},\ and\ \citenamefont {Nagaosa}}]{lu2023tunable}%
  \BibitemOpen
  \bibfield  {author} {\bibinfo {author} {\bibfnamefont {B.}~\bibnamefont {Lu}}, \bibinfo {author} {\bibfnamefont {S.}~\bibnamefont {Ikegaya}}, \bibinfo {author} {\bibfnamefont {P.}~\bibnamefont {Burset}}, \bibinfo {author} {\bibfnamefont {Y.}~\bibnamefont {Tanaka}},\ and\ \bibinfo {author} {\bibfnamefont {N.}~\bibnamefont {Nagaosa}},\ }\bibfield  {title} {\bibinfo {title} {{Tunable Josephson diode effect on the surface of topological insulators}},\ }\href {https://doi.org/10.1103/physrevlett.131.096001} {\bibfield  {journal} {\bibinfo  {journal} {Phys. Rev. Lett.}\ }\textbf {\bibinfo {volume} {131}},\ \bibinfo {pages} {096001} (\bibinfo {year} {2023})}\BibitemShut {NoStop}%
\bibitem [{\citenamefont {Maiani}\ \emph {et~al.}(2023)\citenamefont {Maiani}, \citenamefont {Flensberg}, \citenamefont {Leijnse}, \citenamefont {Schrade}, \citenamefont {Vaitiek{\.e}nas},\ and\ \citenamefont {Souto}}]{maiani2023nonsinusoidal}%
  \BibitemOpen
  \bibfield  {author} {\bibinfo {author} {\bibfnamefont {A.}~\bibnamefont {Maiani}}, \bibinfo {author} {\bibfnamefont {K.}~\bibnamefont {Flensberg}}, \bibinfo {author} {\bibfnamefont {M.}~\bibnamefont {Leijnse}}, \bibinfo {author} {\bibfnamefont {C.}~\bibnamefont {Schrade}}, \bibinfo {author} {\bibfnamefont {S.}~\bibnamefont {Vaitiek{\.e}nas}},\ and\ \bibinfo {author} {\bibfnamefont {R.~S.}\ \bibnamefont {Souto}},\ }\bibfield  {title} {\bibinfo {title} {Nonsinusoidal current-phase relations in semiconductor--superconductor--ferromagnetic insulator devices},\ }\href {https://doi.org/10.1103/PhysRevB.107.245415} {\bibfield  {journal} {\bibinfo  {journal} {Phys. Rev. B}\ }\textbf {\bibinfo {volume} {107}},\ \bibinfo {pages} {245415} (\bibinfo {year} {2023})}\BibitemShut {NoStop}%
\bibitem [{\citenamefont {Hu}\ \emph {et~al.}(2023)\citenamefont {Hu}, \citenamefont {Sun}, \citenamefont {Xie},\ and\ \citenamefont {Law}}]{hu2023josephson}%
  \BibitemOpen
  \bibfield  {author} {\bibinfo {author} {\bibfnamefont {J.-X.}\ \bibnamefont {Hu}}, \bibinfo {author} {\bibfnamefont {Z.-T.}\ \bibnamefont {Sun}}, \bibinfo {author} {\bibfnamefont {Y.-M.}\ \bibnamefont {Xie}},\ and\ \bibinfo {author} {\bibfnamefont {K.~T.}\ \bibnamefont {Law}},\ }\bibfield  {title} {\bibinfo {title} {Josephson diode effect induced by valley polarization in twisted bilayer graphene},\ }\href {https://doi.org/10.1103/physrevlett.130.266003} {\bibfield  {journal} {\bibinfo  {journal} {Phys. Rev. Lett.}\ }\textbf {\bibinfo {volume} {130}},\ \bibinfo {pages} {266003} (\bibinfo {year} {2023})}\BibitemShut {NoStop}%
\bibitem [{\citenamefont {Costa}\ \emph {et~al.}(2023)\citenamefont {Costa}, \citenamefont {Fabian},\ and\ \citenamefont {Kochan}}]{costa2023microscopic}%
  \BibitemOpen
  \bibfield  {author} {\bibinfo {author} {\bibfnamefont {A.}~\bibnamefont {Costa}}, \bibinfo {author} {\bibfnamefont {J.}~\bibnamefont {Fabian}},\ and\ \bibinfo {author} {\bibfnamefont {D.}~\bibnamefont {Kochan}},\ }\bibfield  {title} {\bibinfo {title} {{Microscopic study of the Josephson supercurrent diode effect in Josephson junctions based on two-dimensional electron gas}},\ }\href {https://doi.org/10.1103/physrevb.108.054522} {\bibfield  {journal} {\bibinfo  {journal} {Phys. Rev. B}\ }\textbf {\bibinfo {volume} {108}},\ \bibinfo {pages} {054522} (\bibinfo {year} {2023})}\BibitemShut {NoStop}%
\bibitem [{\citenamefont {Cayao}\ \emph {et~al.}(2024)\citenamefont {Cayao}, \citenamefont {Nagaosa},\ and\ \citenamefont {Tanaka}}]{cayao2024enhancing}%
  \BibitemOpen
  \bibfield  {author} {\bibinfo {author} {\bibfnamefont {J.}~\bibnamefont {Cayao}}, \bibinfo {author} {\bibfnamefont {N.}~\bibnamefont {Nagaosa}},\ and\ \bibinfo {author} {\bibfnamefont {Y.}~\bibnamefont {Tanaka}},\ }\bibfield  {title} {\bibinfo {title} {{Enhancing the Josephson diode effect with Majorana bound states}},\ }\href {https://link.aps.org/pdf/10.1103/PhysRevB.109.L081405} {\bibfield  {journal} {\bibinfo  {journal} {Phys. Rev. B}\ }\textbf {\bibinfo {volume} {109}},\ \bibinfo {pages} {L081405} (\bibinfo {year} {2024})}\BibitemShut {NoStop}%
\bibitem [{\citenamefont {Meyer}\ and\ \citenamefont {Houzet}(2024)}]{meyer2024josephson}%
  \BibitemOpen
  \bibfield  {author} {\bibinfo {author} {\bibfnamefont {J.~S.}\ \bibnamefont {Meyer}}\ and\ \bibinfo {author} {\bibfnamefont {M.}~\bibnamefont {Houzet}},\ }\bibfield  {title} {\bibinfo {title} {Josephson diode effect in a ballistic single-channel nanowire},\ }\href {https://doi.org/10.1063/5.0211491} {\bibfield  {journal} {\bibinfo  {journal} {Appl. Phys. Lett.}\ }\textbf {\bibinfo {volume} {125}} (\bibinfo {year} {2024})}\BibitemShut {NoStop}%
\bibitem [{\citenamefont {Liu}\ \emph {et~al.}(2024{\natexlab{a}})\citenamefont {Liu}, \citenamefont {Huang},\ and\ \citenamefont {Wang}}]{liu2024josephson}%
  \BibitemOpen
  \bibfield  {author} {\bibinfo {author} {\bibfnamefont {Z.}~\bibnamefont {Liu}}, \bibinfo {author} {\bibfnamefont {L.}~\bibnamefont {Huang}},\ and\ \bibinfo {author} {\bibfnamefont {J.}~\bibnamefont {Wang}},\ }\bibfield  {title} {\bibinfo {title} {Josephson diode effect in topological superconductors},\ }\href {https://doi.org/10.1103/physrevb.110.014519} {\bibfield  {journal} {\bibinfo  {journal} {Phys. Rev. B}\ }\textbf {\bibinfo {volume} {110}},\ \bibinfo {pages} {014519} (\bibinfo {year} {2024}{\natexlab{a}})}\BibitemShut {NoStop}%
\bibitem [{\citenamefont {Zazunov}\ \emph {et~al.}(2024)\citenamefont {Zazunov}, \citenamefont {Rech}, \citenamefont {Jonckheere}, \citenamefont {Gr\'emaud}, \citenamefont {Martin},\ and\ \citenamefont {Egger}}]{zazunov2023nonreciprocal}%
  \BibitemOpen
  \bibfield  {author} {\bibinfo {author} {\bibfnamefont {A.}~\bibnamefont {Zazunov}}, \bibinfo {author} {\bibfnamefont {J.}~\bibnamefont {Rech}}, \bibinfo {author} {\bibfnamefont {T.}~\bibnamefont {Jonckheere}}, \bibinfo {author} {\bibfnamefont {B.}~\bibnamefont {Gr\'emaud}}, \bibinfo {author} {\bibfnamefont {T.}~\bibnamefont {Martin}},\ and\ \bibinfo {author} {\bibfnamefont {R.}~\bibnamefont {Egger}},\ }\bibfield  {title} {\bibinfo {title} {{Nonreciprocal charge transport and subharmonic structure in voltage-biased Josephson diodes}},\ }\href {https://doi.org/10.1103/PhysRevB.109.024504} {\bibfield  {journal} {\bibinfo  {journal} {Phys. Rev. B}\ }\textbf {\bibinfo {volume} {109}},\ \bibinfo {pages} {024504} (\bibinfo {year} {2024})}\BibitemShut {NoStop}%
\bibitem [{\citenamefont {Volkov}\ \emph {et~al.}(2024)\citenamefont {Volkov}, \citenamefont {Lantagne-Hurtubise}, \citenamefont {Tummuru}, \citenamefont {Plugge}, \citenamefont {Pixley},\ and\ \citenamefont {Franz}}]{volkov2024josephson}%
  \BibitemOpen
  \bibfield  {author} {\bibinfo {author} {\bibfnamefont {P.~A.}\ \bibnamefont {Volkov}}, \bibinfo {author} {\bibfnamefont {{\'E}.}~\bibnamefont {Lantagne-Hurtubise}}, \bibinfo {author} {\bibfnamefont {T.}~\bibnamefont {Tummuru}}, \bibinfo {author} {\bibfnamefont {S.}~\bibnamefont {Plugge}}, \bibinfo {author} {\bibfnamefont {J.}~\bibnamefont {Pixley}},\ and\ \bibinfo {author} {\bibfnamefont {M.}~\bibnamefont {Franz}},\ }\bibfield  {title} {\bibinfo {title} {Josephson diode effects in twisted nodal superconductors},\ }\href {https://doi.org/10.1103/physrevb.109.094518} {\bibfield  {journal} {\bibinfo  {journal} {Phys. Rev. B}\ }\textbf {\bibinfo {volume} {109}},\ \bibinfo {pages} {094518} (\bibinfo {year} {2024})}\BibitemShut {NoStop}%
\bibitem [{\citenamefont {Lu}\ \emph {et~al.}(2024)\citenamefont {Lu}, \citenamefont {Maeda}, \citenamefont {Ito}, \citenamefont {Yada},\ and\ \citenamefont {Tanaka}}]{lu2024varphi}%
  \BibitemOpen
  \bibfield  {author} {\bibinfo {author} {\bibfnamefont {B.}~\bibnamefont {Lu}}, \bibinfo {author} {\bibfnamefont {K.}~\bibnamefont {Maeda}}, \bibinfo {author} {\bibfnamefont {H.}~\bibnamefont {Ito}}, \bibinfo {author} {\bibfnamefont {K.}~\bibnamefont {Yada}},\ and\ \bibinfo {author} {\bibfnamefont {Y.}~\bibnamefont {Tanaka}},\ }\bibfield  {title} {\bibinfo {title} {{$\varphi$ Josephson junction induced by altermagnetism}},\ }\href {https://doi.org/10.1103/PhysRevLett.133.226002} {\bibfield  {journal} {\bibinfo  {journal} {Phys. Rev. Lett.}\ }\textbf {\bibinfo {volume} {133}},\ \bibinfo {pages} {226002} (\bibinfo {year} {2024})}\BibitemShut {NoStop}%
\bibitem [{\citenamefont {Wang}\ \emph {et~al.}(2024)\citenamefont {Wang}, \citenamefont {Jiang}, \citenamefont {Wang},\ and\ \citenamefont {Liu}}]{wang2024efficient}%
  \BibitemOpen
  \bibfield  {author} {\bibinfo {author} {\bibfnamefont {J.}~\bibnamefont {Wang}}, \bibinfo {author} {\bibfnamefont {Y.}~\bibnamefont {Jiang}}, \bibinfo {author} {\bibfnamefont {J.~J.}\ \bibnamefont {Wang}},\ and\ \bibinfo {author} {\bibfnamefont {J.-F.}\ \bibnamefont {Liu}},\ }\bibfield  {title} {\bibinfo {title} {{Efficient Josephson diode effect on a two-dimensional topological insulator with asymmetric magnetization}},\ }\href {https://doi.org/10.1103/PhysRevB.109.075412} {\bibfield  {journal} {\bibinfo  {journal} {Phys. Rev. B}\ }\textbf {\bibinfo {volume} {109}},\ \bibinfo {pages} {075412} (\bibinfo {year} {2024})}\BibitemShut {NoStop}%
\bibitem [{\citenamefont {Cheng}\ \emph {et~al.}(2024)\citenamefont {Cheng}, \citenamefont {Mao},\ and\ \citenamefont {Sun}}]{cheng2024field}%
  \BibitemOpen
  \bibfield  {author} {\bibinfo {author} {\bibfnamefont {Q.}~\bibnamefont {Cheng}}, \bibinfo {author} {\bibfnamefont {Y.}~\bibnamefont {Mao}},\ and\ \bibinfo {author} {\bibfnamefont {Q.-F.}\ \bibnamefont {Sun}},\ }\bibfield  {title} {\bibinfo {title} {{Field-free Josephson diode effect in altermagnet/normal metal/altermagnet junctions}},\ }\href {https://doi.org/10.1103/physrevb.110.014518} {\bibfield  {journal} {\bibinfo  {journal} {Phys. Rev. B}\ }\textbf {\bibinfo {volume} {110}},\ \bibinfo {pages} {014518} (\bibinfo {year} {2024})}\BibitemShut {NoStop}%
\bibitem [{\citenamefont {Shaffer}\ \emph {et~al.}(2025)\citenamefont {Shaffer}, \citenamefont {Li}, \citenamefont {Hasan}, \citenamefont {Titov},\ and\ \citenamefont {Levchenko}}]{shaffer2025josephson}%
  \BibitemOpen
  \bibfield  {author} {\bibinfo {author} {\bibfnamefont {D.}~\bibnamefont {Shaffer}}, \bibinfo {author} {\bibfnamefont {S.}~\bibnamefont {Li}}, \bibinfo {author} {\bibfnamefont {J.}~\bibnamefont {Hasan}}, \bibinfo {author} {\bibfnamefont {M.}~\bibnamefont {Titov}},\ and\ \bibinfo {author} {\bibfnamefont {A.}~\bibnamefont {Levchenko}},\ }\bibfield  {title} {\bibinfo {title} {Josephson diode effect from nonequilibrium current in a superconducting interferometer},\ }\href {https://doi.org/10.1103/n9y8-31gc} {\bibfield  {journal} {\bibinfo  {journal} {Phys. Rev. B}\ }\textbf {\bibinfo {volume} {112}},\ \bibinfo {pages} {094509} (\bibinfo {year} {2025})}\BibitemShut {NoStop}%
\bibitem [{\citenamefont {Zhuang}\ \emph {et~al.}()\citenamefont {Zhuang}, \citenamefont {Shaffer}, \citenamefont {Hasan},\ and\ \citenamefont {Levchenko}}]{zhuang2025helical}%
  \BibitemOpen
  \bibfield  {author} {\bibinfo {author} {\bibfnamefont {Z.}~\bibnamefont {Zhuang}}, \bibinfo {author} {\bibfnamefont {D.}~\bibnamefont {Shaffer}}, \bibinfo {author} {\bibfnamefont {J.}~\bibnamefont {Hasan}},\ and\ \bibinfo {author} {\bibfnamefont {A.}~\bibnamefont {Levchenko}},\ }\href@noop {} {\bibinfo {title} {Helical phases and bogoliubov fermi surfaces probed by superconducting diode effects}},\ \Eprint {https://arxiv.org/abs/2510.18963 (2025)} {arXiv:2510.18963 (2025)} \BibitemShut {NoStop}%
\bibitem [{\citenamefont {Patil}\ \emph {et~al.}(2025)\citenamefont {Patil}, \citenamefont {Tang},\ and\ \citenamefont {Belzig}}]{patil2025spin}%
  \BibitemOpen
  \bibfield  {author} {\bibinfo {author} {\bibfnamefont {S.}~\bibnamefont {Patil}}, \bibinfo {author} {\bibfnamefont {G.}~\bibnamefont {Tang}},\ and\ \bibinfo {author} {\bibfnamefont {W.}~\bibnamefont {Belzig}},\ }\bibfield  {title} {\bibinfo {title} {{Spin-split Andreev bound states and diode effect in an Ising superconductor Josephson junction}},\ }\href {https://doi.org/10.1103/physrevb.111.l060502} {\bibfield  {journal} {\bibinfo  {journal} {Phys. Rev. B}\ }\textbf {\bibinfo {volume} {111}},\ \bibinfo {pages} {L060502} (\bibinfo {year} {2025})}\BibitemShut {NoStop}%
\bibitem [{\citenamefont {Shen}\ and\ \citenamefont {Zhang}(2025)}]{shen2025josephson}%
  \BibitemOpen
  \bibfield  {author} {\bibinfo {author} {\bibfnamefont {Q.-K.}\ \bibnamefont {Shen}}\ and\ \bibinfo {author} {\bibfnamefont {Y.}~\bibnamefont {Zhang}},\ }\bibfield  {title} {\bibinfo {title} {Josephson diodes induced by loop current states},\ }\href {https://doi.org/10.1103/PhysRevB.111.174515} {\bibfield  {journal} {\bibinfo  {journal} {Phys. Rev. B}\ }\textbf {\bibinfo {volume} {111}},\ \bibinfo {pages} {174515} (\bibinfo {year} {2025})}\BibitemShut {NoStop}%
\bibitem [{\citenamefont {Costa}\ \emph {et~al.}(2025)\citenamefont {Costa}, \citenamefont {Kanehira}, \citenamefont {Matsueda},\ and\ \citenamefont {Fabian}}]{costa2025unconventional}%
  \BibitemOpen
  \bibfield  {author} {\bibinfo {author} {\bibfnamefont {A.}~\bibnamefont {Costa}}, \bibinfo {author} {\bibfnamefont {O.}~\bibnamefont {Kanehira}}, \bibinfo {author} {\bibfnamefont {H.}~\bibnamefont {Matsueda}},\ and\ \bibinfo {author} {\bibfnamefont {J.}~\bibnamefont {Fabian}},\ }\bibfield  {title} {\bibinfo {title} {{Unconventional Josephson supercurrent diode effect induced by chiral spin-orbit coupling}},\ }\href {https://doi.org/10.1103/PhysRevB.111.L140506} {\bibfield  {journal} {\bibinfo  {journal} {Phys. Rev. B}\ }\textbf {\bibinfo {volume} {111}},\ \bibinfo {pages} {L140506} (\bibinfo {year} {2025})}\BibitemShut {NoStop}%
\bibitem [{\citenamefont {Cuozzo}\ and\ \citenamefont {L{\'e}onard}()}]{cuozzo2025perfect}%
  \BibitemOpen
  \bibfield  {author} {\bibinfo {author} {\bibfnamefont {J.~J.}\ \bibnamefont {Cuozzo}}\ and\ \bibinfo {author} {\bibfnamefont {F.}~\bibnamefont {L{\'e}onard}},\ }\href@noop {} {\bibinfo {title} {{Perfect supercurrent diode efficiency in chiral nanotube-based Josephson junctions}}},\ \Eprint {https://arxiv.org/abs/2504.02948 (2025)} {arXiv:2504.02948 (2025)} \BibitemShut {NoStop}%
\bibitem [{\citenamefont {Boruah}\ \emph {et~al.}()\citenamefont {Boruah}, \citenamefont {Acharjee},\ and\ \citenamefont {Saikia}}]{boruah2025field}%
  \BibitemOpen
  \bibfield  {author} {\bibinfo {author} {\bibfnamefont {A.}~\bibnamefont {Boruah}}, \bibinfo {author} {\bibfnamefont {S.}~\bibnamefont {Acharjee}},\ and\ \bibinfo {author} {\bibfnamefont {P.~K.}\ \bibnamefont {Saikia}},\ }\href@noop {} {\bibinfo {title} {{Field free Josephson diode effect in Ising superconductor/altermagnet Josephson junction}}},\ \Eprint {https://arxiv.org/abs/2504.00917 (2025)} {arXiv:2504.00917 (2025)} \BibitemShut {NoStop}%
\bibitem [{\citenamefont {Shaffer}\ and\ \citenamefont {Levchenko}()}]{shaffer2025theories}%
  \BibitemOpen
  \bibfield  {author} {\bibinfo {author} {\bibfnamefont {D.}~\bibnamefont {Shaffer}}\ and\ \bibinfo {author} {\bibfnamefont {A.}~\bibnamefont {Levchenko}},\ }\href@noop {} {\bibinfo {title} {Theories of superconducting diode effects}},\ \Eprint {https://arxiv.org/abs/2510.25864 (2025)} {arXiv:2510.25864 (2025)} \BibitemShut {NoStop}%
\bibitem [{\citenamefont {Steiner}\ \emph {et~al.}(2023)\citenamefont {Steiner}, \citenamefont {Melischek}, \citenamefont {Trahms}, \citenamefont {Franke},\ and\ \citenamefont {von Oppen}}]{steiner2023diode}%
  \BibitemOpen
  \bibfield  {author} {\bibinfo {author} {\bibfnamefont {J.~F.}\ \bibnamefont {Steiner}}, \bibinfo {author} {\bibfnamefont {L.}~\bibnamefont {Melischek}}, \bibinfo {author} {\bibfnamefont {M.}~\bibnamefont {Trahms}}, \bibinfo {author} {\bibfnamefont {K.~J.}\ \bibnamefont {Franke}},\ and\ \bibinfo {author} {\bibfnamefont {F.}~\bibnamefont {von Oppen}},\ }\bibfield  {title} {\bibinfo {title} {{Diode effects in current-biased Josephson junctions}},\ }\href {https://doi.org/10.1103/PhysRevLett.130.177002} {\bibfield  {journal} {\bibinfo  {journal} {Phys. Rev. Lett.}\ }\textbf {\bibinfo {volume} {130}},\ \bibinfo {pages} {177002} (\bibinfo {year} {2023})}\BibitemShut {NoStop}%
\bibitem [{\citenamefont {Stewart}(1968)}]{stewart1968current}%
  \BibitemOpen
  \bibfield  {author} {\bibinfo {author} {\bibfnamefont {W.~C.}\ \bibnamefont {Stewart}},\ }\bibfield  {title} {\bibinfo {title} {{Current-voltage characteristics of Josephson junctions}},\ }\href {https://doi.org/10.1063/1.1651991} {\bibfield  {journal} {\bibinfo  {journal} {Appl. Phys. Lett.}\ }\textbf {\bibinfo {volume} {12}},\ \bibinfo {pages} {277} (\bibinfo {year} {1968})}\BibitemShut {NoStop}%
\bibitem [{\citenamefont {McCumber}(1968)}]{mccumber1968effect}%
  \BibitemOpen
  \bibfield  {author} {\bibinfo {author} {\bibfnamefont {D.~E.}\ \bibnamefont {McCumber}},\ }\bibfield  {title} {\bibinfo {title} {Effect of ac impedance on dc voltage‐current characteristics of superconductor weak‐link junctions},\ }\href {https://doi.org/10.1063/1.1656743} {\bibfield  {journal} {\bibinfo  {journal} {J. Appl. Phys.}\ }\textbf {\bibinfo {volume} {39}},\ \bibinfo {pages} {3113} (\bibinfo {year} {1968})}\BibitemShut {NoStop}%
\bibitem [{\citenamefont {Stewart}(1974)}]{stewart1974current}%
  \BibitemOpen
  \bibfield  {author} {\bibinfo {author} {\bibfnamefont {W.}~\bibnamefont {Stewart}},\ }\bibfield  {title} {\bibinfo {title} {Current-voltage characteristics of superconducting tunnel junctions},\ }\href {https://doi.org/10.1063/1.1663001} {\bibfield  {journal} {\bibinfo  {journal} {J. Appl. Phys.}\ }\textbf {\bibinfo {volume} {45}},\ \bibinfo {pages} {452} (\bibinfo {year} {1974})}\BibitemShut {NoStop}%
\bibitem [{\citenamefont {Kautz}\ and\ \citenamefont {Martinis}(1990)}]{kautz1990noise}%
  \BibitemOpen
  \bibfield  {author} {\bibinfo {author} {\bibfnamefont {R.~L.}\ \bibnamefont {Kautz}}\ and\ \bibinfo {author} {\bibfnamefont {J.~M.}\ \bibnamefont {Martinis}},\ }\bibfield  {title} {\bibinfo {title} {{Noise-affected I-V curves in small hysteretic Josephson junctions}},\ }\href {https://doi.org/10.1103/PhysRevB.42.9903} {\bibfield  {journal} {\bibinfo  {journal} {Phys. Rev. B}\ }\textbf {\bibinfo {volume} {42}},\ \bibinfo {pages} {9903} (\bibinfo {year} {1990})}\BibitemShut {NoStop}%
\bibitem [{\citenamefont {Ambegaokar}\ \emph {et~al.}(1982)\citenamefont {Ambegaokar}, \citenamefont {Eckern},\ and\ \citenamefont {Sch\"on}}]{ambegaokar1982quantum}%
  \BibitemOpen
  \bibfield  {author} {\bibinfo {author} {\bibfnamefont {V.}~\bibnamefont {Ambegaokar}}, \bibinfo {author} {\bibfnamefont {U.}~\bibnamefont {Eckern}},\ and\ \bibinfo {author} {\bibfnamefont {G.}~\bibnamefont {Sch\"on}},\ }\bibfield  {title} {\bibinfo {title} {Quantum dynamics of tunneling between superconductors},\ }\href {https://doi.org/10.1103/PhysRevLett.48.1745} {\bibfield  {journal} {\bibinfo  {journal} {Phys. Rev. Lett.}\ }\textbf {\bibinfo {volume} {48}},\ \bibinfo {pages} {1745} (\bibinfo {year} {1982})}\BibitemShut {NoStop}%
\bibitem [{\citenamefont {Eckern}\ \emph {et~al.}(1984)\citenamefont {Eckern}, \citenamefont {Sch\"on},\ and\ \citenamefont {Ambegaokar}}]{eckern1984quantum}%
  \BibitemOpen
  \bibfield  {author} {\bibinfo {author} {\bibfnamefont {U.}~\bibnamefont {Eckern}}, \bibinfo {author} {\bibfnamefont {G.}~\bibnamefont {Sch\"on}},\ and\ \bibinfo {author} {\bibfnamefont {V.}~\bibnamefont {Ambegaokar}},\ }\bibfield  {title} {\bibinfo {title} {Quantum dynamics of a superconducting tunnel junction},\ }\href {https://doi.org/10.1103/PhysRevB.30.6419} {\bibfield  {journal} {\bibinfo  {journal} {Phys. Rev. B}\ }\textbf {\bibinfo {volume} {30}},\ \bibinfo {pages} {6419} (\bibinfo {year} {1984})}\BibitemShut {NoStop}%
\bibitem [{\citenamefont {Andreev}(1964)}]{andreev1964thermal}%
  \BibitemOpen
  \bibfield  {author} {\bibinfo {author} {\bibfnamefont {A.~F.}\ \bibnamefont {Andreev}},\ }\bibfield  {title} {\bibinfo {title} {Thermal conductivity of the intermediate state of superconductors},\ }\href {http://www.jetp.ras.ru/cgi-bin/e/index/e/19/5/p1228?a=list} {\bibfield  {journal} {\bibinfo  {journal} {Sov. Phys. JETP}\ }\textbf {\bibinfo {volume} {19}},\ \bibinfo {pages} {1228} (\bibinfo {year} {1964})}\BibitemShut {NoStop}%
\bibitem [{\citenamefont {Klapwijk}\ \emph {et~al.}(1982)\citenamefont {Klapwijk}, \citenamefont {Blonder},\ and\ \citenamefont {Tinkham}}]{klapwijk1982explanation}%
  \BibitemOpen
  \bibfield  {author} {\bibinfo {author} {\bibfnamefont {T.}~\bibnamefont {Klapwijk}}, \bibinfo {author} {\bibfnamefont {G.}~\bibnamefont {Blonder}},\ and\ \bibinfo {author} {\bibfnamefont {M.}~\bibnamefont {Tinkham}},\ }\bibfield  {title} {\bibinfo {title} {Explanation of subharmonic energy gap structure in superconducting contacts},\ }\href {https://doi.org/https://doi.org/10.1016/0378-4363(82)90189-9} {\bibfield  {journal} {\bibinfo  {journal} {Physica B+C}\ }\textbf {\bibinfo {volume} {109-110}},\ \bibinfo {pages} {1657} (\bibinfo {year} {1982})}\BibitemShut {NoStop}%
\bibitem [{\citenamefont {Octavio}\ \emph {et~al.}(1983)\citenamefont {Octavio}, \citenamefont {Tinkham}, \citenamefont {Blonder},\ and\ \citenamefont {Klapwijk}}]{octavio1983subharmonic}%
  \BibitemOpen
  \bibfield  {author} {\bibinfo {author} {\bibfnamefont {M.}~\bibnamefont {Octavio}}, \bibinfo {author} {\bibfnamefont {M.}~\bibnamefont {Tinkham}}, \bibinfo {author} {\bibfnamefont {G.}~\bibnamefont {Blonder}},\ and\ \bibinfo {author} {\bibfnamefont {T.}~\bibnamefont {Klapwijk}},\ }\bibfield  {title} {\bibinfo {title} {Subharmonic energy-gap structure in superconducting constrictions},\ }\href {https://doi.org/10.1103/physrevb.27.6739} {\bibfield  {journal} {\bibinfo  {journal} {Phys. Rev. B}\ }\textbf {\bibinfo {volume} {27}},\ \bibinfo {pages} {6739} (\bibinfo {year} {1983})}\BibitemShut {NoStop}%
\bibitem [{\citenamefont {Averin}\ and\ \citenamefont {Bardas}(1995)}]{averin1995ac}%
  \BibitemOpen
  \bibfield  {author} {\bibinfo {author} {\bibfnamefont {D.}~\bibnamefont {Averin}}\ and\ \bibinfo {author} {\bibfnamefont {A.}~\bibnamefont {Bardas}},\ }\bibfield  {title} {\bibinfo {title} {{ac Josephson effect in a single quantum channel}},\ }\href {https://doi.org/10.1103/physrevlett.75.1831} {\bibfield  {journal} {\bibinfo  {journal} {Phys. Rev. Lett.}\ }\textbf {\bibinfo {volume} {75}},\ \bibinfo {pages} {1831} (\bibinfo {year} {1995})}\BibitemShut {NoStop}%
\bibitem [{\citenamefont {Cuevas}\ \emph {et~al.}(1996)\citenamefont {Cuevas}, \citenamefont {Mart\'{\i}n-Rodero},\ and\ \citenamefont {Yeyati}}]{cuevas1996hamiltonian}%
  \BibitemOpen
  \bibfield  {author} {\bibinfo {author} {\bibfnamefont {J.~C.}\ \bibnamefont {Cuevas}}, \bibinfo {author} {\bibfnamefont {A.}~\bibnamefont {Mart\'{\i}n-Rodero}},\ and\ \bibinfo {author} {\bibfnamefont {A.~L.}\ \bibnamefont {Yeyati}},\ }\bibfield  {title} {\bibinfo {title} {Hamiltonian approach to the transport properties of superconducting quantum point contacts},\ }\href {https://doi.org/10.1103/PhysRevB.54.7366} {\bibfield  {journal} {\bibinfo  {journal} {Phys. Rev. B}\ }\textbf {\bibinfo {volume} {54}},\ \bibinfo {pages} {7366} (\bibinfo {year} {1996})}\BibitemShut {NoStop}%
\bibitem [{\citenamefont {Cuevas}\ \emph {et~al.}(1999)\citenamefont {Cuevas}, \citenamefont {Mart\'{\i}n-Rodero},\ and\ \citenamefont {Yeyati}}]{cuevas1999shot}%
  \BibitemOpen
  \bibfield  {author} {\bibinfo {author} {\bibfnamefont {J.~C.}\ \bibnamefont {Cuevas}}, \bibinfo {author} {\bibfnamefont {A.}~\bibnamefont {Mart\'{\i}n-Rodero}},\ and\ \bibinfo {author} {\bibfnamefont {A.~L.}\ \bibnamefont {Yeyati}},\ }\bibfield  {title} {\bibinfo {title} {Shot noise and coherent multiple charge transfer in superconducting quantum point contacts},\ }\href {https://doi.org/10.1103/PhysRevLett.82.4086} {\bibfield  {journal} {\bibinfo  {journal} {Phys. Rev. Lett.}\ }\textbf {\bibinfo {volume} {82}},\ \bibinfo {pages} {4086} (\bibinfo {year} {1999})}\BibitemShut {NoStop}%
\bibitem [{\citenamefont {Tobiska}\ and\ \citenamefont {Nazarov}(2005)}]{tobiska2005inelastic}%
  \BibitemOpen
  \bibfield  {author} {\bibinfo {author} {\bibfnamefont {J.}~\bibnamefont {Tobiska}}\ and\ \bibinfo {author} {\bibfnamefont {Y.~V.}\ \bibnamefont {Nazarov}},\ }\bibfield  {title} {\bibinfo {title} {Inelastic interaction corrections and universal relations for full counting statistics in a quantum contact},\ }\href {https://doi.org/10.1103/PhysRevB.72.235328} {\bibfield  {journal} {\bibinfo  {journal} {Phys. Rev. B}\ }\textbf {\bibinfo {volume} {72}},\ \bibinfo {pages} {235328} (\bibinfo {year} {2005})}\BibitemShut {NoStop}%
\bibitem [{\citenamefont {Saito}\ and\ \citenamefont {Utsumi}(2008)}]{saito2008symmetry}%
  \BibitemOpen
  \bibfield  {author} {\bibinfo {author} {\bibfnamefont {K.}~\bibnamefont {Saito}}\ and\ \bibinfo {author} {\bibfnamefont {Y.}~\bibnamefont {Utsumi}},\ }\bibfield  {title} {\bibinfo {title} {Symmetry in full counting statistics, fluctuation theorem, and relations among nonlinear transport coefficients in the presence of a magnetic field},\ }\href {https://doi.org/10.1103/PhysRevB.78.115429} {\bibfield  {journal} {\bibinfo  {journal} {Phys. Rev. B}\ }\textbf {\bibinfo {volume} {78}},\ \bibinfo {pages} {115429} (\bibinfo {year} {2008})}\BibitemShut {NoStop}%
\bibitem [{\citenamefont {Esposito}\ \emph {et~al.}(2009)\citenamefont {Esposito}, \citenamefont {Harbola},\ and\ \citenamefont {Mukamel}}]{esposito2009nonequilibrium}%
  \BibitemOpen
  \bibfield  {author} {\bibinfo {author} {\bibfnamefont {M.}~\bibnamefont {Esposito}}, \bibinfo {author} {\bibfnamefont {U.}~\bibnamefont {Harbola}},\ and\ \bibinfo {author} {\bibfnamefont {S.}~\bibnamefont {Mukamel}},\ }\bibfield  {title} {\bibinfo {title} {Nonequilibrium fluctuations, fluctuation theorems, and counting statistics in quantum systems},\ }\href {https://doi.org/10.1103/RevModPhys.81.1665} {\bibfield  {journal} {\bibinfo  {journal} {Rev. Mod. Phys.}\ }\textbf {\bibinfo {volume} {81}},\ \bibinfo {pages} {1665} (\bibinfo {year} {2009})}\BibitemShut {NoStop}%
\bibitem [{\citenamefont {Campisi}\ \emph {et~al.}(2011)\citenamefont {Campisi}, \citenamefont {H\"anggi},\ and\ \citenamefont {Talkner}}]{campisi2011colloquium}%
  \BibitemOpen
  \bibfield  {author} {\bibinfo {author} {\bibfnamefont {M.}~\bibnamefont {Campisi}}, \bibinfo {author} {\bibfnamefont {P.}~\bibnamefont {H\"anggi}},\ and\ \bibinfo {author} {\bibfnamefont {P.}~\bibnamefont {Talkner}},\ }\bibfield  {title} {\bibinfo {title} {Quantum fluctuation relations: Foundations and applications},\ }\href {https://doi.org/10.1103/RevModPhys.83.771} {\bibfield  {journal} {\bibinfo  {journal} {Rev. Mod. Phys.}\ }\textbf {\bibinfo {volume} {83}},\ \bibinfo {pages} {771} (\bibinfo {year} {2011})}\BibitemShut {NoStop}%
\bibitem [{\citenamefont {Werthamer}(1966)}]{werthamer1966nonlinear}%
  \BibitemOpen
  \bibfield  {author} {\bibinfo {author} {\bibfnamefont {N.~R.}\ \bibnamefont {Werthamer}},\ }\bibfield  {title} {\bibinfo {title} {{Nonlinear self-coupling of Josephson radiation in superconducting tunnel junctions}},\ }\href {https://doi.org/10.1103/PhysRev.147.255} {\bibfield  {journal} {\bibinfo  {journal} {Phys. Rev.}\ }\textbf {\bibinfo {volume} {147}},\ \bibinfo {pages} {255} (\bibinfo {year} {1966})}\BibitemShut {NoStop}%
\bibitem [{\citenamefont {Rossignol}\ \emph {et~al.}(2019)\citenamefont {Rossignol}, \citenamefont {Kloss},\ and\ \citenamefont {Waintal}}]{rossignol2019role}%
  \BibitemOpen
  \bibfield  {author} {\bibinfo {author} {\bibfnamefont {B.}~\bibnamefont {Rossignol}}, \bibinfo {author} {\bibfnamefont {T.}~\bibnamefont {Kloss}},\ and\ \bibinfo {author} {\bibfnamefont {X.}~\bibnamefont {Waintal}},\ }\bibfield  {title} {\bibinfo {title} {{Role of quasiparticles in an electric circuit with Josephson junctions}},\ }\href {https://link.aps.org/accepted/10.1103/PhysRevLett.122.207702} {\bibfield  {journal} {\bibinfo  {journal} {Phys. Rev. Lett.}\ }\textbf {\bibinfo {volume} {122}},\ \bibinfo {pages} {207702} (\bibinfo {year} {2019})}\BibitemShut {NoStop}%
\bibitem [{\citenamefont {Choi}\ and\ \citenamefont {Trauzettel}(2022)}]{choi2022microscopic}%
  \BibitemOpen
  \bibfield  {author} {\bibinfo {author} {\bibfnamefont {S.-J.}\ \bibnamefont {Choi}}\ and\ \bibinfo {author} {\bibfnamefont {B.}~\bibnamefont {Trauzettel}},\ }\bibfield  {title} {\bibinfo {title} {Microscopic theory of the current-voltage characteristics of josephson tunnel junctions},\ }\href {https://doi.org/10.1103/PhysRevLett.128.126801} {\bibfield  {journal} {\bibinfo  {journal} {Phys. Rev. Lett.}\ }\textbf {\bibinfo {volume} {128}},\ \bibinfo {pages} {126801} (\bibinfo {year} {2022})}\BibitemShut {NoStop}%
\bibitem [{\citenamefont {Lahiri}\ \emph {et~al.}(2023)\citenamefont {Lahiri}, \citenamefont {Choi},\ and\ \citenamefont {Trauzettel}}]{lahiri2023nonequilibrium}%
  \BibitemOpen
  \bibfield  {author} {\bibinfo {author} {\bibfnamefont {A.}~\bibnamefont {Lahiri}}, \bibinfo {author} {\bibfnamefont {S.-J.}\ \bibnamefont {Choi}},\ and\ \bibinfo {author} {\bibfnamefont {B.}~\bibnamefont {Trauzettel}},\ }\bibfield  {title} {\bibinfo {title} {{Nonequilibrium fractional Josephson effect}},\ }\href {https://doi.org/10.1103/PhysRevLett.131.126301} {\bibfield  {journal} {\bibinfo  {journal} {Phys. Rev. Lett.}\ }\textbf {\bibinfo {volume} {131}},\ \bibinfo {pages} {126301} (\bibinfo {year} {2023})}\BibitemShut {NoStop}%
\bibitem [{\citenamefont {Lahiri}\ \emph {et~al.}(2025)\citenamefont {Lahiri}, \citenamefont {Choi},\ and\ \citenamefont {Trauzettel}}]{lahiri2025origin}%
  \BibitemOpen
  \bibfield  {author} {\bibinfo {author} {\bibfnamefont {A.}~\bibnamefont {Lahiri}}, \bibinfo {author} {\bibfnamefont {S.-J.}\ \bibnamefont {Choi}},\ and\ \bibinfo {author} {\bibfnamefont {B.}~\bibnamefont {Trauzettel}},\ }\bibfield  {title} {\bibinfo {title} {{Origin of the subharmonic gap structure of dc current-biased Josephson junctions}},\ }\href {https://doi.org/10.1103/zbd5-1cc2} {\bibfield  {journal} {\bibinfo  {journal} {Phys. Rev. B}\ }\textbf {\bibinfo {volume} {111}},\ \bibinfo {pages} {L220504} (\bibinfo {year} {2025})}\BibitemShut {NoStop}%
\bibitem [{\citenamefont {Nazarov}(2007)}]{nazarov2007full}%
  \BibitemOpen
  \bibfield  {author} {\bibinfo {author} {\bibfnamefont {Y.}~\bibnamefont {Nazarov}},\ }\bibfield  {title} {\bibinfo {title} {Full counting statistics and field theory},\ }\href {https://doi.org/https://doi.org/10.1002/andp.200751910-1106} {\bibfield  {journal} {\bibinfo  {journal} {Ann. Phys. (Leipzig)}\ }\textbf {\bibinfo {volume} {519}},\ \bibinfo {pages} {720} (\bibinfo {year} {2007})}\BibitemShut {NoStop}%
\bibitem [{\citenamefont {Ingold}\ and\ \citenamefont {Nazarov}(1992)}]{ingold1992single}%
  \BibitemOpen
  \bibfield  {author} {\bibinfo {author} {\bibfnamefont {G.-L.}\ \bibnamefont {Ingold}}\ and\ \bibinfo {author} {\bibfnamefont {Y.}~\bibnamefont {Nazarov}},\ }\bibfield  {title} {\bibinfo {title} {Single charge tunneling},\ }\href {https://arxiv.org/pdf/cond-mat/0508728} {\bibfield  {journal} {\bibinfo  {journal} {NATO ASI Series B}\ }\textbf {\bibinfo {volume} {294}},\ \bibinfo {pages} {21} (\bibinfo {year} {1992})}\BibitemShut {NoStop}%
\bibitem [{Note1()}]{Note1}%
  \BibitemOpen
  \bibinfo {note} {It may be possible to relax this assumption by adapting insights from Nazarov's Keldysh action for full counting statistics in terms of the scattering matrix \cite {snyman2008keldysh}.}\BibitemShut {Stop}%
\bibitem [{\citenamefont {Kamenev}(2011)}]{kamenev2011field}%
  \BibitemOpen
  \bibfield  {author} {\bibinfo {author} {\bibfnamefont {A.}~\bibnamefont {Kamenev}},\ }\href {https://doi.org/10.1017/cbo9781139003667} {\emph {\bibinfo {title} {Field Theory of Non-Equilibrium Systems}}}\ (\bibinfo  {publisher} {Cambridge University Press, Cambridge},\ \bibinfo {year} {2011})\BibitemShut {NoStop}%
\bibitem [{Note2()}]{Note2}%
  \BibitemOpen
  \bibinfo {note} {This is equivalent to the introduction of a source field through $H_\eta = H \pm _\protect \mathcal {C} \protect \tfrac {1}{2}\eta (t)\protect \mathcal {I}_\protect \mathrm {tun}$ up to contact terms which arise from repeated action of derivatives on the exponential factors $\exp {\mp \pm _\protect \mathcal {C}\protect \tfrac {i}{2} e\eta }$. The latter arise only for cumulants of third and higher order, which are dropped below.}\BibitemShut {Stop}%
\bibitem [{\citenamefont {Schmid}(1982)}]{schmid1982quasiclassical}%
  \BibitemOpen
  \bibfield  {author} {\bibinfo {author} {\bibfnamefont {A.}~\bibnamefont {Schmid}},\ }\bibfield  {title} {\bibinfo {title} {{On a quasiclassical Langevin equation}},\ }\href {https://doi.org/https://doi.org/10.1007/BF00681904} {\bibfield  {journal} {\bibinfo  {journal} {J.\ Low Temp.\ Phys.}\ }\textbf {\bibinfo {volume} {49}},\ \bibinfo {pages} {609} (\bibinfo {year} {1982})}\BibitemShut {NoStop}%
\bibitem [{Note3()}]{Note3}%
  \BibitemOpen
  \bibinfo {note} {In the overdamped regime, the velocity is proportional to the instantaneous force. This leads to $\protect \ddot { \varphi }/\protect \dot {\varphi } \sim \omega _\protect \textrm {pl}^2 \tau _{RC}$ which should be small compared to $\omega _\protect \textrm {pl}$ in the overdamped regime. However, $\omega _\protect \textrm {pl}^2 \tau _{RC} \sim E_J R / (e^2 /h) \sim \Delta R/R_N$ independent of $C$, with $R_N$ the resistance in the normal state. This leads to a contradiction unless $R$ is much smaller than $R_N$. This is rather unrealistic if $R$ represents the resistance from quasiparticle tunneling only but plausible if the external electromagnetic environment is included \cite {kautz1990noise}.}\BibitemShut {Stop}%
\bibitem [{\citenamefont {Liu}\ \emph {et~al.}(2024{\natexlab{b}})\citenamefont {Liu}, \citenamefont {Smith}, \citenamefont {Andreev},\ and\ \citenamefont {Spivak}}]{liu2024giant}%
  \BibitemOpen
  \bibfield  {author} {\bibinfo {author} {\bibfnamefont {T.}~\bibnamefont {Liu}}, \bibinfo {author} {\bibfnamefont {M.}~\bibnamefont {Smith}}, \bibinfo {author} {\bibfnamefont {A.~V.}\ \bibnamefont {Andreev}},\ and\ \bibinfo {author} {\bibfnamefont {B.~Z.}\ \bibnamefont {Spivak}},\ }\bibfield  {title} {\bibinfo {title} {Giant nonreciprocity of current-voltage characteristics of noncentrosymmetric superconductor--normal metal--superconductor junctions},\ }\href {https://doi.org/10.1103/PhysRevB.109.L020501} {\bibfield  {journal} {\bibinfo  {journal} {Phys. Rev. B}\ }\textbf {\bibinfo {volume} {109}},\ \bibinfo {pages} {L020501} (\bibinfo {year} {2024}{\natexlab{b}})}\BibitemShut {NoStop}%
\bibitem [{\citenamefont {Martin}\ and\ \citenamefont {Mozyrsky}(2014)}]{martin2014nonequilibrium}%
  \BibitemOpen
  \bibfield  {author} {\bibinfo {author} {\bibfnamefont {I.}~\bibnamefont {Martin}}\ and\ \bibinfo {author} {\bibfnamefont {D.}~\bibnamefont {Mozyrsky}},\ }\bibfield  {title} {\bibinfo {title} {Nonequilibrium theory of tunneling into a localized state in a superconductor},\ }\href {https://doi.org/10.1103/PhysRevB.90.100508} {\bibfield  {journal} {\bibinfo  {journal} {Phys. Rev. B}\ }\textbf {\bibinfo {volume} {90}},\ \bibinfo {pages} {100508} (\bibinfo {year} {2014})}\BibitemShut {NoStop}%
\bibitem [{\citenamefont {Ruby}\ \emph {et~al.}(2015)\citenamefont {Ruby}, \citenamefont {Pientka}, \citenamefont {Peng}, \citenamefont {von Oppen}, \citenamefont {Heinrich},\ and\ \citenamefont {Franke}}]{ruby2015tunneling}%
  \BibitemOpen
  \bibfield  {author} {\bibinfo {author} {\bibfnamefont {M.}~\bibnamefont {Ruby}}, \bibinfo {author} {\bibfnamefont {F.}~\bibnamefont {Pientka}}, \bibinfo {author} {\bibfnamefont {Y.}~\bibnamefont {Peng}}, \bibinfo {author} {\bibfnamefont {F.}~\bibnamefont {von Oppen}}, \bibinfo {author} {\bibfnamefont {B.~W.}\ \bibnamefont {Heinrich}},\ and\ \bibinfo {author} {\bibfnamefont {K.~J.}\ \bibnamefont {Franke}},\ }\bibfield  {title} {\bibinfo {title} {Tunneling processes into localized subgap states in superconductors},\ }\href {https://doi.org/https://doi.org/10.1103/PhysRevLett.115.087001} {\bibfield  {journal} {\bibinfo  {journal} {Phys. Rev. Lett.}\ }\textbf {\bibinfo {volume} {115}},\ \bibinfo {pages} {087001} (\bibinfo {year} {2015})}\BibitemShut {NoStop}%
\bibitem [{\citenamefont {Yu}(1965)}]{yu1965bound}%
  \BibitemOpen
  \bibfield  {author} {\bibinfo {author} {\bibfnamefont {L.}~\bibnamefont {Yu}},\ }\bibfield  {title} {\bibinfo {title} {Bound state in superconductors with paramagnetic impurities},\ }\href {https://doi.org/10.7498/aps.21.75} {\bibfield  {journal} {\bibinfo  {journal} {Acta Phys. Sin.}\ }\textbf {\bibinfo {volume} {21}},\ \bibinfo {pages} {75} (\bibinfo {year} {1965})}\BibitemShut {NoStop}%
\bibitem [{\citenamefont {Shiba}(1968)}]{shiba1968classical}%
  \BibitemOpen
  \bibfield  {author} {\bibinfo {author} {\bibfnamefont {H.}~\bibnamefont {Shiba}},\ }\bibfield  {title} {\bibinfo {title} {{Classical Spins in Superconductors}},\ }\href {https://doi.org/10.1143/PTP.40.435} {\bibfield  {journal} {\bibinfo  {journal} {Prog. Theor. Phys.}\ }\textbf {\bibinfo {volume} {40}},\ \bibinfo {pages} {435} (\bibinfo {year} {1968})}\BibitemShut {NoStop}%
\bibitem [{\citenamefont {Rusinov}(1969)}]{rusinov1969superconcductivity}%
  \BibitemOpen
  \bibfield  {author} {\bibinfo {author} {\bibfnamefont {A.~I.}\ \bibnamefont {Rusinov}},\ }\bibfield  {title} {\bibinfo {title} {{Superconductivity near a paramagnetic impurity}},\ }\href {http://jetpletters.ru/ps/1658/article_25295.shtml} {\bibfield  {journal} {\bibinfo  {journal} {JETP Lett.}\ }\textbf {\bibinfo {volume} {9}},\ \bibinfo {pages} {85} (\bibinfo {year} {1969})}\BibitemShut {NoStop}%
\bibitem [{\citenamefont {Misaki}\ and\ \citenamefont {Nagaosa}(2021)}]{misaki2021theory}%
  \BibitemOpen
  \bibfield  {author} {\bibinfo {author} {\bibfnamefont {K.}~\bibnamefont {Misaki}}\ and\ \bibinfo {author} {\bibfnamefont {N.}~\bibnamefont {Nagaosa}},\ }\bibfield  {title} {\bibinfo {title} {{Theory of the nonreciprocal Josephson effect}},\ }\href {https://doi.org/10.1103/physrevb.103.245302} {\bibfield  {journal} {\bibinfo  {journal} {Phys. Rev. B}\ }\textbf {\bibinfo {volume} {103}},\ \bibinfo {pages} {245302} (\bibinfo {year} {2021})}\BibitemShut {NoStop}%
\bibitem [{\citenamefont {Seleznev}\ and\ \citenamefont {Fominov}(2024)}]{seleznev2024influence}%
  \BibitemOpen
  \bibfield  {author} {\bibinfo {author} {\bibfnamefont {G.~S.}\ \bibnamefont {Seleznev}}\ and\ \bibinfo {author} {\bibfnamefont {Y.~V.}\ \bibnamefont {Fominov}},\ }\bibfield  {title} {\bibinfo {title} {{Influence of capacitance and thermal fluctuations on the Josephson diode effect in asymmetric higher-harmonic SQUIDs}},\ }\href {https://doi.org/10.1103/PhysRevB.110.104508} {\bibfield  {journal} {\bibinfo  {journal} {Phys. Rev. B}\ }\textbf {\bibinfo {volume} {110}},\ \bibinfo {pages} {104508} (\bibinfo {year} {2024})}\BibitemShut {NoStop}%
\bibitem [{\citenamefont {Seoane~Souto}\ \emph {et~al.}(2024)\citenamefont {Seoane~Souto}, \citenamefont {Leijnse}, \citenamefont {Schrade}, \citenamefont {Valentini}, \citenamefont {Katsaros},\ and\ \citenamefont {Danon}}]{souto2024tuning}%
  \BibitemOpen
  \bibfield  {author} {\bibinfo {author} {\bibfnamefont {R.}~\bibnamefont {Seoane~Souto}}, \bibinfo {author} {\bibfnamefont {M.}~\bibnamefont {Leijnse}}, \bibinfo {author} {\bibfnamefont {C.}~\bibnamefont {Schrade}}, \bibinfo {author} {\bibfnamefont {M.}~\bibnamefont {Valentini}}, \bibinfo {author} {\bibfnamefont {G.}~\bibnamefont {Katsaros}},\ and\ \bibinfo {author} {\bibfnamefont {J.}~\bibnamefont {Danon}},\ }\bibfield  {title} {\bibinfo {title} {Tuning the josephson diode response with an ac current},\ }\href {https://doi.org/10.1103/PhysRevResearch.6.L022002} {\bibfield  {journal} {\bibinfo  {journal} {Phys. Rev. Res.}\ }\textbf {\bibinfo {volume} {6}},\ \bibinfo {pages} {L022002} (\bibinfo {year} {2024})}\BibitemShut {NoStop}%
\bibitem [{\citenamefont {Monroe}\ \emph {et~al.}(2024)\citenamefont {Monroe}, \citenamefont {Shen}, \citenamefont {Tringali}, \citenamefont {Alidoust}, \citenamefont {Zhou},\ and\ \citenamefont {{\v{Z}}uti{\'c}}}]{monroe2024phase}%
  \BibitemOpen
  \bibfield  {author} {\bibinfo {author} {\bibfnamefont {D.}~\bibnamefont {Monroe}}, \bibinfo {author} {\bibfnamefont {C.}~\bibnamefont {Shen}}, \bibinfo {author} {\bibfnamefont {D.}~\bibnamefont {Tringali}}, \bibinfo {author} {\bibfnamefont {M.}~\bibnamefont {Alidoust}}, \bibinfo {author} {\bibfnamefont {T.}~\bibnamefont {Zhou}},\ and\ \bibinfo {author} {\bibfnamefont {I.}~\bibnamefont {{\v{Z}}uti{\'c}}},\ }\bibfield  {title} {\bibinfo {title} {Phase jumps in josephson junctions with time-dependent spin--orbit coupling},\ }\href {https://doi.org/10.1063/5.0211562} {\bibfield  {journal} {\bibinfo  {journal} {Applied Physics Letters}\ }\textbf {\bibinfo {volume} {125}} (\bibinfo {year} {2024})}\BibitemShut {NoStop}%
\bibitem [{\citenamefont {Wang}\ \emph {et~al.}()\citenamefont {Wang}, \citenamefont {Wang},\ and\ \citenamefont {Wu}}]{wang2025josephson}%
  \BibitemOpen
  \bibfield  {author} {\bibinfo {author} {\bibfnamefont {D.}~\bibnamefont {Wang}}, \bibinfo {author} {\bibfnamefont {Q.-H.}\ \bibnamefont {Wang}},\ and\ \bibinfo {author} {\bibfnamefont {C.}~\bibnamefont {Wu}},\ }\href@noop {} {\bibinfo {title} {Josephson diode effect: a phenomenological perspective}},\ \Eprint {https://arxiv.org/abs/2506.23200 (2025)} {arXiv:2506.23200 (2025)} \BibitemShut {NoStop}%
\bibitem [{\citenamefont {Snyman}\ and\ \citenamefont {Nazarov}(2008)}]{snyman2008keldysh}%
  \BibitemOpen
  \bibfield  {author} {\bibinfo {author} {\bibfnamefont {I.}~\bibnamefont {Snyman}}\ and\ \bibinfo {author} {\bibfnamefont {Y.~V.}\ \bibnamefont {Nazarov}},\ }\bibfield  {title} {\bibinfo {title} {Keldysh action of a multiterminal time-dependent scatterer},\ }\href {https://doi.org/10.1103/PhysRevB.77.165118} {\bibfield  {journal} {\bibinfo  {journal} {Phys. Rev. B}\ }\textbf {\bibinfo {volume} {77}},\ \bibinfo {pages} {165118} (\bibinfo {year} {2008})}\BibitemShut {NoStop}%
\end{thebibliography}%

\end{document}